\documentclass[letterpaper,10pt]{article}
\usepackage{graphicx}	% For figures/pictures
\usepackage{caption}		% For captioning non-floats (multicol doesn't play w/floats)
\usepackage{amsmath}	% For math/equations
\usepackage{cite}		% For bibtex citing
\usepackage{multicol} 	% For two column environment
\usepackage{tabularx}
\usepackage{array}
\usepackage{geometry}
\usepackage{titlesec}
\titleformat{\section}{\large\bfseries}{\thesection}{0.1cm}{}
\titleformat{\subsection}{\normalsize\itshape}{\thesubsection}{0.3cm}{}
\titleformat{\subsubsection}{\normalsize\itshape}{\thesubsubsection}{0.05cm}{}

\titlespacing*{\section}{0pt}{22pt}{4pt} %12pt (template) + 10pt (extra space)
\titlespacing*{\subsection}{0pt}{16pt}{3pt} %6pt (template) + 10pt (extra space)

\renewcommand{\thesection}{\Roman{section}.} 
\renewcommand{\thesubsection}{\Alph{subsection}.}
\renewcommand{\thesubsubsection}{\Alph{subsection}.\arabic{subsubsection}.}

\newcolumntype{L}{>{\raggedright\arraybackslash}X}
\usepackage[T1]{fontenc}
\usepackage{newtxmath,newtxtext}
\Large
\title{\vspace{-2cm}\textbf{Capturing dynamics of post-earnings-announcement drift using genetic algorithm-optimised supervised learning}}

\normalsize{
\author{Zhengxin Joseph Ye and Bj\"orn W.\ Schuller \\
GLAM, Department of Computing \\
Imperial College London\\
London, UK}
}
\date{} % Show no date

\begin{document}

\maketitle

\noindent\rule{\textwidth}{2pt}\\[\dimexpr-\baselineskip+2pt]
\rule{\textwidth}{0.4pt}\\
\begin{changemargin}{0.5in}{0.5in}
\large \textbf{Abstract} \\
\normalsize
Post-Earnings-Announcement Drift (PEAD) is a stock market phenomenon when a stock's cumulative abnormal return has a tendency to drift in the direction of an earnings surprise in the near term following an earnings announcement. Although it is one of the most studied stock market anomalies, the current literature is often limited in explaining this phenomenon by a small number of factors using simpler regression methods. In this paper, we use a machine learning based approach instead, and aim to capture the PEAD dynamics using data from a large group of stocks and a wide range of both fundamental and technical factors. Our model is built around the Extreme Gradient Boosting (XGBoost) and uses a long list of engineered input features based on quarterly financial announcement data from 1\,106 companies in the Russell 1\,000 index between 1997 and 2018. We perform numerous experiments on PEAD predictions and analysis and have the following contributions to the literature. First, we show how Post-Earnings-Announcement Drift can be analysed using machine learning methods and demonstrate such methods' prowess in producing credible forecasting on the drift direction. It is the first time PEAD dynamics are studied using XGBoost. We show that the drift direction is in fact driven by different factors for stocks from different industrial sectors and in different quarters and XGBoost is effective in understanding the changing drivers. Second, we show that an XGBoost well optimised by a Genetic Algorithm can help allocate out-of-sample stocks to form portfolios with higher positive returns to long and portfolios with lower negative returns to short, 
%BS: is there something missing after "long" and "short"?
a finding that could be adopted in the process of developing market neutral strategies. Third, we show how theoretical event-driven stock strategies have to grapple with ever changing market prices in reality, reducing their effectiveness. We present a tactic to remedy the difficulty of buying into a moving market when dealing with PEAD signals. \\ 
\\

\end{changemargin}
\rule{\textwidth}{2pt}\\[\dimexpr-\baselineskip+2pt]
\rule{\textwidth}{0.4pt}

%%%%%%%%%%%%%%%%%
%%%%%%%%%%%%%%%%%
% Begin the Two Column Part
%%%%%%%%%%%%%%%%%
%%%%%%%%%%%%%%%%%
\begin{multicols}{2}

\section{Introduction}

The stock market is characterised by nonlinearities, discontinuities, and multi-polynomial components because it continuously interacts with many factors such as individual companys' news, political events, macro economic conditions, and general supply and demand, etc.\  \cite{Gocken2016}. The non-stationary nature of the stock market is supported by a widely accepted, but still hotly contested economic theory \textit{Efficient Market Hypothesis} which states that asset prices fully reflect all available information and the market only moves by reacting to new information. Such a theory implies that the stock market behaves like a martingale and knowledge of all past prices is not informative regarding the expectation of future prices.

Ball and Brown \cite{Ball1968} were the first to note that after earnings are announced, estimated cumulative abnormal returns continue to drift up for firms that are perceived to have reported good financial results for the preceding quarter and drift down for firms whose results have turned out worse than the market had expected. The discovery of Post Earnings Announcement Drift (PEAD), which is a violation of a semi-strong Efficient Market Hypothesis, seems to suggest that while stock markets are generally efficient, there may be information leakages around the announcement dates, coupled with post-earnings drift, resulting in price movement anomalies. It also seems to suggest that past stock price information or other past economic or financial information can potentially be used to predict price movement following a significant economic event such as an earnings announcement. 

Researches on Post Earnings Announcement Drift
%BS: You have to re-introduce abbreviations after the abstract...
proliferated in the late 1980s and 1990s. Fama and French \cite{Fama1993} show that average stock returns co-vary with three factors, namely, the market risk factor, the book-to-market factor, and the size factor. Bhushan suggests that the existence of sophisticated and unsophisticated investors, transaction costs, and economies of scale in managing money can explain the market's delayed response to earnings \cite{Bhushan1994}. We notice nearly all previous researches pooled companies with negative and positive earnings surprises when measuring the effect of earnings surprises on abnormal returns and regress the absolute value of earnings surprise as well as other factors against the absolute value of abnormal return \cite{Qiu2014}. However, we have found that stock markets do not
%BS: no colloquial style, please - don't --> do not, etc.
just react symmetrically to negative and positive earnings surprises and there are a lot more factors in play that drive the near term risk adjusted returns of a stock following an earnings release. 

Rather than trying to analyse the link between PEAD and economic and accounting factors as commonly seen in the literature, by using machine learning models, we manage to leap straight to the more important goal of predicting the direction of PEAD. In this process we have overcome a number of constraints commonly seen in previous researches: we are including a much wider range of factors including both fundamental and technical/momentum factors; we achieve a higher level of generality without having to pre-group companies by the value of their earnings surprises or other attributes prior to the analysis or prediction (\textit{subsample analysis}) which is common in the literature \cite{Baker2016}. Additionally we have chosen 1\,106 stocks that are or once existed as components of the Russell 1\,000 index (which tracks approximately the 1\,000 largest public companies in the US) during the chosen time period between 1997 and 2018. Our selection includes companies that either went bankrupt or dropped out of Russell 1\,000, significantly reducing survivorship bias in our training data. This test population is larger than most earlier studies of similar nature. For example, Beyaz and colleagues only chose 140 stocks from S\&P500 when they attempted to forecast stock prices both six months and a year out based on fundamental analysis and technical analysis \cite{Beyaz2018}, and Bradbury used a sample of only 172 firms to research the relationships among voluntary semi-annual earnings disclosures, earnings volatility, unexpected earnings, and firm size \cite{Bradbury1992}. Our results should generalise better with the universe of stocks on the US markets.

Recognising the highly nonlinear nature of stock price movements, we have chosen to run our experiments using XGBoost which is a state-of-the-art supervised model. We divide the training data into in-sample and out-of-sample periods of varying lengths and use the in-sample data set to optimise a model's hyperparameters before training it. Our earlier experiments show that grid search as a traditional way of finding an optimal parameter set is inexhaustive and can be very slow. Instead we have chosen to use the highly adaptable Genetic Algorithm (GA) to optimise our models \cite{Deng2013}. We recognise hyperparameter optimisation is a delicate step and searching with a limited set of parameters will result in a non-optimal model which will not able to fit the essential structure of the training data set. To avoid this potential problem, we have chosen to use a broad value range and a small granular step for each of the hyperparameters. A 5-fold cross validation (CV) is employed within each GA iteration during the optimisation.
%BS: 5-fold CV is a problem - can it be reproduced by others? EVERYTHING should be reproducible - otherwise, we cannot go to good journals and will not find many citations... 

% As inspired by Chung's work \cite{Chung2011} who successfully evidenced that short term return predictability captures other factors, such as volatility, information asymmetry, investor sophistication, volume, size, and trading costs that affect arbitrage activities and the extent to which information is impounded in prices, we decided to explore the possibility of using post-earnings return predictability as a measure of broader market inefficiency. 
Our machine learning-based approach is in direct contrast to most earlier financial research work in the literature as typified by \cite{Kim2003},  which sought to devise different portfolios \textit{a priori} by different factor characteristics and tried to analyse and make sense of the link between portfolio returns and the corresponding economic factors that segregated the portfolios. Instead, our model automatically learns the intrinsic link between the input feature space and stock price returns with no \textit{a priori} assumptions. We find that stocks that belong to different industrial sectors can have their PEAD movements driven by different primary factors and such factors can also change from quarter to quarter. Despite such differences and changes in the driving factors, a GA optimised XGBoost model is able to pick up the underlying signals embedded in our engineered features and forecast the 30 day PEAD direction with reasonable accuracy. We also study the possibility of grouping stocks into portfolios according to their predicted levels of Cumulative Abnormal Return (CAR). 
%BS: Please introduce CAR spelt out once before using the abbreviation.
We have found that ranking the out-of-sample stocks by their predicted returns help form portfolios which consistently offer higher positive returns and lower negative returns, a result that could potentially form the basis of further usage in market neutral long-short trading strategies. In the end, we also look at the challenges of applying predictive models in real life markets due to ever changing market prices and asymmetrical level of information access by certain market participants. We share a tactic that can turn a model's forecasts into actionable signals. 

\section{Related Work}
% \subsection{Splitter/Combiner}

Since the discovery of Post Earnings Announcement Drift as a stock market anomaly by Ball and Brown \cite{Ball1968} who documented the return predictability for up to two months after the annual earnings announcements, extensive research has been carried out in the literature though with varying results. For example, Foster, Olsen, and Shevlin \cite{Foster1984} found that systematic post-announcement drifts in security returns are only observed for a subset of earnings expectations models when testing drifts in the [+1, +60] trading day period. In recent years, the literature has become less limited to the specific study of PEAD and instead put more focus on the direct predictions of stock price movement using stocks' fundamental and/or technical information, again with varying rates of success. Malkiel studied the impact of price/earnings (P/E) ratios and dividend yields on stock prices using the Campbell-Shiller model. He conceded his work demonstrating that exploitable arbitrage did not exist for investors to earn excess risk-adjusted returns and he could not find a market timing strategy capable of producing returns exceeding a buy-and-hold on a broad market index \cite{Malkiel2004}. Olson and Mossman on the other hand not only showed that an artificial neural network (ANN) outperforms traditional regression based methods when forecasting 12-month returns by examining 61 financial ratios for 2352 Canadian stocks, but, more importantly, shows that by using fundamental metrics sourced from earning reports, they were able to achieve excessive risk-adjusted returns \cite{Olson2003}.

Other authors went beyond metrics from earnings reports and attempted stock forecast using both fundamental and technical analysis. Sheta et al.\ explored the use of ANNs, Support Vector Machines (SVMs), and Multiple Linear Regression for prediction of S\&P500 market index. They selected 27 technical indicators as well as macro economic indicators and reported that SVM contributed to better predictions than the other models tested \cite{Sheta2015}. Hafezi et al.\ considered both fundamental and technical analyses in a novel model called Bat-neural Network Multi-agent System when forecasting stock returns. The resulted mean absolute percentage error showed that the new model performed better than a typical Neural Network coupled with a GA \cite{Hafezi2015}. Alternative data are becoming popular, too. Solberg and Karlsen investigated the possibility to predict the direction of stock prices using scripts of earnings conference calls. By analysing 29\,330 different earnings call scripts between 2014 and 2017 using four different machine learning algorithms they managed to achieve a classification error rate of 43.8\,\% using logistic regression and beat the S\&P500 benchmark using both logistic regression and gradient boosting. Their results showed that earnings calls contain predictive power for the next day's stock price direction post earnings release \cite{Solberg2018}.

% When it comes to selecting machine learning models for event driven stock price forecast the literature has looked a lot at Support Vector Machines. Zhang constructed a novel ensemble method integrated with AdaBoost algorithm, probabilistic Support Vector Machine and Genetic Algorithm and verified its performance over 20 shares from the SZSE and 16 stocks from NASDAQ. He showed the new ensemble method achieved preferable profit in simulation of stock investment \cite{Zhang2016}. Madge used daily closing price for 34 technology stocks on a SVM model with radial kernel to calculate price volatility and momentum for individual stocks and for the overall sector. The model attempts to predict whether a stock price sometime in the future will be higher or lower than it is on a given day. They found little predictive ability in the short-run but definite predictive ability in the long-run \cite{Madge2015}. Tsai and Cheng focused on testing the impact of feature selection while using GA to optimize SVR which is the regression version of SVM. They formulated stock price prediction as a time series problem, used a variety of technical indicators and other time series data as model inputs and evaluated a number of methods including Kernel Ridge Regression and Multivariate Adaptive Regression Splines etc for feature selection. They were able to show some feature selection methods were better than others which in turn meant certain inputs were more impactful than others \cite{tsai2018}.

Researchers also studied how machine learning would directly benefit financial trading. Through a series of applications involving hundreds of predictors and stocks, Huck looked at how to apply some of the state-of-the-art machine learning techniques to manage a long-short portfolio. In that process he also explored a series of practical questions with regard to the predictor data and was able to show that the techniques he examined generated useful trading signals for portfolios with short holding periods \cite{Nicolas2019}. Sant'Anna and Caldeira applied Lasso regression for index tracking and long-short investing strategies. They used stocks from three benchmarks, S\&P100, Russell 1000, and the Ibovespa Index from Brazil from 2010 to 2017 to assess the quality of Lasso-based tracking portfolios. By using co-integration as a benchmark method to solve the same problems, they demonstrated that the Lasso regression based approach was able to form portfolios that produced similar returns compared to using co-integration, but incurred significantly less transaction costs \cite{Leonardo2019}.

As a model that has only recently burst on the scene, there is limited study of XGBoost in financial applications. Chatzis et al.\  \cite{Sotirios2018} evaluate the possibility of a market crash over a 1-day and 20-day horizon across the global markets by forecasting 1-day and 20-day stock market returns and see if they will have dropped below a low quantile of historical distribution of stock market returns. By using a vast set of data from global stock markets, bond markets and FX markets, the paper explores a large set of supervised learning models including Logistic Regression, Decision Trees, Random Forest, Support Vector Machines, Deep Neural Networks, and XGBoost. The paper draws conclusions by declaring the superiority of certain models including XGBoost over others by examining the forecast results on stock returns through a list of statistical measurement metrics. Li and Zhang \cite{Li2018} use XGBoost to dynamically predict the value of a set of seven factors that contribute a stock selection process. Dynamically generated factors are then used to select 
a portfolio of different stocks whose return is measured over a multiple year period. Portfolios of dynamically selected stocks are shown to perform better than benchmark portfolios.

\section{Model Features Generation}
We have chosen to use 1106 US companies in the Russell-1000 index in total. The data time frame is between the first financial quarter of 1997 (1997 Q1) and the fourth financial quarter of 2018 (2018 Q4). The model output is the $30$ day Cumulative Abnormal Return post earnings release of each individual stock and the input space consists of the following set of unadjusted data which we have sourced from Bloomberg:
\begin{itemize}
\item Financial statements data
%BS: Why is "Suprise" written with capital "S"?
\item Earnings Surprise data
\item Momentum indicator data
\item Short interest data
\end{itemize}

In total, we have sourced 97\,901 quarterly financial statements from our chosen companies over the test time frame. The final population of valid data points used for training and testing whose input features include both financial statement metrics and other economic metrics stands close to 50,000, depending on the test cases. There are a number of reasons for the reduced population: (a) there are no Earnings data, Short interest data or other input feature data on Bloomberg for a good number of historical financial quarters within the test time frame; (b) we have discarded companies in certain historical quarters when the earnings reports suffered badly from missing data; (c) We have been very careful with whether an earnings report was released before market opened, after market closed or during trading hours as such a difference is significant as we would
%BS: never use colloquial style such as "we'd", please.
need to alter the forecast starting point accordingly. Bloomberg is missing the release time of day for some financial quarters in earlier years, and we have discarded those quarters.

\subsection{\label{sec:level2}Financial Statements data}

\begin{table*}[ht]
\centering
\begin{tabular}{|p{5cm}|p{5cm}|}
\hline
Cash   & Operating Margin                           \\ \hline
Cash from Operating Activities                   & Price to Book Ratios                            \\ \hline
Cost of Revenue                & Price to Cashflow Ratios                  \\ \hline
Current Ratio                     & Price to Sales Ratios                   \\ \hline
Dividend Payout Ratio & Quick Ratio              \\ \hline
Dividend Yield               & Return On Assets                      \\ \hline
Free Cash Flow               & Return On Common Equity                       \\ \hline
Gross Profit                 & Revenue                  \\ \hline
Income from Continued Operations                  & Short Term Debt                           \\ \hline
Inventory Turnover                        & Total Asset                   \\ \hline
Net Debt to EBIT    & Total Asset \\ \hline
Net Income                & Total Debt to Total Assets                        \\ \hline
Operating Expenses                & Total Debt to Total Equity                        \\ \hline
Operating Income                & Total Inventory                        \\ \hline
                & Total Liabilities                         \\ \hline
\end{tabular}
\caption{\label{tab:FeatureList}Earnings report metrics chosen as input features}
\end{table*}

As shown in Table \ref{tab:FeatureList}, twenty nine metrics from earnings reports have been chosen to create training data.

Based on the reported value of these metrics, we have engineered new features as quarterly change and yearly change of each of these financial metrics. 

\subsection{\label{sec:level2}Earnings Surprise data}

Earnings Surprise represents how much a company's actual reported Earnings Per Share (EPS) is more (or less) than the average of a selected group of stock analysts' estimates on the same quarter's EPS. We are not calculating Earnings Surprise as a \%change between the reported EPS and market estimated EPS because (a) \%change is very volatile when a EPS level is close to zero and a small change can lead to a misleadingly large \%change, and (b) we would like to avoid the change-of-signs problem.

We have subsequently engineered the following three features related to Earnings Surprise:

\begin{itemize}
\item Current quarter's Earnings Surprise (reported EPS minus market estimated EPS); 
\item Difference between current quarter's Earnings Surprise and that of the previous quarter;
\item Difference between current quarter's Earnings Surprise and the average Earnings surprise of the preceding three quarters;
\end{itemize}

\subsection{\label{sec:level2}Momentum Indicators}
We have chosen the following technical/momentum indicator values calculated on the same day an individual company's quarterly earnings data was released:
\begin{itemize}
\item 9-day Relative Strength Index (RSI)
\item 30-day Relative Strength Index
\item 5-day Moving Average / 50-day Moving Average
\item 5-day Moving Average /200-day Moving Average
\item 50-day Moving Average / 200-day Moving Average
\end{itemize}
We believe all these indicators should in a way measure how a stock's recent short term movements compare to its historical movements further back in time. The inclusion of momentum indicators is 
%BS: added: 
motivated by the intention 
to allow the prediction process of future stock movements to take into account a stock's recent movement trend as information leakage does happen prior to financial reportings. We have engineered the three ratios of short term moving averages to near or long term moving averages as proxies to the \textit{golden cross} indicators.

\subsection{\label{sec:level2}Short Interest data}
Short interest ratio is released for most companies twice a month and is calculated by dividing the number of shares short in a stock by the stock's average daily trading volume. The short interest ratio is a good gauge on how heavily shorted a stock may be versus its trading volume. The most recent short interest ratio for each company prior to its earnings release is sourced as an input feature to the model for that company.

\section{Data Pre-processing}

With totally 1106 companies involved over 21 years, there is a lot of data representing input features for each company at each quarter. In order for them to be understood by the model, we put them into a matrix-like data structure $A \in M_{m \times n}( \mathbb{R} )$, where each of the $m$ rows represents a $n$ dimensional training data point, indexed by the pairing of a company name and a historical quarter, and each column holds data of the same feature from all the data points.

% \begin{figure}[h!]
\begin{figure*}[ht]
  \centering
  \includegraphics[width=0.7\textwidth]{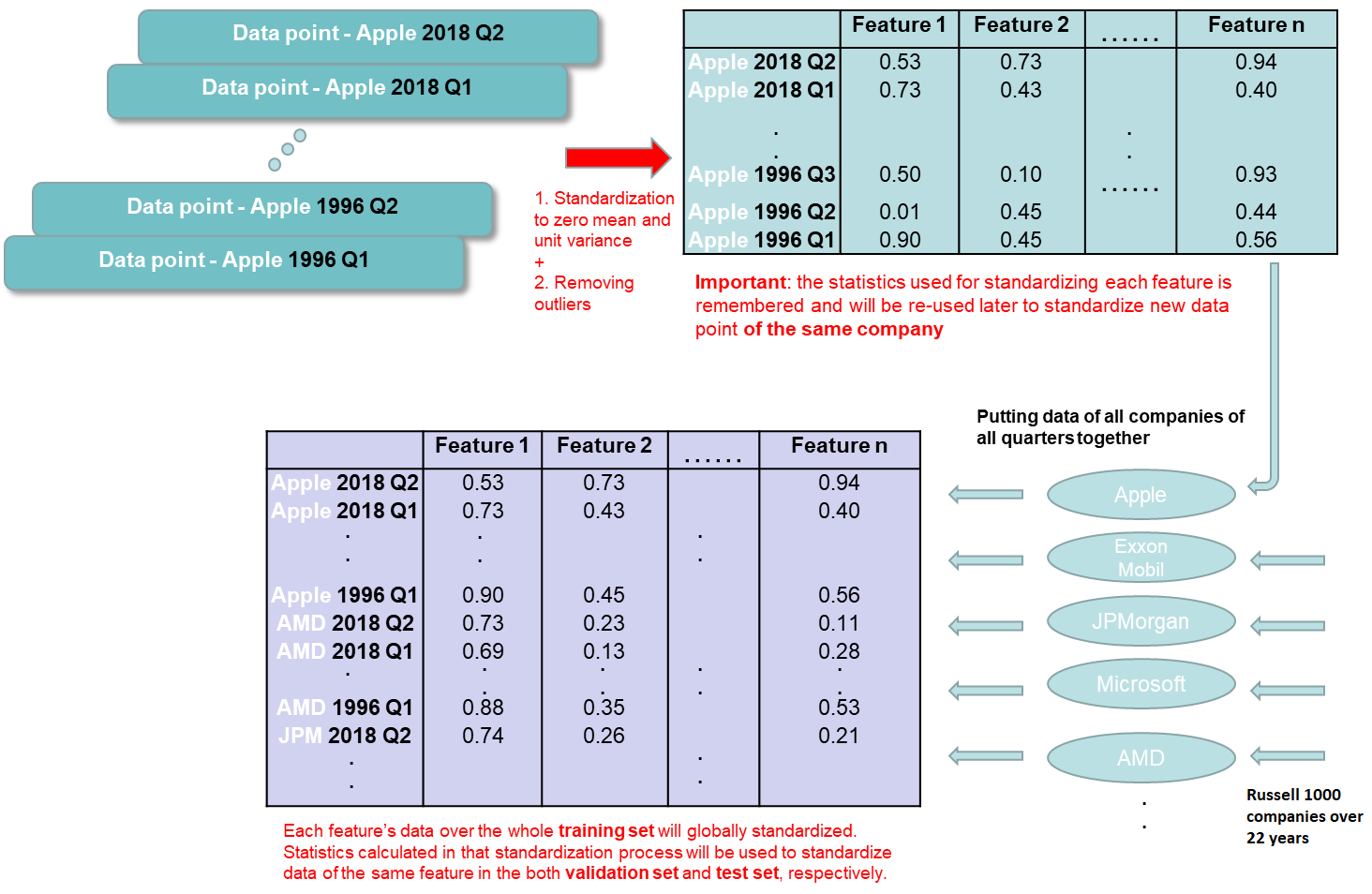}
  \caption{Steps of Data Pre-processing}
  \label{fig:preprocessing}
\end{figure*}

Before we put the data of all the companies and of all the quarters into a matrix, we pre-process each company's data to deal with outliers and to standardise data of every company. Firstly, we employ Winsorisation \cite{Bin1998} to reduce the number of outliers present in the input features. This is carried out on the feature data of each individual company. Secondly, we standardise a selective group of features of each company. Every company's standardised features will then be stacked back into a full training data set. The pre-processing process is illustrated in Figure \ref{fig:preprocessing}.

\section{Models and Methods}

% A recommended Latex math formula editor: https://hostmath.com/

% Using align* can align equations in the same block. The ampersand symbol & determines where the alignment happens
% https://www.overleaf.com/learn/latex/Aligning%20equations%20with%20amsmath

% To include texts as part of a math formula, simply use \mbox{}
% https://www.overleaf.com/learn/latex/Questions/Including_text_within_equations_in_LaTeX

% Here is a collection of SVM kernel formulae: https://data-flair.training/blogs/svm-kernel-functions/

\subsection{\label{sec:level2}Extreme Gradient Boosting} 

%https://arxiv.org/pdf/1603.02754.pdf
%https://xgboost.readthedocs.io/en/latest/tutorials/model.html
%https://medium.com/@vikeshsingh37/math-behind-gbm-and-xgboost-d00e8536b7de
%https://www.kdnuggets.com/2018/08/unveiling-mathematics-behind-xgboost.html

Extreme Gradient Boosting (XGBoost) is a scalable machine learning system for tree boosting invented by Tianqi Chen \cite{ChenTianqi2016}, which has gained much prominence in recent years. It distinguishes itself from other existing tree boosting methods \cite{Tyree2011} \cite{Ye2009} by having cache-aware and sparsity-aware learnings. The former technology gives the system twice the speed against running a non-cache-aware but otherwise identical greedy tree splitting algorithm, and the latter gives an amazing 50 times speed boosting against a naive implementation handling an Allstate-10k dataset \cite{ChenTianqi2016}. More importantly, XGBoost has achieved algorithmic optimisations by introducing a regularised learning objective within a tree structure which helps achieve smart tree splitting and branch pruning.

For a data set in matrix form $A \in M_{m \times n}( \mathbb{R} )$ with $m$ data points and $n$ features, a tree ensemble model uses $K$ base leaner functions to predict the output:

\begin{equation}
\widetilde{y_{i}} = \phi(x_{i})=\sum_{k=1}^K f_{k} (x_{i}), f_{k}\in \mathbb{F},
\end{equation}

where $\mathbb{F}$ is the space of regression trees. Each hypothesis $f_{k}$ corresponds to an independent tree structure $q$ with leaf scores $\omega$. XGBoost utilises regression trees each of which contains a score on each of its leaves. These scores help form the decision rules in the trees to classify each set of inputs into leaves and calculate the final predicted output by summing up the scores in the related leaves. Unlike other standard gradient boosting models such as AdaBoost and GBM which do not intrinsically perform regularisation, XGBoost minimises a $regularised$ loss function in order to learn the set of functions:

\begin{equation}
L(\phi)=\sum_{i} \ell(\widehat{y_{i}},y_{i})+\sum_{k}\Omega(f_{k}). 
\label{func:xgboost_loss}
\end{equation}

Here, $\ell$ is a differentiable convex loss function for the model output and the regularisation term is defined as (though not limited to) $\Omega(f)=\gamma T + \frac{1}{2}\lambda {||\omega||}^{2}$,  which reduces the chance of overfitting. As in a typical gradient tree boosting model, a new base learner regression tree $f_{i}$ which most minimises the loss function in equation \ref{func:xgboost_loss} is greedily and iteratively added to the final loss function. Let $\widehat{y}_{i,t}$ be the model output of the $i$-th instance at the $t$-th iteration the loss function can be re-written as

\begin{equation}
L_{t}(\phi)=\sum_{i,k} \ell(y_{i}, \widehat{y}_{i,t-1} + f_t(x_i))+\sum_{k}\Omega(f_{k,t}).  
\end{equation}

By taking the Taylor expansion on this loss function up to the second order and removing the constant terms as a result of the expansion the loss function can be simplified to:
\begin{equation}
L_{t}(\phi)=\sum_{j=1}^T [G_j \omega_j + \frac{1}{2}(H_j + \lambda) \omega_j^2] + \lambda T,
\end{equation}

where

\begin{align*}
& G_j = \sum_{i\in I_j} g_i \\  
& H_j = \sum_{i\in I_j} h_i \\  
& I_j=\left\{ i|q(x_i)=j \right\} \\  
& g_i =  \partial_{\widehat{y}_{i, t-1}} \ell(y_i, \widehat{y}_{i, t-1}) \\
& h_i =  \partial_{\widehat{y}_{i, t-1}}^2 \ell(y_i, \widehat{y}_{i, t-1}).
\end{align*}

Here, $T$ is the number of leaves in the tree. With $\omega_j$ being independent with respect to others, Tianqi \cite{ChenTianqi2016} has proven that the best $\omega_j$ for a given tree structure $q(x)$ should be

\begin{equation}
\omega_j^* = - \frac{G_j}{H_j + \lambda},
\end{equation}

which in turn makes the objective function come to its final form:
\begin{equation}
L_j^* =-\frac{1}{2}\sum_{j=1}^T \frac{G_j^2}{H_j + \lambda}+\gamma T.
\end{equation}

Ideally, the model would enumerate all possible tree structures with a quality score and pick the best one to be added iteratively. In reality, this is intractable and optimisation has to be executed one tree level at a time. This is made available by the final form of the loss function, as the model uses it as a scoring function to decide on the optimal leaf splitting point. Assume that $I_L$ and $I_R$ are the instance sets of left and right nodes after the split. Letting $I = I_L  \cup  I_R$, the scoring function for leaf splitting is

\begin{equation}
L_{split} =\frac{1}{2} \left[ \frac{(\sum_{i\in I_{L}}g_i)^{2}}{\sum_{i\in I_{L}}h_i + \lambda}  + \frac{(\sum_{i\in I_{R}}g_i)^{2}}{\sum_{i\in I_{R}}h_i + \lambda} + \frac{(\sum_{i\in I}g_i)^{2}}{\sum_{i\in I}h_i + \lambda} \right] - \gamma.
\end{equation}

These scores are then used by a method called the $exact$ $greedy$ $algorithm$ to enumerate all the possible splits for continuous features, allowing each level of a tree to be optimised and the overall loss function to be minimised in the process. When deployed on a distributed platform XGBoost employs approximate algorithms instead to alleviate the huge memory consumption demanded by the exact greedy algorithm although this is not needed in our experiments which run on a single machine.

\subsection{\label{sec:level2}Hyperparameter Optimisation}

Model optimisation is one of the two most important steps (the other being data cleansing) in ensuring the model output can meaningfully capture the underlying dynamics of the dependent variable. In search of optimal hyperparameter sets, we initially experimented a more straightforward approach of grid search but found it less effective in its performance and inexhuastive in the search results. GA as an adaptable and easily extensible heuristic optimisation method is chosen instead to carry out this task. Table \ref{tab:ModelHyperparameters} gives the list of model hyperparameters we have put through GA. Before we start the optimisation process, we first split the population of data into training data and test data. Selection of the out-of-sample test data varies and depends on the nature of a test which will be explained in subsequent sections.  
%BS: Is this splitting reproducible by others? If not, can we provide a link to the instance lists per fold?
It is the training data that we use to optimise the model. We use 5-fold cross validation to calculate the fitness value on a particular set of hyperparameters examined by the GA. To do that we split the training data into five equal groups, use four groups to train the model and calculate the fitness value using the last group (validation). This process is repeated five times iteratively on each of the five groups and the final fitness value is the averaged fitness of the five iterations.
%BS: stratified CV?

\begin{table*}[]
\centering
\begin{tabular}{|c|}
\hline
\textbf{Hyper Parameters} \\ \hline
Gamma                     \\ \hline
Max depth                 \\ \hline
Sub sample                \\ \hline
Learning Rate             \\ \hline
Minimum child weight      \\ \hline
Column sample by tree     \\ \hline
\end{tabular}
\caption{XGBoost hyperparameters optimised by GA + CV}
\label{tab:ModelHyperparameters}
\end{table*}

% \begin{figure}[h!]
\begin{figure*}[ht]
  \centering
  \includegraphics[width=0.48\textwidth]{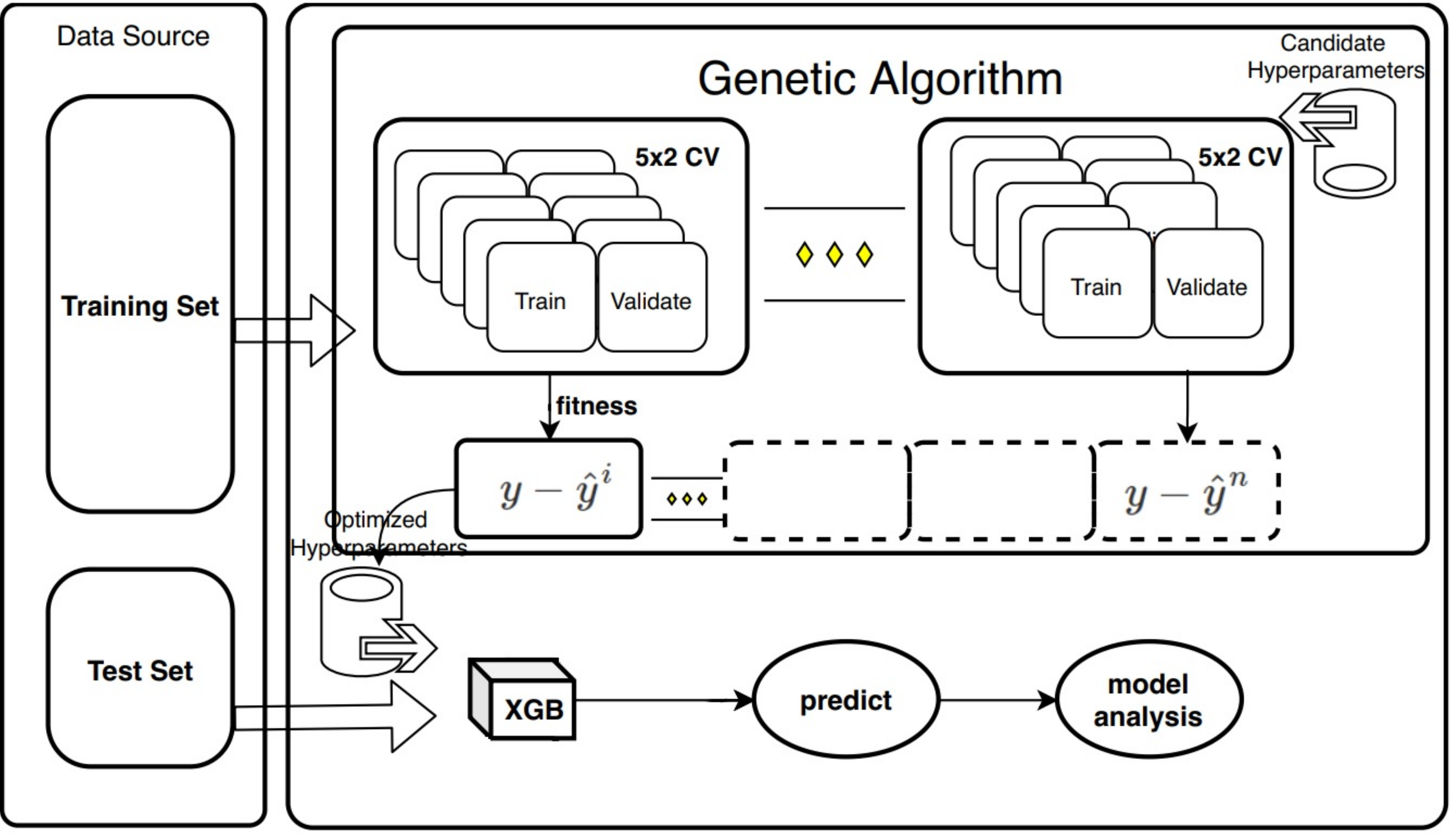}
  \caption{Hyperparameter Optimisation using GA + CV}
  \label{fig:HyperTuning}
% \end{figure}
\end{figure*}

%https://www.overleaf.com/learn/latex/Inserting_Images

To optimise the model, each hyperparameter is randomly
%BS: again: reproducible? We could, eg, name the random seed? 
initialised according to its own valid range of values. This initialisation is repeated 40 times so that we have 40 sets of randomly initialised hyperparameters to start the GA process with. Each set is called a \textit{population}, and each hyperparameter within a set is called a \textit{chromosome}. All of the 40 populations are considered to be part of the current \textit{generation}. The GA process carries out a 5-fold cross-validation on a model using each of the 40 populations of parameters and when finished, keeps the 20 populations that have produced the smallest fitness values in the cross validation step. These 20 sets or \textit{populations} of hyperparameters are considered to have performed better in forecasting post-announcement drifts with the current model than the 20 discarded ones. The 20 better populations are then used to \textit{cross-breed} into 20 new populations and in this process \textit{mutation} is allowed to happen to the cross-bred populations, i.\,e., chromosomes in the 20 newly created populations are allowed to randomly change value following a predefined level of probability. At the end of this process, we have produced a new and potentially better set of 40 populations of hyperparameters and we call them the \textit{new generation}. The new generation are then fed through a second iteration of the GA process until eventually the minimum fitness value produced by the cross-validation step no longer changes its value within tolerance and at this point we have arrived at the optimal set of hyperparameters which produces the smallest fitness value when being used in the current model.
Figure \ref{fig:HyperTuning}  shows how GA and Cross Validation work together to produce the set of hyperparameters of each model which result in the highest prediction accuracy (smallest fitness value) on the validation set.

\section{Results}

All of our experiments centre around the 30 day post-earnings Cumulative Abnormal Return (CAR) as a measure of risk adjusted stock price return. An abnormal return is between the actual return of a security and its expected rate of return.

\begin{equation}
AR_{i} = r_{i} - E(r_{i}),
\end{equation}

where $AR_{i}$ is the one-day abnormal return for company $i$ on day $t$, $r_{i}$ is the actual one-day stock return and $E(r_{i})$ is the expected one-day return of stock $i$. As explored by Kim \cite{Kim2003}, there is a variety of ways of evaluating the expected return including using quantitative models such as the one-factor CAPM model and the Fama French three-factor model \cite{Fama1993}. In our experiments, we choose to use the S\&P500 index return to represent the broader market's return and use that to proxy a stock's expected return. Consequently, our model output for stock $i$, which is the cumulative abnormal return from $T_{1}$ to $T_{2}$, is defined as: 

\begin{equation}
CAR_{i}(T_{1},T_{2}) =\sum_{t=T_{1}}^{T_{2}} (AR_{it})=\sum_{t=T_{1}}^{T_{2}} (r_{it} - r_{S\&P500}).
\end{equation}

\subsection{\label{sec:level2} Single Stock Forecast}

In this experiment, we have chosen stocks that filed for earnings with SEC in the four quarters in each financial year from 2014 to 2018 as our out-of-sample test population. That means, we first run a forecast on movement direction of all the stocks that reported earnings in 2014 while using all the data prior to 2014 as training data. Once done, we move on to repeat the same exercise on stocks that reported in 2015, etc. It should be noted that a company that filed in each of the 4 quarters of a financial year is considered as four independent data points since the only data consumed by the XGBoost + GA model to predict the PEAD direction of a stock at any quarter are the near term momentum signals and financial statement data of this stock in that particular quarter.

\begin{figure*}[ht]
  \centering
  \includegraphics[scale = 0.12]{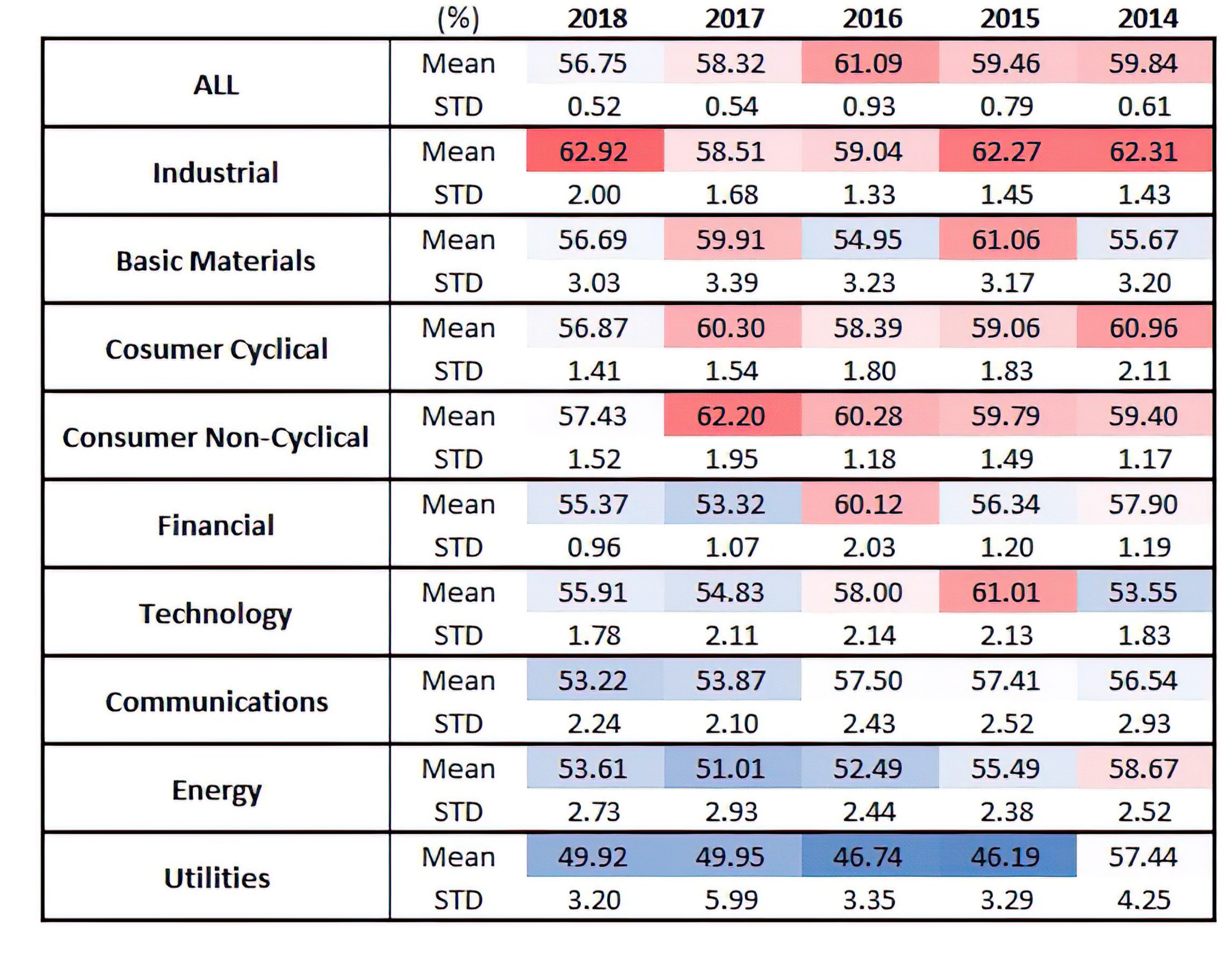}
  \caption{Accuracy rate of predicting 30 day PEAD movement}
  \label{Fig:Classification rate - all quarters 2014-2018}
\end{figure*}

Separately, the same test as described above is also repeated on stocks belonging to a particular industrial sector. Bloomberg categorises US companies into seven sectors: \textit{Industrial, Basic Materials, Consumer Cyclical, Consumer Non-Cyclical, Financial, Technology, Communications, Energy, and Utilities}.
%BS: what is the sorting order? Relevance? Otherwise, sort alphabetically?
Our chosen companies and their data are divided up into seven groups by industrial sector so that we can run the same tests per industrial group.

Each of such tests whether on all of the stocks or stocks belonging to a particular industrial sector are run 100 times. In each run, the same set of training data are used to train the model whose performance is verified using the same set of out-of-sample test data. This generates 100 sets of results for each test from which stats are calculated. 

Results and stats of predicting 30 day PEAD direction in each year from 2014 to 2018 and over the seven industrial sectors are presented in figure \ref{Fig:Classification rate - all quarters 2014-2018}. Our results clearly suggest that our model has strong prediction power and is able to pick up the patterns in the input data space when there is any driver in it. It is particularly interesting to see the model performs much better with stocks from some industrial sectors than others. Given that the same set of input features is used across the board, there is clear evidence that our data is more impactful to some sectors than others. There are probably two reasons that can explain this observation. First, there are other data that are not included in our feature space that does affect stock movements following earnings release. Second, stocks from different sectors are subject to different drivers, i.\,e.,  investment personnel/computer trading algorithms look for signals in different financial metrics for different sectors. Even if the same driving features are examined, the implicit feature weighting must be different for different sectors.  

The first reason is true, as there are impactful data that have yet to be included in our research, such as management's guidance, recent revisions of analysts' price forecast, other text information carried in financial reports, and meeting minutes with analysts, etc. It is entirely possible that certain stocks, or stocks from certain sectors are more susceptible to those data and the absence of such data reduces the model's prediction accuracy on those stocks. 

We have taken a closer look at the second possible cause and we find indeed stocks from different sectors are driven by different factors. With the 100 groups of tests we have carried out on individual sectors in each financial year from 2014 to 2018, we have counted the appearance of the three factors that appear most often as the top five driving forces. The results are recorded in the Appendix section of this paper and present some very interesting findings. First, most of the time, it is the three EPS related metrics that feature heavily on the top five spots of most influential factors. This finding is consistent with market practice and Earning Per Share surprise/disappointment is indeed one of the most important factors that investors examine. We need to point out that two of these features are engineered by us, which represent how the current quarter's reported EPS compares to those at the preceding quarters. The fact that these two factors also dominate shows that  investors look for more complex movements in financial metrics. Second, over the years, we consistently see important albeit less strong features appearing on the top five list for some of the sectors. For instance, the quarterly change in Return On Assets, Price-to-Sales Ratio, and Dividend Payout Ratio are consistently making up the top five spots driving PEAD of stocks from the \textit{Industrial}, \textit{Financial}, and \textit{Basic Materials} sectors respectively. We have carried out separate forecasting tests using the three EPS features only, and did not obtain good results which show that the model cannot be driven purely by a handful of key features and other, less strong, but also impactful features must not be ignored. Third, in the years when our model produces better prediction results for a particular sector, we are frequently seeing features that are more consistently dominant. This is represented by higher occurrence counts observed for the dominant features. This can be observed in the results for the \textit{Industrial}, \textit{Consumer Cyclical}, and \textit{Consumer Non-Cyclical} sectors. 
%BS: in all these listings, perhaps you can indicate the order's motivation?
The opposite is also true. With \textit{Energy} and \textit{Utilities} being the most difficult sectors to predict, the model is returning an inconsistent set of top drivers from the 5 yearly tests among which the occurrence count is also comparatively lower. Without strong and consistent drivers among our feature data for such sectors, the prediction result is unsurprisingly poorer.

The results of our experiments in this section help us conclude that Post Earnings-Announcement Drift is not merely a market anomaly, but a characteristics of the markets whose direction can be materially predicted. The strength of signal may vary in time and from sector to sector, but machine learning models --- especially an XGBoost well optimised by GA --- are able to pick up on them. However, the fact that the model performs well with stocks from certain sectors but not on others suggest that there may be limitations in our input feature space or the way the special features have been engineered. The input space may not have captured enough driving factors for certain sectors. We acknowledge that, when a company makes an earnings announcement, information come out in many different forms such as in financial metric numbers, textual information embedded in the documents filed with SEC, earnings calls with a selected group of equity analysts, let alone information leakage prior to announcement or even insider trading. Trying to capture and take advantage of more forms of drivers on Post Earnings Announcement Drifts is a future research topic.

\subsection{\label{sec:level2} PEAD Analysis on Portfolio of Stocks }

We believe it is meaningful to evaluate the dynamics of post-earnings drifts in the context of portfolios. In the subsequent series of tests, we use the model to forecast the actual level of $30$ day post earning cumulative abnormal returns instead of only the movement direction. We rank out-of-sample stocks according to each stock's \textit{predicted returns} from high to low, group stocks together into small portfolios and examine the \textit{actual returns} of the portfolios.

\begin{figure*}[ht]
  \centering
  \includegraphics[width=0.5\textwidth]{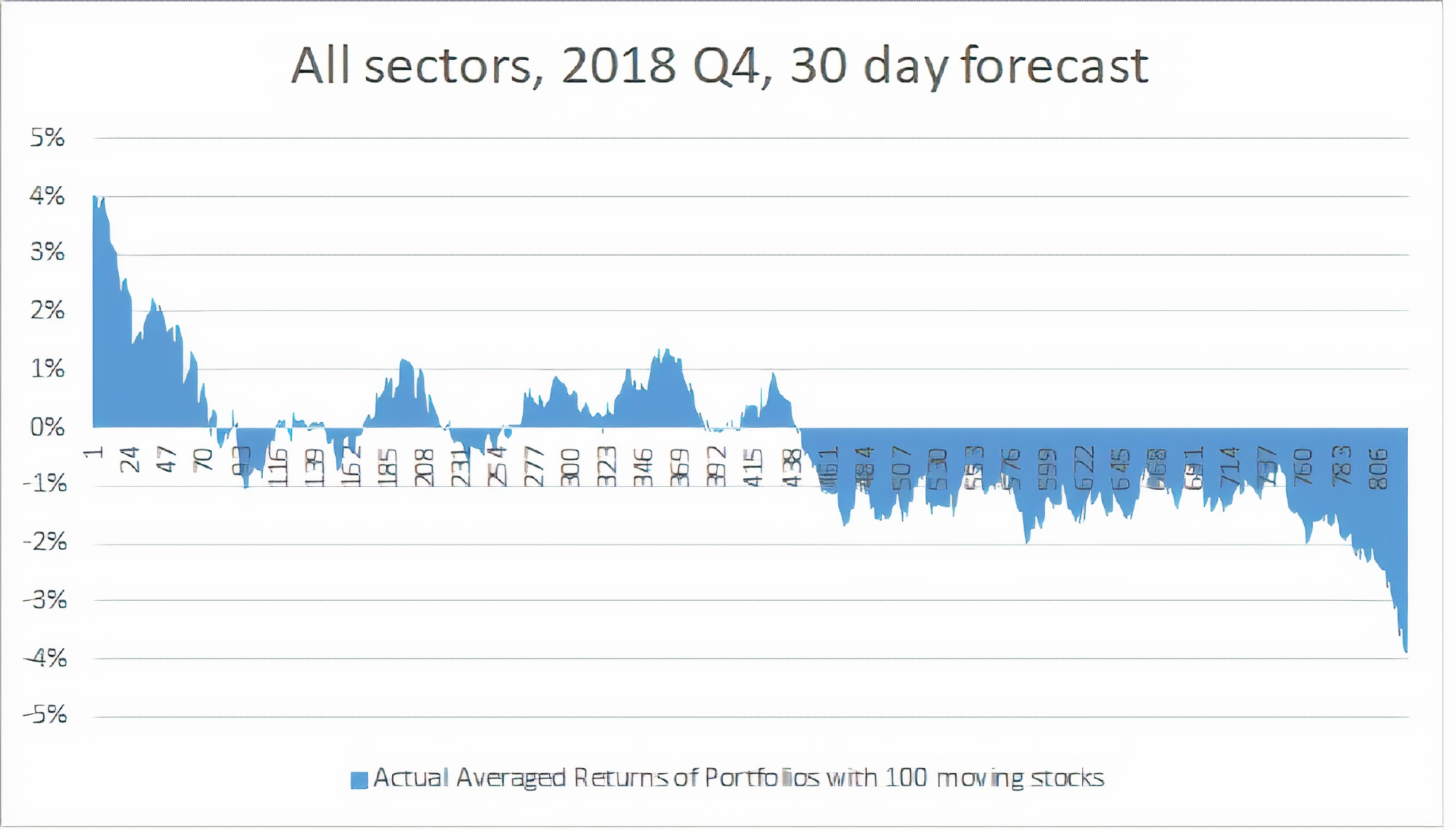}
  \caption{2018 Q4 test result. Actual returns of moving portfolios consisting of 100 stocks. All stocks have been pre-ranked by the model-predicted stock returns from high to low.}
  \label{Fig:2018Q4_30d_All}
\end{figure*}

\begin{figure*}[ht]
  \centering
  \includegraphics[width=0.5\textwidth]{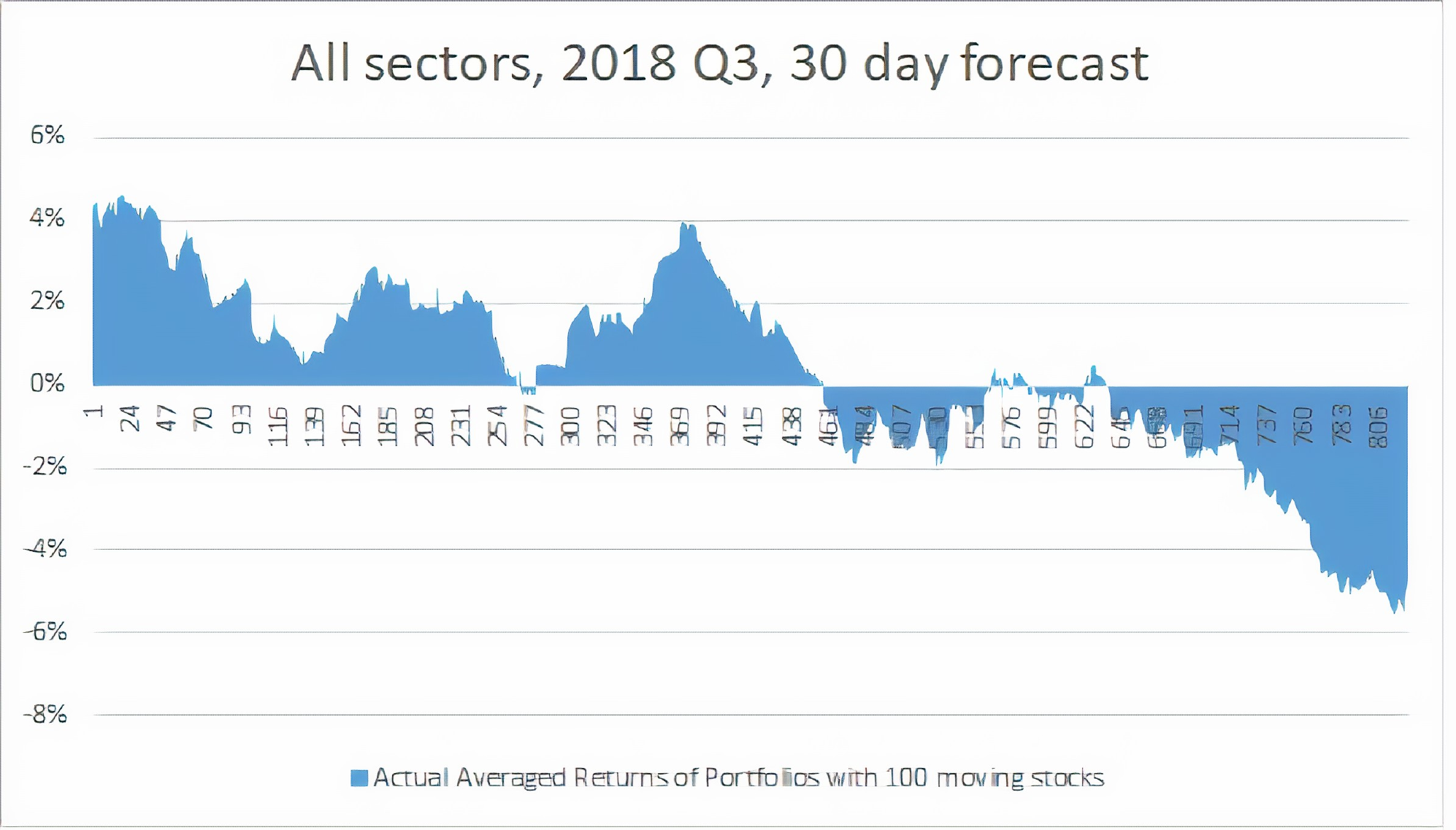}
  \caption{2018 Q3 test result. Actual returns of moving portfolios consisting of 100 stocks. All stocks have been pre-ranked by the model-predicted stock returns from high to low.}
  \label{Fig:2018Q3_30d_All}
\end{figure*}

% \begin{figure}[ht]
%   \centering
%   \includegraphics[width=0.5\textwidth]{"Results - 2018 Q4 - 30D - All".jpg}
%   \caption{2018 Q4 test result. Actual returns of moving portfolios consisting of 100 stocks. All stocks have been pre-ranked by the model-predicted stock returns from high to low.}
%   \label{Fig:2018Q4_30d_All}
% \end{figure}

% \begin{figure}[ht]
%   \centering
%   \includegraphics[width=0.5\textwidth]{"Results - 2018 Q3 - 30D - All".jpg}
%   \caption{2018 Q3 test result. Actual returns of moving portfolios consisting of 100 stocks. All stocks have been pre-ranked by the model-predicted stock returns from high to low.}
%   \label{Fig:2018Q3_30d_All}
% \end{figure}

\subsubsection{\label{sec:level3}Stocks that reported earnings in the same financial quarter}

We've mainly examined stocks that reported earnings in the same financial quarter in the years between 2014 and 2018. In each test we use all the data prior to the test quarter for training and carry out stock return prediction on stocks in the test quarter. In all the tests once the stocks being examined have been ranked according to their model-predicted post earnings returns, we are consistently observing that portfolios which include stocks from top quantiles of the ranked list are producing higher positive returns, whereas portfolios which include stocks from bottom quantiles are producing lower negative returns. Theoretically, a long-short market neutral strategy \cite{vandehave2017} could be formed through longing the top quantile portfolios and shorting the bottom ones. 

We've selected the Q3 2018 and Q4 2018 earning seasons to demonstrate the results . One of the reasons for choosing these two quarters is that the US stock market went through two polar opposite phases of development in these two quarters with the S\&P500 shedding 20\,\% in the last quarter of 2018 (around the time most Q3 2018 earnings were reported in the US) amid fear of Fed rate rises and US-China trade war escalation among other things but gaining a major rebound in the first quarter of 2019 (when most US companies reported Q4 2018 earnings). Our intention is to evaluate if our model can successfully capture those very different PEAD dynamics given very different macro conditions and different company specific accounts.

Each point on figures \ref{Fig:2018Q4_30d_All} and \ref{Fig:2018Q3_30d_All} corresponds to the actual return of a portfolio consisting of 100 stocks when we move down the list of out-of-sample stocks which have now been ranked by their predicted 30-day risk-adjusted returns following earnings releases. For instance, the first point is the actual return of a portfolio consisting of the 1st to the 100th stocks and the second point is the actual return of a portfolio including the 2nd to the 101st stocks, etc. We consistently produce similar figures with a downward slope when we continuously run the same tests. Our model has captured an unseen $collective$ trend of movement by groups of stocks as triggered by their earnings release and other relevant economic factors.

\begin{figure*}[ht]
  \centering
  \includegraphics[width=0.5\textwidth]{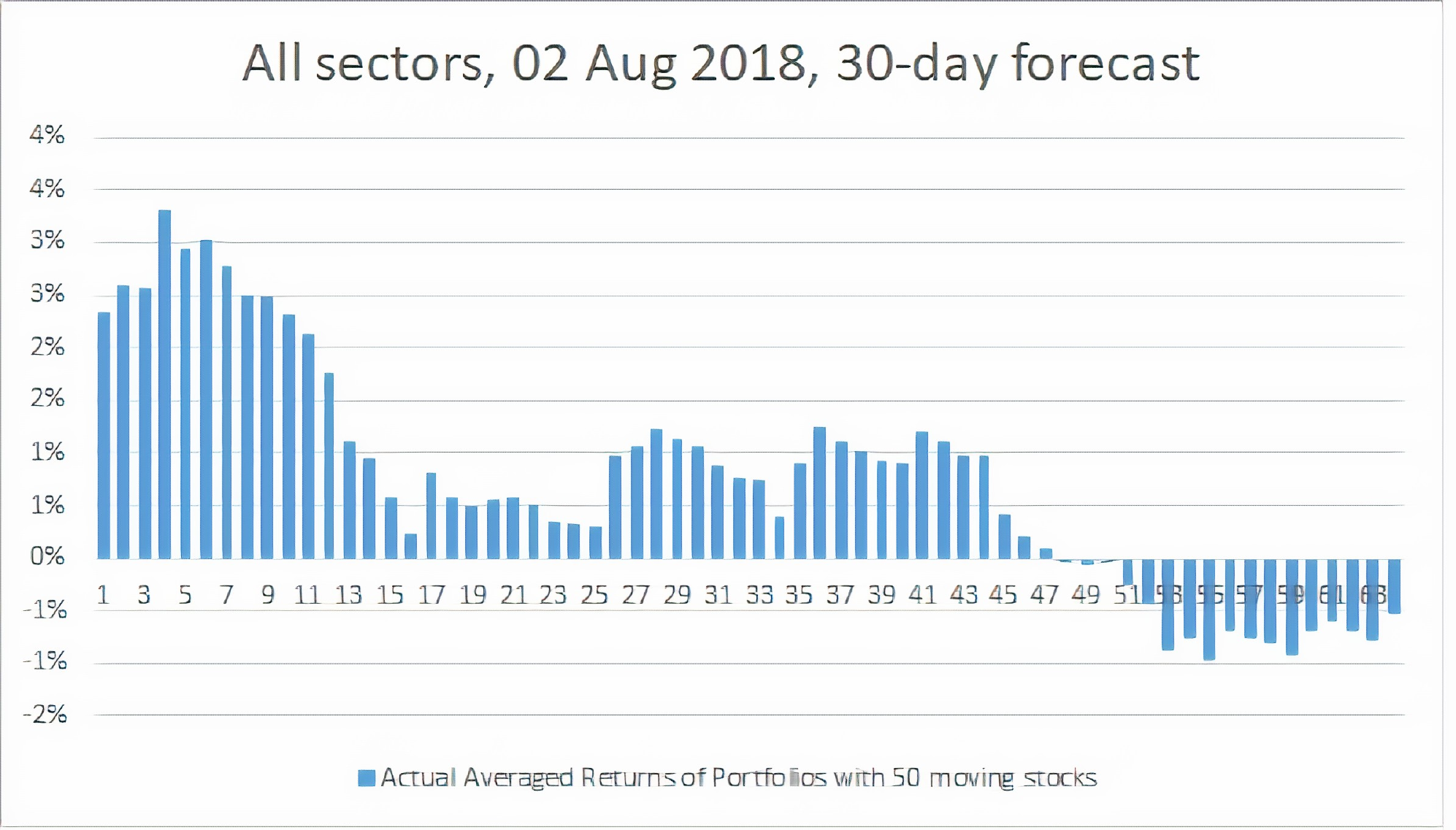}
  \caption{02 Aug 2018. Actual returns of moving portfolios consisting of 50 stocks from all sectors that reported earnings on this date. All stocks have been pre-ranked by the model-predicted stock returns from high to low.}
  \label{Fig:02AUG2018_30d_All}
\end{figure*}

\begin{figure*}[ht]
  \centering
  \includegraphics[width=0.5\textwidth]{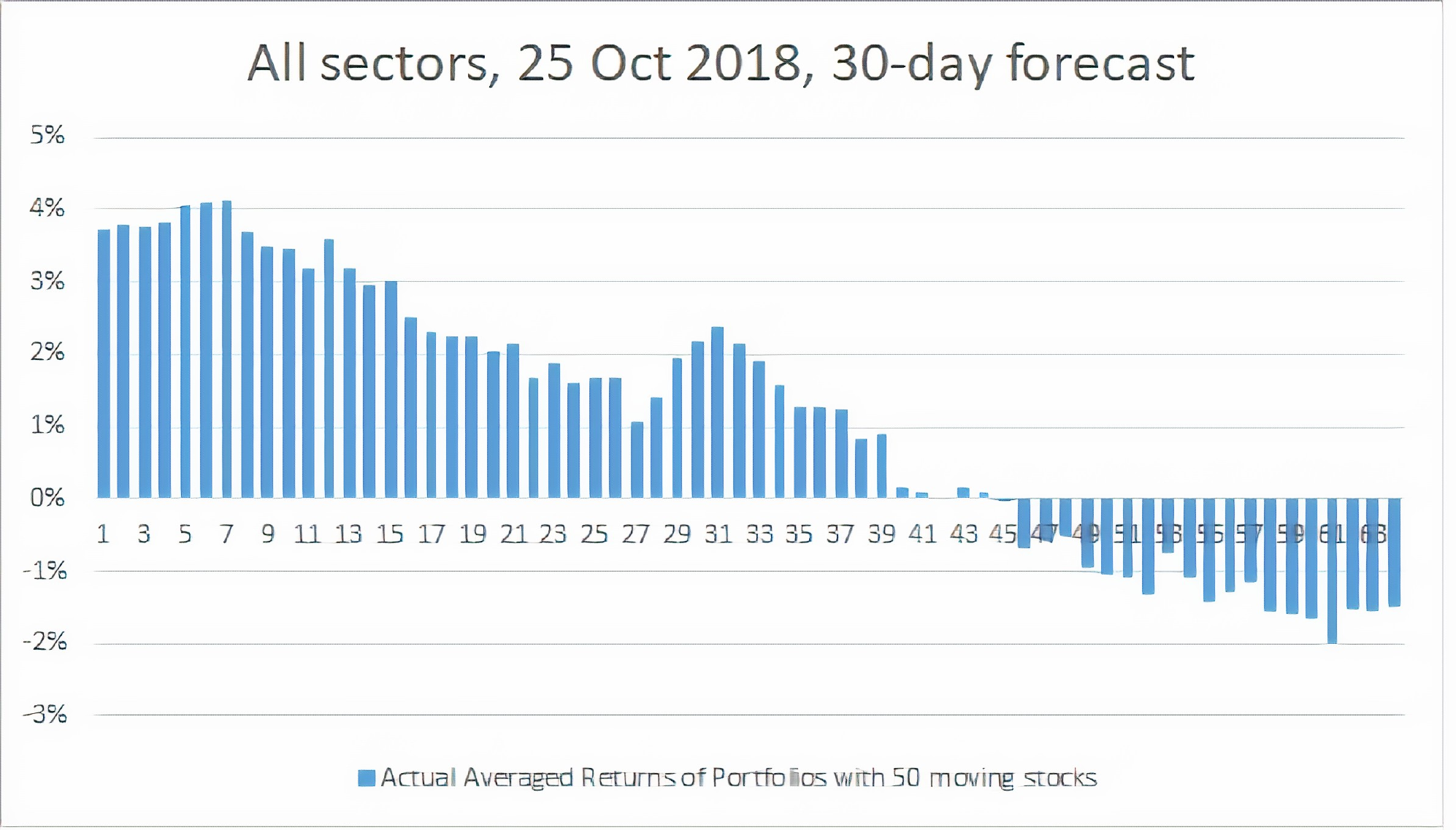}
  \caption{25 Oct 2018. Actual returns of moving portfolios consisting of 50 stocks from all sectors that reported earnings on this date. All stocks have been pre-ranked by the model-predicted stock returns from high to low.}
  \label{Fig:25OCT2018_30d_All}
\end{figure*}

\begin{table*}[]
\centering
\begin{tabular}{llllll}
\hline
\multicolumn{1}{|l|}{\textbf{\begin{tabular}[c]{@{}l@{}}Out-of-sample\\   Time Frame\end{tabular}}} & \multicolumn{1}{l|}{\textbf{Industries}} & \multicolumn{1}{l|}{\textbf{\begin{tabular}[c]{@{}l@{}}Forecast\\   Holding Period\end{tabular}}} & \multicolumn{1}{l|}{\textbf{\begin{tabular}[c]{@{}l@{}}Top Quantile\\   Portfolio Return\end{tabular}}} & \multicolumn{1}{l|}{\textbf{\begin{tabular}[c]{@{}l@{}}Average Actual\\   Return\end{tabular}}} & \multicolumn{1}{l|}{\textbf{\begin{tabular}[c]{@{}l@{}}Bottom Quantile\\   Portfolio Return\end{tabular}}} \\ \hline
\multicolumn{1}{|l|}{Q4 2018} & \multicolumn{1}{l|}{All}                 & \multicolumn{1}{l|}{30 days} & \multicolumn{1}{l|}{3.90\%} & \multicolumn{1}{l|}{-0.29\,\%} & \multicolumn{1}{l|}{-3.76\,\%} \\ \hline
\multicolumn{1}{|l|}{Q3 2018} & \multicolumn{1}{l|}{All}                 & \multicolumn{1}{l|}{30 days} & \multicolumn{1}{l|}{4.09\,\%} & \multicolumn{1}{l|}{0.36\%} & \multicolumn{1}{l|}{-4.78\,\%}  \\ \hline
 & & & & & \\ \hline
% \multicolumn{1}{|l|}{Q4 2018} & \multicolumn{1}{l|}{Financials}          & \multicolumn{1}{l|}{30 days} & \multicolumn{1}{l|}{0.92\,\%} & \multicolumn{1}{l|}{-1.13\%} & \multicolumn{1}{l|}{-3.35\,\%} \\ \hline
% \multicolumn{1}{|l|}{Q3 2018} & \multicolumn{1}{l|}{Financials}          & \multicolumn{1}{l|}{30 days} & \multicolumn{1}{l|}{2.97\%} & \multicolumn{1}{l|}{1.46\,\%} & \multicolumn{1}{l|}{-0.56\,\%} \\ \hline
%  & & & & & \\ \hline
\multicolumn{1}{|l|}{02-Aug-18} & \multicolumn{1}{l|}{All}                 & \multicolumn{1}{l|}{30 days} & \multicolumn{1}{l|}{2.75\,\%} & \multicolumn{1}{l|}{1.16\,\%} & \multicolumn{1}{l|}{0.54\,\%} \\ \hline
\multicolumn{1}{|l|}{25-Oct-18} & \multicolumn{1}{l|}{All}                 & \multicolumn{1}{l|}{30 days} & \multicolumn{1}{l|}{3.83\%} & \multicolumn{1}{l|}{1.06\,\%} & \multicolumn{1}{l|}{-1.64\,\%} \\ \hline
\end{tabular}
\caption{\label{tab:portfolioreturnstats}Actual returns of portfolios consisting of top and bottom quantile stocks. Stocks have been ranked by their predicted returns}
\end{table*}

\subsubsection{\label{sec:level3}Stocks that reported earnings on the same date}

If we were to construct market neutral portfolios, practically speaking it would only make sense if we could execute the buying and short-selling of model-chosen stocks within a short time frame, such as within a day or ideally less. Here, we run the same portfolio test on stocks which filed for earnings with SEC on the same date. Two dates in 2018 with busy earnings release activities were chosen for demonstration. Figures \ref{Fig:02AUG2018_30d_All} and \ref{Fig:25OCT2018_30d_All} are returns of portfolios created using stocks that reported earnings on each date. The stocks have been ranked by their model-predicted 30-day post earning CAR before being grouped into portfolios. Once again, the combination of the XGBoost-GA model and our engineered input features is producing the kind of results which can be used to rank stocks and construct portfolios, which would produce higher positive returns or lower negative returns. 

Table \ref{tab:portfolioreturnstats} gives the stats on how the top quantile portfolios and bottom quantile portfolios are performing compared against the average return of all the out-of-sample stocks. In some cases, returns from portfolios consisting of top/bottom quantile stocks are considerably higher/lower than the out-of-sample stock population's average.
%BS: never use "significantly" w/o reporting test method and p-value - changed to "considerably"
Such patterns of portfolio returns could have
%BS: no colloquial style! (could've...)
theoretically made them good candidates for a long-short strategy capitalising on the events of earnings release. This, however, is made difficult in reality, as we will discuss more about the timeliness of the signals.

\subsection{\label{sec:level2}Trading on Earning Event Signals}

\begin{table*}[]
\centering
%\begin{tabular}{|c|c|c|c|c|l|}
\begin{tabular}{| >{\centering\arraybackslash}m{1in} | >{\centering\arraybackslash}m{1.2in} | >{\centering\arraybackslash}m{1.2in} | >{\centering\arraybackslash}m{1in} | >{\centering\arraybackslash}m{1.7in}| }
\hline
 Year tested    & No.\ of out-of-samples before filtering & No.\ of out-of-samples after filtering& $t_{0}$ to $t_{30}$ accuracy by model & \textbf{Inferred accuracy $t_{1}$ to $t_{30}$ for remaining stocks}  \\ \hline
2016 & 3635 & 254 & 71.40\,\% & \textbf{58.30}\,\%   \\ \hline
2017 & 3715 & 151 & 70.80\,\% & \textbf{60.30}\,\%   \\ \hline
2018 & 3728 & 261 & 69.90\,\% & \textbf{57.50}\,\%   \\ \hline
\end{tabular}
\caption{\label{tab:InferredDirection}Accuracy of inferring stock direction from $t_{1}$ to $t_{30}$ using model predicted direction from $t_{0}$ to $t_{30}$ and the known stock movement from $t_{0}$ to $t_{1}$}
\end{table*}

In the aforementioned experiments we have chosen the last publicly available tradable stock price before an earnings release as the starting point of a 30 day forecasting period. This is an intuitive choice and commonly seen in the literature. For instance, Erlien \cite{Erlien2011} uses the end point of her training window as the beginning of a calculation window for cumulative abnormal returns. Similarly, when examining how numerous factors drives the revision of analysts' consensus forecast on a company's EPS, Ahmed and Irfan \cite{Ahmed2017} collect the final consensus available prior to earnings announcement to start the forecast period.  

However in reality a company's stock price moves on receipt of the first trickle of news. Information is never symmetrical, and some parties always possess greater material knowledge than others. They can and will act on such material information driving the stock price away from the last tradable price before the wider market gains access to the same level of information. Also, the incorporation of earnings information into the latest price is hugely accelerated by the presence of algorithmic trading systems as verified by Frino et al.\ who studied a unique dataset obtained from the Australian Securities Exchange \cite{Frino2016}. Correct forecasting of stock movements upon financial events is not practically useful unless they can be acted upon.

With this in mind, we have attempted to forecast cumulative abnormal returns from 1 day \textit{after} the announcement of news to 30 day after, i.\,e., CAR from $t_{1}$ to $t_{30}$. Our results show that the forecasting is inferior with accuracy of around 50\,\% and sometimes less and cannot be relied upon. This is not at all surprising, because, as per the efficient market hypothesis any granular earnings information embedded in the financial statements and management's guidance, coupled with the market's own interpretations, will have been mostly consumed by the markets and reflected in the latest stock prices not too long after the announcement. A similar observation was already seen by Allen and Karjalainen \cite{Franklin1999} that introducing a one-day delay to trading signals removes most of the forecasting ability when they used GAs to find technical trading rules. 

Since we are not able to accurately forecast the direction of CAR from $t_{t+\Delta t}$ to $t_{t+T}$ (with $\Delta t$ being non-negligible) using newly released financial statements data and a stock's momentum signals prior to announcement, we have devised a tactic to infer the delayed-starting direction. In this case we wait until 1 day after earnings release. It is important to point out that since we intend to trade on a stock's movement from $t_{1}$ to $t_{30}$, our standpoint is now 1 day after the earnings announcement, and we are already in possession of the knowledge of how a stock has moved from $t_{0}$ to $t_{1}$. Standing near the close of the market 1 day after announcement, we would follow the following steps to infer a stock's movement from $t_{1}$ to $t_{30}$:

\begin{enumerate}
\item Stock Exclusion: Exclude all the stocks whose actual movement from $t_{0}$ to $t_{1}$ are within the interval of [-0.05\,\%, 0.05\,\%] (obtained through empirical analysis) so as to eliminate stocks with weak immediate response to earnings announcements; 
\item Re-run the forecast on PEAD direction from $t_{0}$ to $t_{30}$ by also including in the input space a stock's known movement direction from $t_{0}$ to $t_{1}$. The overall accuracy has increased to around 70\,\% due to this new input to the model;
\item Filtering: Select stocks whose real movement from $t_{0}$ to $t_{1}$ is in opposite direction compared with the predicted movement from $t_{0}$ to $t_{30}$, i.\,e., select stocks which have gone up (down) in the first day despite being forecasted to move down (up) over 30 days;
\item For all the remaining stocks, deduce a stock's movement direction from $t_{1}$ to $t_{30}$ to be the same as the predicted movement direction from $t_{0}$ to $t_{30}$.
\end{enumerate}

We test this tactic using data from 2016, 2017, and 2018, respectively. Data from preceding years are used for training. As noted in an earlier section, a company that filed in each of the four quarters of a year is considered as four independent data points. Table \ref{tab:InferredDirection} gives the results of applying this tactic on the three years. After stock exclusion and filtering the number of eligible stocks have come down to lower hundreds. With the remaining stocks we observe that we are consistently achieving close to 60\,\% accuracy in inferring the stock direction from $t_{1}$ to $t_{30}$. The important thing is that, since this tactic is meant to be exercised by a trader at or near the close of market one day after an earnings announcement, this is a signal that can genuinely be acted upon. We also expect the overall accuracy of inferring the stock direction from $t_{t+\Delta t}$ to $t_{t+T}$ to increase once we are able to further increase the PEAD prediction accuracy by the model in general. We believe this is possible, as there are other sources of impactful information that have yet to be included in the feature space, such as management's guidance, equity analyst's price revisions, other text data carried in financial reports, and meeting minutes with analysts, etc. This is another potential future research direction.

\section{Conclusion}

Post-earnings announcement drift is a well known and well studied stock market anomaly when a stock's risk adjusted price can continue in the direction of an earnings surprise in the near to mid term following an earnings release. Past  research was, however, often limited in using simpler regression based methods to explain this phenomenon, and was often confined to using a limited set of explaining factors. Even fewer research was carried out on how to potentially take advantage of this known anomaly and conduct actionable forecast on stock price movements following such a significant economic event to companies. Attempting to fill this gap in the literature, our experiment is including a much bigger set of carefully selected input factors of various types with some being specifically engineered,  sourcing the data over a longer historical time frame and attempting to forecast the direction of Cumulative Abnormal Returns (CAR) with a machine learning approach. We have adopted the state-of-the-art XGBoost and put it through a rigorous optimisation process. We not only looked at specific forecast success rates, but also examined if there is a $collective$ trend of movement enjoyed by a group of stocks following their individual earnings release. First, our results show that when properly configured using a Generic Algorithm, XGBoost produces meaningful %BS: you can only write significant when you report test method and p-value - cf. above
%BS: Conclusion is usually the (only) part in past-tense!
prediction accuracy on the direction of PEAD. We demonstrated that our selected input features were genuinely driving PEAD with a classification success rate going up to 63\,\%
%BS: classification success is a badly defined metric, and w/o knowing the chance level almost meaningless...
depending on the test scenarios. In a further breakdown, we observed that stocks from different industrial sectors and at a different time can have their PEADs driven by different primary factors. 
%BS: Please avoid show, show, show, get, get, get, etc. - I reworded...
The strengths of the driving factors are well understood by our model with stocks from certain sectors producing excellent/poor forecast results when the underlying factor dominance is more/less pronounced. Second, guided by the model's forecast outputs we found that it is possible to build portfolios which consistently offer higher positive returns and lower negative returns and such an observation could potentially form the basis of market neutral long-short trading strategy. Third, we studied the challenges of applying earning event signals in real trading. Market participants with information advantage can drive the price away before the rest of the markets have an opportunity to act on the signals. Instead of trying to buy in as soon as event data comes out, we have devised a tactic to create opportunities to delay-buy into the market at a later time using the same prediction results by the models as well as public knowledge of market movements immediately following the release of earnings data. Lastly, future efforts will need to also investigate recent methods of deep learning, which, in our preliminary experiments were inferior to the considered approach. However, their partial or combined usage such as for representation learning or data augmentation appears promising. %BS: You NEED an outlook! I added:

%BS: References are missing a lot of info which needs to be added such as page numbers. 

\bibliographystyle{ieeetr}
\bibliography{references}
\end{multicols}

\clearpage

\section{Appendix}
We are listing the most significant driving factors for stocks from each of the seven industrial sectors as indicated by our XGBoost + GA model. The results are created after running 100 tests on each group of stocks. The occurrence of features has been counted. The three features that most frequently appear as the top five driving factors are given along with their occurrence counts. These information is provided for all the years and industrial sectors that have been tested. In all the tables provided in Appendix, F1 to F5 represent the top five most impactful features. Features whose name starts with the name of a financial variable and ends with "Q\_Change" or "Y\_Change" represents the quarterly change or yearly change of the same variable. URL to the source code and source data will be provided upon acceptance of the paper.

\begin{table*}[ht]
\centering
\resizebox{0.9\textwidth}{!}{%
\begin{tabular}{lllllll}
\hline
\multicolumn{1}{|l|}{\textbf{2018}} &
  \multicolumn{1}{l|}{\textbf{Highest Occurance}} &
  \multicolumn{1}{l|}{\textbf{Count}} &
  \multicolumn{1}{l|}{\textbf{\begin{tabular}[c]{@{}l@{}}Second Highest\\   Occurance\end{tabular}}} &
  \multicolumn{1}{l|}{\textbf{Count}} &
  \multicolumn{1}{l|}{\textbf{Third Highest Occurance}} &
  \multicolumn{1}{l|}{\textbf{Count}} \\ \hline
\multicolumn{1}{|l|}{\textbf{F1}} &
  \multicolumn{1}{l|}{EPS\_Earnings\_Surprise\_Backward\_Ave\_Diff} &
  \multicolumn{1}{l|}{45} &
  \multicolumn{1}{l|}{EPS\_EarningsSurprise} &
  \multicolumn{1}{l|}{2} &
  \multicolumn{1}{l|}{EPS\_EarningsSurprise} &
  \multicolumn{1}{l|}{2} \\ \hline
\multicolumn{1}{|l|}{\textbf{F2}} &
  \multicolumn{1}{l|}{EPS\_EarningsSurprise} &
  \multicolumn{1}{l|}{31} &
  \multicolumn{1}{l|}{EPS\_Earnings\_Surprise\_Backward\_Diff} &
  \multicolumn{1}{l|}{12} &
  \multicolumn{1}{l|}{EPS\_Earnings\_Surprise\_Backward\_Ave\_Diff} &
  \multicolumn{1}{l|}{4} \\ \hline
\multicolumn{1}{|l|}{\textbf{F3}} &
  \multicolumn{1}{l|}{Total\_Liabilities\_Q\_Change} &
  \multicolumn{1}{l|}{19} &
  \multicolumn{1}{l|}{EPS\_EarningsSurprise} &
  \multicolumn{1}{l|}{12} &
  \multicolumn{1}{l|}{EPS\_Earnings\_Surprise\_Backward\_Diff} &
  \multicolumn{1}{l|}{8} \\ \hline
\multicolumn{1}{|l|}{\textbf{F4}} &
  \multicolumn{1}{l|}{Total\_Liabilities\_Q\_Change} &
  \multicolumn{1}{l|}{19} &
  \multicolumn{1}{l|}{Return\_On\_Common\_Equity} &
  \multicolumn{1}{l|}{13} &
  \multicolumn{1}{l|}{EPS\_Earnings\_Surprise\_Backward\_Diff} &
  \multicolumn{1}{l|}{8} \\ \hline
\multicolumn{1}{|l|}{\textbf{F5}} &
  \multicolumn{1}{l|}{Operating\_Income\_Y\_Change} &
  \multicolumn{1}{l|}{13} &
  \multicolumn{1}{l|}{Return\_On\_Common\_Equity} &
  \multicolumn{1}{l|}{12} &
  \multicolumn{1}{l|}{Total\_Liabilities\_Q\_Change} &
  \multicolumn{1}{l|}{8} \\ \hline
 &
   &
   &
   &
   &
   &
   \\ \hline
\multicolumn{1}{|l|}{\textbf{2017}} &
  \multicolumn{1}{l|}{\textbf{Highest Occurance}} &
  \multicolumn{1}{l|}{\textbf{Count}} &
  \multicolumn{1}{l|}{\textbf{Second Highest Occurance}} &
  \multicolumn{1}{l|}{\textbf{Count}} &
  \multicolumn{1}{l|}{\textbf{Third Highest Occurance}} &
  \multicolumn{1}{l|}{\textbf{Count}} \\ \hline
\multicolumn{1}{|l|}{\textbf{F1}} &
  \multicolumn{1}{l|}{EPS\_Earnings\_Surprise\_Backward\_Ave\_Diff} &
  \multicolumn{1}{l|}{45} &
  \multicolumn{1}{l|}{EPS\_EarningsSurprise} &
  \multicolumn{1}{l|}{4} &
  \multicolumn{1}{l|}{EPS\_Earnings\_Surprise\_Backward\_Diff} &
  \multicolumn{1}{l|}{1} \\ \hline
\multicolumn{1}{|l|}{\textbf{F2}} &
  \multicolumn{1}{l|}{EPS\_EarningsSurprise} &
  \multicolumn{1}{l|}{43} &
  \multicolumn{1}{l|}{EPS\_Earnings\_Surprise\_Backward\_Diff} &
  \multicolumn{1}{l|}{4} &
  \multicolumn{1}{l|}{EPS\_Earnings\_Surprise\_Backward\_Ave\_Diff} &
  \multicolumn{1}{l|}{3} \\ \hline
\multicolumn{1}{|l|}{\textbf{F3}} &
  \multicolumn{1}{l|}{EPS\_Earnings\_Surprise\_Backward\_Diff} &
  \multicolumn{1}{l|}{30} &
  \multicolumn{1}{l|}{Total\_Liabilities\_Q\_Change} &
  \multicolumn{1}{l|}{10} &
  \multicolumn{1}{l|}{Operating\_Income\_Y\_Change} &
  \multicolumn{1}{l|}{4} \\ \hline
\multicolumn{1}{|l|}{\textbf{F4}} &
  \multicolumn{1}{l|}{Total\_Liabilities\_Q\_Change} &
  \multicolumn{1}{l|}{19} &
  \multicolumn{1}{l|}{Return\_On\_Common\_Equity} &
  \multicolumn{1}{l|}{13} &
  \multicolumn{1}{l|}{Operating\_Income\_Y\_Change} &
  \multicolumn{1}{l|}{10} \\ \hline
\multicolumn{1}{|l|}{\textbf{F5}} &
  \multicolumn{1}{l|}{Operating\_Income\_Y\_Change} &
  \multicolumn{1}{l|}{14} &
  \multicolumn{1}{l|}{Total\_Liabilities\_Q\_Change} &
  \multicolumn{1}{l|}{13} &
  \multicolumn{1}{l|}{Return\_On\_Common\_Equity} &
  \multicolumn{1}{l|}{12} \\ \hline
 &
   &
   &
   &
   &
   &
   \\ \hline
\multicolumn{1}{|l|}{\textbf{2016}} &
  \multicolumn{1}{l|}{\textbf{Highest Occurance}} &
  \multicolumn{1}{l|}{\textbf{Count}} &
  \multicolumn{1}{l|}{\textbf{Second Highest Occurance}} &
  \multicolumn{1}{l|}{\textbf{Count}} &
  \multicolumn{1}{l|}{\textbf{Third Highest Occurance}} &
  \multicolumn{1}{l|}{\textbf{Count}} \\ \hline
\multicolumn{1}{|l|}{\textbf{F1}} &
  \multicolumn{1}{l|}{EPS\_Earnings\_Surprise\_Backward\_Ave\_Diff} &
  \multicolumn{1}{l|}{36} &
  \multicolumn{1}{l|}{EPS\_EarningsSurprise} &
  \multicolumn{1}{l|}{13} &
  \multicolumn{1}{l|}{EPS\_Earnings\_Surprise\_Backward\_Diff} &
  \multicolumn{1}{l|}{1} \\ \hline
\multicolumn{1}{|l|}{\textbf{F2}} &
  \multicolumn{1}{l|}{EPS\_EarningsSurprise} &
  \multicolumn{1}{l|}{32} &
  \multicolumn{1}{l|}{EPS\_Earnings\_Surprise\_Backward\_Ave\_Diff} &
  \multicolumn{1}{l|}{14} &
  \multicolumn{1}{l|}{EPS\_Earnings\_Surprise\_Backward\_Diff} &
  \multicolumn{1}{l|}{3} \\ \hline
\multicolumn{1}{|l|}{\textbf{F3}} &
  \multicolumn{1}{l|}{EPS\_Earnings\_Surprise\_Backward\_Diff} &
  \multicolumn{1}{l|}{24} &
  \multicolumn{1}{l|}{Total\_Liabilities\_Q\_Change} &
  \multicolumn{1}{l|}{14} &
  \multicolumn{1}{l|}{EPS\_EarningsSurprise} &
  \multicolumn{1}{l|}{3} \\ \hline
\multicolumn{1}{|l|}{\textbf{F4}} &
  \multicolumn{1}{l|}{Total\_Liabilities\_Q\_Change} &
  \multicolumn{1}{l|}{23} &
  \multicolumn{1}{l|}{Return\_On\_Common\_Equity} &
  \multicolumn{1}{l|}{8} &
  \multicolumn{1}{l|}{EPS\_Earnings\_Surprise\_Backward\_Diff} &
  \multicolumn{1}{l|}{7} \\ \hline
\multicolumn{1}{|l|}{\textbf{F5}} &
  \multicolumn{1}{l|}{Operating\_Income\_Y\_Change} &
  \multicolumn{1}{l|}{17} &
  \multicolumn{1}{l|}{Return\_On\_Common\_Equity} &
  \multicolumn{1}{l|}{10} &
  \multicolumn{1}{l|}{Gross\_Profit\_Y\_Change} &
  \multicolumn{1}{l|}{6} \\ \hline
 &
   &
   &
   &
   &
   &
   \\ \hline
\multicolumn{1}{|l|}{\textbf{2015}} &
  \multicolumn{1}{l|}{\textbf{Highest Occurance}} &
  \multicolumn{1}{l|}{\textbf{Count}} &
  \multicolumn{1}{l|}{\textbf{Second Highest Occurance}} &
  \multicolumn{1}{l|}{\textbf{Count}} &
  \multicolumn{1}{l|}{\textbf{Third Highest Occurance}} &
  \multicolumn{1}{l|}{\textbf{Count}} \\ \hline
\multicolumn{1}{|l|}{\textbf{F1}} &
  \multicolumn{1}{l|}{EPS\_Earnings\_Surprise\_Backward\_Ave\_Diff} &
  \multicolumn{1}{l|}{27} &
  \multicolumn{1}{l|}{EPS\_EarningsSurprise} &
  \multicolumn{1}{l|}{18} &
  \multicolumn{1}{l|}{EPS\_Earnings\_Surprise\_Backward\_Diff} &
  \multicolumn{1}{l|}{5} \\ \hline
\multicolumn{1}{|l|}{\textbf{F2}} &
  \multicolumn{1}{l|}{EPS\_EarningsSurprise} &
  \multicolumn{1}{l|}{29} &
  \multicolumn{1}{l|}{EPS\_Earnings\_Surprise\_Backward\_Ave\_Diff} &
  \multicolumn{1}{l|}{9} &
  \multicolumn{1}{l|}{EPS\_Earnings\_Surprise\_Backward\_Ave\_Diff} &
  \multicolumn{1}{l|}{9} \\ \hline
\multicolumn{1}{|l|}{\textbf{F3}} &
  \multicolumn{1}{l|}{Total\_Liabilities\_Q\_Change} &
  \multicolumn{1}{l|}{28} &
  \multicolumn{1}{l|}{EPS\_Earnings\_Surprise\_Backward\_Diff} &
  \multicolumn{1}{l|}{10} &
  \multicolumn{1}{l|}{EPS\_Earnings\_Surprise\_Backward\_Ave\_Diff} &
  \multicolumn{1}{l|}{7} \\ \hline
\multicolumn{1}{|l|}{\textbf{F4}} &
  \multicolumn{1}{l|}{Total\_Liabilities\_Q\_Change} &
  \multicolumn{1}{l|}{17} &
  \multicolumn{1}{l|}{EPS\_Earnings\_Surprise\_Backward\_Diff} &
  \multicolumn{1}{l|}{16} &
  \multicolumn{1}{l|}{Return\_On\_Common\_Equity} &
  \multicolumn{1}{l|}{3} \\ \hline
\multicolumn{1}{|l|}{\textbf{F5}} &
  \multicolumn{1}{l|}{Operating\_Income\_Y\_Change} &
  \multicolumn{1}{l|}{25} &
  \multicolumn{1}{l|}{EPS\_Earnings\_Surprise\_Backward\_Diff} &
  \multicolumn{1}{l|}{6} &
  \multicolumn{1}{l|}{Return\_On\_Common\_Equity} &
  \multicolumn{1}{l|}{4} \\ \hline
 &
   &
   &
   &
   &
   &
   \\ \hline
\multicolumn{1}{|l|}{\textbf{2014}} &
  \multicolumn{1}{l|}{\textbf{Highest Occurance}} &
  \multicolumn{1}{l|}{\textbf{Count}} &
  \multicolumn{1}{l|}{\textbf{Second Highest Occurance}} &
  \multicolumn{1}{l|}{\textbf{Count}} &
  \multicolumn{1}{l|}{\textbf{Third Highest Occurance}} &
  \multicolumn{1}{l|}{\textbf{Count}} \\ \hline
\multicolumn{1}{|l|}{\textbf{F1}} &
  \multicolumn{1}{l|}{EPS\_Earnings\_Surprise\_Backward\_Ave\_Diff} &
  \multicolumn{1}{l|}{44} &
  \multicolumn{1}{l|}{EPS\_EarningsSurprise} &
  \multicolumn{1}{l|}{4} &
  \multicolumn{1}{l|}{EPS\_Earnings\_Surprise\_Backward\_Diff} &
  \multicolumn{1}{l|}{2} \\ \hline
\multicolumn{1}{|l|}{\textbf{F2}} &
  \multicolumn{1}{l|}{EPS\_EarningsSurprise} &
  \multicolumn{1}{l|}{36} &
  \multicolumn{1}{l|}{EPS\_Earnings\_Surprise\_Backward\_Diff} &
  \multicolumn{1}{l|}{6} &
  \multicolumn{1}{l|}{EPS\_Earnings\_Surprise\_Backward\_Ave\_Diff} &
  \multicolumn{1}{l|}{5} \\ \hline
\multicolumn{1}{|l|}{\textbf{F3}} &
  \multicolumn{1}{l|}{EPS\_Earnings\_Surprise\_Backward\_Diff} &
  \multicolumn{1}{l|}{26} &
  \multicolumn{1}{l|}{Total\_Liabilities\_Q\_Change} &
  \multicolumn{1}{l|}{14} &
  \multicolumn{1}{l|}{EPS\_EarningsSurprise} &
  \multicolumn{1}{l|}{5} \\ \hline
\multicolumn{1}{|l|}{\textbf{F4}} &
  \multicolumn{1}{l|}{Total\_Liabilities\_Q\_Change} &
  \multicolumn{1}{l|}{31} &
  \multicolumn{1}{l|}{EPS\_Earnings\_Surprise\_Backward\_Diff} &
  \multicolumn{1}{l|}{6} &
  \multicolumn{1}{l|}{Operating\_Income\_Y\_Change} &
  \multicolumn{1}{l|}{5} \\ \hline
\multicolumn{1}{|l|}{\textbf{F5}} &
  \multicolumn{1}{l|}{Operating\_Income\_Y\_Change} &
  \multicolumn{1}{l|}{20} &
  \multicolumn{1}{l|}{Return\_On\_Common\_Equity} &
  \multicolumn{1}{l|}{16} &
  \multicolumn{1}{l|}{Cost\_Of\_Revenue\_Q\_Change} &
  \multicolumn{1}{l|}{2} \\ \hline
\end{tabular}%
}
\caption{Top five driving factors for all stocks in each financial reporting year from 2014 to 2018. }
\label{tab:FactorOccuranceCount_All}
\end{table*}

% \usepackage{graphicx}
\begin{table*}[ht]
\centering
\resizebox{0.9\textwidth}{!}{%
\begin{tabular}{lllllll}
\hline
\multicolumn{1}{|l|}{\textbf{2018}} &
  \multicolumn{1}{l|}{\textbf{Highest Occurance}} &
  \multicolumn{1}{l|}{\textbf{Count}} &
  \multicolumn{1}{l|}{\textbf{\begin{tabular}[c]{@{}l@{}}Second Highest\\   Occurance\end{tabular}}} &
  \multicolumn{1}{l|}{\textbf{Count}} &
  \multicolumn{1}{l|}{\textbf{Third Highest Occurance}} &
  \multicolumn{1}{l|}{\textbf{Count}} \\ \hline
\multicolumn{1}{|l|}{\textbf{F1}} &
  \multicolumn{1}{l|}{EPS\_Earnings\_Surprise\_Backward\_Ave\_Diff} &
  \multicolumn{1}{l|}{94} &
  \multicolumn{1}{l|}{EPS\_Earnings\_Surprise\_Backward\_Diff} &
  \multicolumn{1}{l|}{6} &
  \multicolumn{1}{l|}{EPS\_EarningsSurprise} &
  \multicolumn{1}{l|}{0} \\ \hline
\multicolumn{1}{|l|}{\textbf{F2}} &
  \multicolumn{1}{l|}{EPS\_Earnings\_Surprise\_Backward\_Diff} &
  \multicolumn{1}{l|}{67} &
  \multicolumn{1}{l|}{EPS\_EarningsSurprise} &
  \multicolumn{1}{l|}{20} &
  \multicolumn{1}{l|}{EPS\_Earnings\_Surprise\_Backward\_Ave\_Diff} &
  \multicolumn{1}{l|}{6} \\ \hline
\multicolumn{1}{|l|}{\textbf{F3}} &
  \multicolumn{1}{l|}{EPS\_EarningsSurprise} &
  \multicolumn{1}{l|}{64} &
  \multicolumn{1}{l|}{EPS\_Earnings\_Surprise\_Backward\_Diff} &
  \multicolumn{1}{l|}{19} &
  \multicolumn{1}{l|}{Free\_Cash\_Flow\_Q\_Change} &
  \multicolumn{1}{l|}{7} \\ \hline
\multicolumn{1}{|l|}{\textbf{F4}} &
  \multicolumn{1}{l|}{Free\_Cash\_Flow\_Q\_Change} &
  \multicolumn{1}{l|}{55} &
  \multicolumn{1}{l|}{Return\_On\_Assets\_Q\_Change} &
  \multicolumn{1}{l|}{9} &
  \multicolumn{1}{l|}{Return\_On\_Assets\_Q\_Change} &
  \multicolumn{1}{l|}{9} \\ \hline
\multicolumn{1}{|l|}{\textbf{F5}} &
  \multicolumn{1}{l|}{Return\_On\_Assets\_Q\_Change} &
  \multicolumn{1}{l|}{20} &
  \multicolumn{1}{l|}{Return\_On\_Assets\_Y\_Change} &
  \multicolumn{1}{l|}{12} &
  \multicolumn{1}{l|}{Free\_Cash\_Flow\_Q\_Change} &
  \multicolumn{1}{l|}{11} \\ \hline
 &
   &
   &
   &
   &
   &
   \\ \hline
\multicolumn{1}{|l|}{\textbf{2017}} &
  \multicolumn{1}{l|}{\textbf{Highest Occurance}} &
  \multicolumn{1}{l|}{\textbf{Count}} &
  \multicolumn{1}{l|}{\textbf{Second Highest Occurance}} &
  \multicolumn{1}{l|}{\textbf{Count}} &
  \multicolumn{1}{l|}{\textbf{Third Highest Occurance}} &
  \multicolumn{1}{l|}{\textbf{Count}} \\ \hline
\multicolumn{1}{|l|}{\textbf{F1}} &
  \multicolumn{1}{l|}{EPS\_Earnings\_Surprise\_Backward\_Ave\_Diff} &
  \multicolumn{1}{l|}{52} &
  \multicolumn{1}{l|}{EPS\_Earnings\_Surprise\_Backward\_Diff} &
  \multicolumn{1}{l|}{39} &
  \multicolumn{1}{l|}{EPS\_EarningsSurprise} &
  \multicolumn{1}{l|}{9} \\ \hline
\multicolumn{1}{|l|}{\textbf{F2}} &
  \multicolumn{1}{l|}{EPS\_Earnings\_Surprise\_Backward\_Diff} &
  \multicolumn{1}{l|}{52} &
  \multicolumn{1}{l|}{EPS\_Earnings\_Surprise\_Backward\_Ave\_Diff} &
  \multicolumn{1}{l|}{34} &
  \multicolumn{1}{l|}{EPS\_EarningsSurprise} &
  \multicolumn{1}{l|}{12} \\ \hline
\multicolumn{1}{|l|}{\textbf{F3}} &
  \multicolumn{1}{l|}{EPS\_EarningsSurprise} &
  \multicolumn{1}{l|}{66} &
  \multicolumn{1}{l|}{EPS\_Earnings\_Surprise\_Backward\_Ave\_Diff} &
  \multicolumn{1}{l|}{9} &
  \multicolumn{1}{l|}{Return\_On\_Assets\_Q\_Change} &
  \multicolumn{1}{l|}{8} \\ \hline
\multicolumn{1}{|l|}{\textbf{F4}} &
  \multicolumn{1}{l|}{Return\_On\_Assets\_Q\_Change} &
  \multicolumn{1}{l|}{36} &
  \multicolumn{1}{l|}{PC\_Ratios\_Y\_Change} &
  \multicolumn{1}{l|}{16} &
  \multicolumn{1}{l|}{Current\_Ratio\_Q\_Change} &
  \multicolumn{1}{l|}{11} \\ \hline
\multicolumn{1}{|l|}{\textbf{F5}} &
  \multicolumn{1}{l|}{Return\_On\_Assets\_Q\_Change} &
  \multicolumn{1}{l|}{24} &
  \multicolumn{1}{l|}{RSI-30D} &
  \multicolumn{1}{l|}{11} &
  \multicolumn{1}{l|}{Free\_Cash\_Flow\_Q\_Change} &
  \multicolumn{1}{l|}{10} \\ \hline
 &
   &
   &
   &
   &
   &
   \\ \hline
\multicolumn{1}{|l|}{\textbf{2016}} &
  \multicolumn{1}{l|}{\textbf{Highest Occurance}} &
  \multicolumn{1}{l|}{\textbf{Count}} &
  \multicolumn{1}{l|}{\textbf{Second Highest Occurance}} &
  \multicolumn{1}{l|}{\textbf{Count}} &
  \multicolumn{1}{l|}{\textbf{Third Highest Occurance}} &
  \multicolumn{1}{l|}{\textbf{Count}} \\ \hline
\multicolumn{1}{|l|}{\textbf{F1}} &
  \multicolumn{1}{l|}{EPS\_Earnings\_Surprise\_Backward\_Ave\_Diff} &
  \multicolumn{1}{l|}{64} &
  \multicolumn{1}{l|}{EPS\_Earnings\_Surprise\_Backward\_Diff} &
  \multicolumn{1}{l|}{24} &
  \multicolumn{1}{l|}{EPS\_EarningsSurprise} &
  \multicolumn{1}{l|}{10} \\ \hline
\multicolumn{1}{|l|}{\textbf{F2}} &
  \multicolumn{1}{l|}{EPS\_Earnings\_Surprise\_Backward\_Diff} &
  \multicolumn{1}{l|}{51} &
  \multicolumn{1}{l|}{EPS\_Earnings\_Surprise\_Backward\_Ave\_Diff} &
  \multicolumn{1}{l|}{22} &
  \multicolumn{1}{l|}{EPS\_EarningsSurprise} &
  \multicolumn{1}{l|}{18} \\ \hline
\multicolumn{1}{|l|}{\textbf{F3}} &
  \multicolumn{1}{l|}{EPS\_EarningsSurprise} &
  \multicolumn{1}{l|}{46} &
  \multicolumn{1}{l|}{Return\_On\_Assets\_Q\_Change} &
  \multicolumn{1}{l|}{18} &
  \multicolumn{1}{l|}{EPS\_Earnings\_Surprise\_Backward\_Diff} &
  \multicolumn{1}{l|}{17} \\ \hline
\multicolumn{1}{|l|}{\textbf{F4}} &
  \multicolumn{1}{l|}{Return\_On\_Assets\_Q\_Change} &
  \multicolumn{1}{l|}{27} &
  \multicolumn{1}{l|}{EPS\_EarningsSurprise} &
  \multicolumn{1}{l|}{14} &
  \multicolumn{1}{l|}{EPS\_EarningsSurprise} &
  \multicolumn{1}{l|}{14} \\ \hline
\multicolumn{1}{|l|}{\textbf{F5}} &
  \multicolumn{1}{l|}{Return\_On\_Assets\_Q\_Change} &
  \multicolumn{1}{l|}{23} &
  \multicolumn{1}{l|}{Free\_Cash\_Flow\_Q\_Change} &
  \multicolumn{1}{l|}{20} &
  \multicolumn{1}{l|}{Current\_Ratio\_Q\_Change} &
  \multicolumn{1}{l|}{17} \\ \hline
 &
   &
   &
   &
   &
   &
   \\ \hline
\multicolumn{1}{|l|}{\textbf{2015}} &
  \multicolumn{1}{l|}{\textbf{Highest Occurance}} &
  \multicolumn{1}{l|}{\textbf{Count}} &
  \multicolumn{1}{l|}{\textbf{Second Highest Occurance}} &
  \multicolumn{1}{l|}{\textbf{Count}} &
  \multicolumn{1}{l|}{\textbf{Third Highest Occurance}} &
  \multicolumn{1}{l|}{\textbf{Count}} \\ \hline
\multicolumn{1}{|l|}{\textbf{F1}} &
  \multicolumn{1}{l|}{EPS\_Earnings\_Surprise\_Backward\_Ave\_Diff} &
  \multicolumn{1}{l|}{72} &
  \multicolumn{1}{l|}{EPS\_Earnings\_Surprise\_Backward\_Diff} &
  \multicolumn{1}{l|}{15} &
  \multicolumn{1}{l|}{EPS\_EarningsSurprise} &
  \multicolumn{1}{l|}{12} \\ \hline
\multicolumn{1}{|l|}{\textbf{F2}} &
  \multicolumn{1}{l|}{EPS\_Earnings\_Surprise\_Backward\_Diff} &
  \multicolumn{1}{l|}{39} &
  \multicolumn{1}{l|}{EPS\_EarningsSurprise} &
  \multicolumn{1}{l|}{36} &
  \multicolumn{1}{l|}{EPS\_Earnings\_Surprise\_Backward\_Ave\_Diff} &
  \multicolumn{1}{l|}{16} \\ \hline
\multicolumn{1}{|l|}{\textbf{F3}} &
  \multicolumn{1}{l|}{EPS\_EarningsSurprise} &
  \multicolumn{1}{l|}{46} &
  \multicolumn{1}{l|}{EPS\_Earnings\_Surprise\_Backward\_Diff} &
  \multicolumn{1}{l|}{31} &
  \multicolumn{1}{l|}{EPS\_Earnings\_Surprise\_Backward\_Ave\_Diff} &
  \multicolumn{1}{l|}{8} \\ \hline
\multicolumn{1}{|l|}{\textbf{F4}} &
  \multicolumn{1}{l|}{Free\_Cash\_Flow\_Q\_Change} &
  \multicolumn{1}{l|}{27} &
  \multicolumn{1}{l|}{Return\_On\_Assets\_Y\_Change} &
  \multicolumn{1}{l|}{14} &
  \multicolumn{1}{l|}{Return\_On\_Assets\_Q\_Change} &
  \multicolumn{1}{l|}{11} \\ \hline
\multicolumn{1}{|l|}{\textbf{F5}} &
  \multicolumn{1}{l|}{Return\_On\_Assets\_Y\_Change} &
  \multicolumn{1}{l|}{19} &
  \multicolumn{1}{l|}{PC\_Ratios\_Y\_Change} &
  \multicolumn{1}{l|}{15} &
  \multicolumn{1}{l|}{Return\_On\_Assets\_Q\_Change} &
  \multicolumn{1}{l|}{14} \\ \hline
 &
   &
   &
   &
   &
   &
   \\ \hline
\multicolumn{1}{|l|}{\textbf{2014}} &
  \multicolumn{1}{l|}{\textbf{Highest Occurance}} &
  \multicolumn{1}{l|}{\textbf{Count}} &
  \multicolumn{1}{l|}{\textbf{Second Highest Occurance}} &
  \multicolumn{1}{l|}{\textbf{Count}} &
  \multicolumn{1}{l|}{\textbf{Third Highest Occurance}} &
  \multicolumn{1}{l|}{\textbf{Count}} \\ \hline
\multicolumn{1}{|l|}{\textbf{F1}} &
  \multicolumn{1}{l|}{EPS\_Earnings\_Surprise\_Backward\_Ave\_Diff} &
  \multicolumn{1}{l|}{65} &
  \multicolumn{1}{l|}{EPS\_Earnings\_Surprise\_Backward\_Diff} &
  \multicolumn{1}{l|}{19} &
  \multicolumn{1}{l|}{EPS\_EarningsSurprise} &
  \multicolumn{1}{l|}{14} \\ \hline
\multicolumn{1}{|l|}{\textbf{F2}} &
  \multicolumn{1}{l|}{EPS\_Earnings\_Surprise\_Backward\_Diff} &
  \multicolumn{1}{l|}{54} &
  \multicolumn{1}{l|}{EPS\_EarningsSurprise} &
  \multicolumn{1}{l|}{19} &
  \multicolumn{1}{l|}{EPS\_Earnings\_Surprise\_Backward\_Ave\_Diff} &
  \multicolumn{1}{l|}{15} \\ \hline
\multicolumn{1}{|l|}{\textbf{F3}} &
  \multicolumn{1}{l|}{EPS\_EarningsSurprise} &
  \multicolumn{1}{l|}{53} &
  \multicolumn{1}{l|}{EPS\_Earnings\_Surprise\_Backward\_Diff} &
  \multicolumn{1}{l|}{21} &
  \multicolumn{1}{l|}{EPS\_Earnings\_Surprise\_Backward\_Ave\_Diff} &
  \multicolumn{1}{l|}{10} \\ \hline
\multicolumn{1}{|l|}{\textbf{F4}} &
  \multicolumn{1}{l|}{Free\_Cash\_Flow\_Q\_Change} &
  \multicolumn{1}{l|}{18} &
  \multicolumn{1}{l|}{Current\_Ratio\_Q\_Change} &
  \multicolumn{1}{l|}{15} &
  \multicolumn{1}{l|}{PC\_Ratios\_Y\_Change} &
  \multicolumn{1}{l|}{14} \\ \hline
\multicolumn{1}{|l|}{\textbf{F5}} &
  \multicolumn{1}{l|}{Free\_Cash\_Flow\_Q\_Change} &
  \multicolumn{1}{l|}{24} &
  \multicolumn{1}{l|}{Return\_On\_Assets\_Q\_Change} &
  \multicolumn{1}{l|}{14} &
  \multicolumn{1}{l|}{RSI-30D} &
  \multicolumn{1}{l|}{9} \\ \hline
\end{tabular}%
}
\caption{Top five driving factors for the \textbf{Industrial} stocks in each financial reporting year from 2014 to 2018.}
\label{tab:FactorOccuranceCount_Industrial}
\end{table*}

% \usepackage{graphicx}
\begin{table*}[ht]
\centering
\resizebox{0.9\textwidth}{!}{%
\begin{tabular}{lllllll}
\hline
\multicolumn{1}{|l|}{\textbf{2018}} &
  \multicolumn{1}{l|}{\textbf{Highest Occurance}} &
  \multicolumn{1}{l|}{\textbf{Count}} &
  \multicolumn{1}{l|}{\textbf{\begin{tabular}[c]{@{}l@{}}Second Highest\\   Occurance\end{tabular}}} &
  \multicolumn{1}{l|}{\textbf{Count}} &
  \multicolumn{1}{l|}{\textbf{Third Highest Occurance}} &
  \multicolumn{1}{l|}{\textbf{Count}} \\ \hline
\multicolumn{1}{|l|}{\textbf{F1}} &
  \multicolumn{1}{l|}{EPS\_Earnings\_Surprise\_Backward\_Ave\_Diff} &
  \multicolumn{1}{l|}{36} &
  \multicolumn{1}{l|}{Total\_Liabilities\_Q\_Change} &
  \multicolumn{1}{l|}{11} &
  \multicolumn{1}{l|}{PC\_Ratios} &
  \multicolumn{1}{l|}{6} \\ \hline
\multicolumn{1}{|l|}{\textbf{F2}} &
  \multicolumn{1}{l|}{EPS\_Earnings\_Surprise\_Backward\_Ave\_Diff} &
  \multicolumn{1}{l|}{17} &
  \multicolumn{1}{l|}{Dividend\_Payout\_Ratio} &
  \multicolumn{1}{l|}{12} &
  \multicolumn{1}{l|}{Total\_Liabilities\_Q\_Change} &
  \multicolumn{1}{l|}{9} \\ \hline
\multicolumn{1}{|l|}{\textbf{F3}} &
  \multicolumn{1}{l|}{EPS\_Earnings\_Surprise\_Backward\_Ave\_Diff} &
  \multicolumn{1}{l|}{10} &
  \multicolumn{1}{l|}{DMA\_50D/200D} &
  \multicolumn{1}{l|}{9} &
  \multicolumn{1}{l|}{Dividend\_Payout\_Ratio} &
  \multicolumn{1}{l|}{7} \\ \hline
\multicolumn{1}{|l|}{\textbf{F4}} &
  \multicolumn{1}{l|}{EPS\_EarningsSurprise} &
  \multicolumn{1}{l|}{9} &
  \multicolumn{1}{l|}{EPS\_EarningsSurprise} &
  \multicolumn{1}{l|}{9} &
  \multicolumn{1}{l|}{PC\_Ratios} &
  \multicolumn{1}{l|}{7} \\ \hline
\multicolumn{1}{|l|}{\textbf{F5}} &
  \multicolumn{1}{l|}{EPS\_EarningsSurprise} &
  \multicolumn{1}{l|}{8} &
  \multicolumn{1}{l|}{PC\_Ratios} &
  \multicolumn{1}{l|}{6} &
  \multicolumn{1}{l|}{PC\_Ratios} &
  \multicolumn{1}{l|}{6} \\ \hline
 &
   &
   &
   &
   &
   &
   \\ \hline
\multicolumn{1}{|l|}{\textbf{2017}} &
  \multicolumn{1}{l|}{\textbf{Highest Occurance}} &
  \multicolumn{1}{l|}{\textbf{Count}} &
  \multicolumn{1}{l|}{\textbf{Second Highest Occurance}} &
  \multicolumn{1}{l|}{\textbf{Count}} &
  \multicolumn{1}{l|}{\textbf{Third Highest Occurance}} &
  \multicolumn{1}{l|}{\textbf{Count}} \\ \hline
\multicolumn{1}{|l|}{\textbf{F1}} &
  \multicolumn{1}{l|}{Dividend\_Payout\_Ratio} &
  \multicolumn{1}{l|}{31} &
  \multicolumn{1}{l|}{DMA\_50D/200D} &
  \multicolumn{1}{l|}{9} &
  \multicolumn{1}{l|}{EPS\_EarningsSurprise} &
  \multicolumn{1}{l|}{3} \\ \hline
\multicolumn{1}{|l|}{\textbf{F2}} &
  \multicolumn{1}{l|}{Dividend\_Payout\_Ratio} &
  \multicolumn{1}{l|}{14} &
  \multicolumn{1}{l|}{EPS\_Earnings\_Surprise\_Backward\_Ave\_Diff} &
  \multicolumn{1}{l|}{8} &
  \multicolumn{1}{l|}{DMA\_50D/200D} &
  \multicolumn{1}{l|}{7} \\ \hline
\multicolumn{1}{|l|}{\textbf{F3}} &
  \multicolumn{1}{l|}{DMA\_50D/200D} &
  \multicolumn{1}{l|}{11} &
  \multicolumn{1}{l|}{Dividend\_Payout\_Ratio} &
  \multicolumn{1}{l|}{10} &
  \multicolumn{1}{l|}{EPS\_Earnings\_Surprise\_Backward\_Ave\_Diff} &
  \multicolumn{1}{l|}{9} \\ \hline
\multicolumn{1}{|l|}{\textbf{F4}} &
  \multicolumn{1}{l|}{EPS\_Earnings\_Surprise\_Backward\_Ave\_Diff} &
  \multicolumn{1}{l|}{10} &
  \multicolumn{1}{l|}{DMA\_50D/200D} &
  \multicolumn{1}{l|}{6} &
  \multicolumn{1}{l|}{EPS\_EarningsSurprise} &
  \multicolumn{1}{l|}{3} \\ \hline
\multicolumn{1}{|l|}{\textbf{F5}} &
  \multicolumn{1}{l|}{PE\_Ratios\_Q\_Change} &
  \multicolumn{1}{l|}{8} &
  \multicolumn{1}{l|}{EPS\_Earnings\_Surprise\_Backward\_Ave\_Diff} &
  \multicolumn{1}{l|}{6} &
  \multicolumn{1}{l|}{EPS\_EarningsSurprise} &
  \multicolumn{1}{l|}{5} \\ \hline
 &
   &
   &
   &
   &
   &
   \\ \hline
\multicolumn{1}{|l|}{\textbf{2016}} &
  \multicolumn{1}{l|}{\textbf{Highest Occurance}} &
  \multicolumn{1}{l|}{\textbf{Count}} &
  \multicolumn{1}{l|}{\textbf{Second Highest Occurance}} &
  \multicolumn{1}{l|}{\textbf{Count}} &
  \multicolumn{1}{l|}{\textbf{Third Highest Occurance}} &
  \multicolumn{1}{l|}{\textbf{Count}} \\ \hline
\multicolumn{1}{|l|}{\textbf{F1}} &
  \multicolumn{1}{l|}{EPS\_EarningsSurprise} &
  \multicolumn{1}{l|}{15} &
  \multicolumn{1}{l|}{EPS\_EarningsSurprise} &
  \multicolumn{1}{l|}{15} &
  \multicolumn{1}{l|}{EPS\_Earnings\_Surprise\_Backward\_Ave\_Diff} &
  \multicolumn{1}{l|}{13} \\ \hline
\multicolumn{1}{|l|}{\textbf{F2}} &
  \multicolumn{1}{l|}{EPS\_Earnings\_Surprise\_Backward\_Diff} &
  \multicolumn{1}{l|}{13} &
  \multicolumn{1}{l|}{Cost\_Of\_Revenue\_Q\_Change} &
  \multicolumn{1}{l|}{10} &
  \multicolumn{1}{l|}{Cost\_Of\_Revenue\_Q\_Change} &
  \multicolumn{1}{l|}{10} \\ \hline
\multicolumn{1}{|l|}{\textbf{F3}} &
  \multicolumn{1}{l|}{Inventory\_Turnover} &
  \multicolumn{1}{l|}{8} &
  \multicolumn{1}{l|}{Cost\_Of\_Revenue\_Q\_Change} &
  \multicolumn{1}{l|}{7} &
  \multicolumn{1}{l|}{Total\_Liabilities\_Q\_Change} &
  \multicolumn{1}{l|}{6} \\ \hline
\multicolumn{1}{|l|}{\textbf{F4}} &
  \multicolumn{1}{l|}{EPS\_Earnings\_Surprise\_Backward\_Diff} &
  \multicolumn{1}{l|}{7} &
  \multicolumn{1}{l|}{EPS\_Earnings\_Surprise\_Backward\_Diff} &
  \multicolumn{1}{l|}{7} &
  \multicolumn{1}{l|}{EPS\_Earnings\_Surprise\_Backward\_Ave\_Diff} &
  \multicolumn{1}{l|}{6} \\ \hline
\multicolumn{1}{|l|}{\textbf{F5}} &
  \multicolumn{1}{l|}{EPS\_Earnings\_Surprise\_Backward\_Diff} &
  \multicolumn{1}{l|}{10} &
  \multicolumn{1}{l|}{EPS\_Earnings\_Surprise\_Backward\_Ave\_Diff} &
  \multicolumn{1}{l|}{7} &
  \multicolumn{1}{l|}{Cost\_Of\_Revenue\_Q\_Change} &
  \multicolumn{1}{l|}{5} \\ \hline
 &
   &
   &
   &
   &
   &
   \\ \hline
\multicolumn{1}{|l|}{\textbf{2015}} &
  \multicolumn{1}{l|}{\textbf{Highest Occurance}} &
  \multicolumn{1}{l|}{\textbf{Count}} &
  \multicolumn{1}{l|}{\textbf{Second Highest Occurance}} &
  \multicolumn{1}{l|}{\textbf{Count}} &
  \multicolumn{1}{l|}{\textbf{Third Highest Occurance}} &
  \multicolumn{1}{l|}{\textbf{Count}} \\ \hline
\multicolumn{1}{|l|}{\textbf{F1}} &
  \multicolumn{1}{l|}{EPS\_Earnings\_Surprise\_Backward\_Ave\_Diff} &
  \multicolumn{1}{l|}{41} &
  \multicolumn{1}{l|}{EPS\_Earnings\_Surprise\_Backward\_Diff} &
  \multicolumn{1}{l|}{6} &
  \multicolumn{1}{l|}{Dividend\_Payout\_Ratio} &
  \multicolumn{1}{l|}{4} \\ \hline
\multicolumn{1}{|l|}{\textbf{F2}} &
  \multicolumn{1}{l|}{EPS\_Earnings\_Surprise\_Backward\_Diff} &
  \multicolumn{1}{l|}{16} &
  \multicolumn{1}{l|}{EPS\_Earnings\_Surprise\_Backward\_Ave\_Diff} &
  \multicolumn{1}{l|}{13} &
  \multicolumn{1}{l|}{Dividend\_Payout\_Ratio} &
  \multicolumn{1}{l|}{7} \\ \hline
\multicolumn{1}{|l|}{\textbf{F3}} &
  \multicolumn{1}{l|}{EPS\_Earnings\_Surprise\_Backward\_Diff} &
  \multicolumn{1}{l|}{10} &
  \multicolumn{1}{l|}{Dividend\_Payout\_Ratio} &
  \multicolumn{1}{l|}{7} &
  \multicolumn{1}{l|}{Dividend\_Payout\_Ratio} &
  \multicolumn{1}{l|}{7} \\ \hline
\multicolumn{1}{|l|}{\textbf{F4}} &
  \multicolumn{1}{l|}{EPS\_Earnings\_Surprise\_Backward\_Diff} &
  \multicolumn{1}{l|}{14} &
  \multicolumn{1}{l|}{Dividend\_Payout\_Ratio} &
  \multicolumn{1}{l|}{9} &
  \multicolumn{1}{l|}{EPS\_Earnings\_Surprise\_Backward\_Ave\_Diff} &
  \multicolumn{1}{l|}{7} \\ \hline
\multicolumn{1}{|l|}{\textbf{F5}} &
  \multicolumn{1}{l|}{Dividend\_Payout\_Ratio} &
  \multicolumn{1}{l|}{11} &
  \multicolumn{1}{l|}{Total\_Liabilities\_Q\_Change} &
  \multicolumn{1}{l|}{7} &
  \multicolumn{1}{l|}{EPS\_Earnings\_Surprise\_Backward\_Diff} &
  \multicolumn{1}{l|}{6} \\ \hline
 &
   &
   &
   &
   &
   &
   \\ \hline
\multicolumn{1}{|l|}{\textbf{2014}} &
  \multicolumn{1}{l|}{\textbf{Highest Occurance}} &
  \multicolumn{1}{l|}{\textbf{Count}} &
  \multicolumn{1}{l|}{\textbf{Second Highest Occurance}} &
  \multicolumn{1}{l|}{\textbf{Count}} &
  \multicolumn{1}{l|}{\textbf{Third Highest Occurance}} &
  \multicolumn{1}{l|}{\textbf{Count}} \\ \hline
\multicolumn{1}{|l|}{\textbf{F1}} &
  \multicolumn{1}{l|}{Dividend\_Payout\_Ratio} &
  \multicolumn{1}{l|}{13} &
  \multicolumn{1}{l|}{EPS\_Earnings\_Surprise\_Backward\_Diff} &
  \multicolumn{1}{l|}{12} &
  \multicolumn{1}{l|}{EPS\_EarningsSurprise} &
  \multicolumn{1}{l|}{10} \\ \hline
\multicolumn{1}{|l|}{\textbf{F2}} &
  \multicolumn{1}{l|}{Dividend\_Payout\_Ratio} &
  \multicolumn{1}{l|}{11} &
  \multicolumn{1}{l|}{EPS\_EarningsSurprise} &
  \multicolumn{1}{l|}{9} &
  \multicolumn{1}{l|}{Total\_Liabilities\_Q\_Change} &
  \multicolumn{1}{l|}{7} \\ \hline
\multicolumn{1}{|l|}{\textbf{F3}} &
  \multicolumn{1}{l|}{Dividend\_Payout\_Ratio} &
  \multicolumn{1}{l|}{15} &
  \multicolumn{1}{l|}{EPS\_EarningsSurprise} &
  \multicolumn{1}{l|}{8} &
  \multicolumn{1}{l|}{EPS\_EarningsSurprise} &
  \multicolumn{1}{l|}{8} \\ \hline
\multicolumn{1}{|l|}{\textbf{F4}} &
  \multicolumn{1}{l|}{EPS\_Earnings\_Surprise\_Backward\_Diff} &
  \multicolumn{1}{l|}{14} &
  \multicolumn{1}{l|}{PC\_Ratios\_Y\_Change} &
  \multicolumn{1}{l|}{6} &
  \multicolumn{1}{l|}{PC\_Ratios\_Y\_Change} &
  \multicolumn{1}{l|}{6} \\ \hline
\multicolumn{1}{|l|}{\textbf{F5}} &
  \multicolumn{1}{l|}{Dividend\_Payout\_Ratio} &
  \multicolumn{1}{l|}{8} &
  \multicolumn{1}{l|}{EPS\_Earnings\_Surprise\_Backward\_Diff} &
  \multicolumn{1}{l|}{6} &
  \multicolumn{1}{l|}{EPS\_EarningsSurprise} &
  \multicolumn{1}{l|}{5} \\ \hline
\end{tabular}%
}
\caption{Top five driving factors for the \textbf{Basic Materials} stocks in each financial reporting year from 2014 to 2018. }
\label{tab:FactorOccuranceCount_BasicMaterials}
\end{table*}

% \usepackage{graphicx}
\begin{table*}[]
\centering
\resizebox{0.9\textwidth}{!}{%
\begin{tabular}{lllllll}
\hline
\multicolumn{1}{|l|}{\textbf{2018}} &
  \multicolumn{1}{l|}{\textbf{Highest Occurance}} &
  \multicolumn{1}{l|}{\textbf{Count}} &
  \multicolumn{1}{l|}{\textbf{\begin{tabular}[c]{@{}l@{}}Second Highest\\   Occurance\end{tabular}}} &
  \multicolumn{1}{l|}{\textbf{Count}} &
  \multicolumn{1}{l|}{\textbf{Third Highest Occurance}} &
  \multicolumn{1}{l|}{\textbf{Count}} \\ \hline
\multicolumn{1}{|l|}{\textbf{F1}} &
  \multicolumn{1}{l|}{EPS\_EarningsSurprise} &
  \multicolumn{1}{l|}{65} &
  \multicolumn{1}{l|}{EPS\_Earnings\_Surprise\_Backward\_Ave\_Diff} &
  \multicolumn{1}{l|}{21} &
  \multicolumn{1}{l|}{EPS\_Earnings\_Surprise\_Backward\_Diff} &
  \multicolumn{1}{l|}{7} \\ \hline
\multicolumn{1}{|l|}{\textbf{F2}} &
  \multicolumn{1}{l|}{EPS\_Earnings\_Surprise\_Backward\_Ave\_Diff} &
  \multicolumn{1}{l|}{41} &
  \multicolumn{1}{l|}{EPS\_EarningsSurprise} &
  \multicolumn{1}{l|}{20} &
  \multicolumn{1}{l|}{EPS\_Earnings\_Surprise\_Backward\_Diff} &
  \multicolumn{1}{l|}{15} \\ \hline
\multicolumn{1}{|l|}{\textbf{F3}} &
  \multicolumn{1}{l|}{Return\_On\_Common\_Equity} &
  \multicolumn{1}{l|}{29} &
  \multicolumn{1}{l|}{Return\_On\_Common\_Equity} &
  \multicolumn{1}{l|}{29} &
  \multicolumn{1}{l|}{EPS\_Earnings\_Surprise\_Backward\_Ave\_Diff} &
  \multicolumn{1}{l|}{16} \\ \hline
\multicolumn{1}{|l|}{\textbf{F4}} &
  \multicolumn{1}{l|}{Return\_On\_Common\_Equity} &
  \multicolumn{1}{l|}{31} &
  \multicolumn{1}{l|}{EPS\_Earnings\_Surprise\_Backward\_Diff} &
  \multicolumn{1}{l|}{26} &
  \multicolumn{1}{l|}{Net\_Income\_Y\_Change} &
  \multicolumn{1}{l|}{13} \\ \hline
\multicolumn{1}{|l|}{\textbf{F5}} &
  \multicolumn{1}{l|}{Net\_Income\_Y\_Change} &
  \multicolumn{1}{l|}{35} &
  \multicolumn{1}{l|}{Return\_On\_Common\_Equity} &
  \multicolumn{1}{l|}{14} &
  \multicolumn{1}{l|}{Total\_Liabilities\_Q\_Change} &
  \multicolumn{1}{l|}{7} \\ \hline
 &
   &
   &
   &
   &
   &
   \\ \hline
\multicolumn{1}{|l|}{\textbf{2017}} &
  \multicolumn{1}{l|}{\textbf{Highest Occurance}} &
  \multicolumn{1}{l|}{\textbf{Count}} &
  \multicolumn{1}{l|}{\textbf{Second Highest Occurance}} &
  \multicolumn{1}{l|}{\textbf{Count}} &
  \multicolumn{1}{l|}{\textbf{Third Highest Occurance}} &
  \multicolumn{1}{l|}{\textbf{Count}} \\ \hline
\multicolumn{1}{|l|}{\textbf{F1}} &
  \multicolumn{1}{l|}{EPS\_Earnings\_Surprise\_Backward\_Ave\_Diff} &
  \multicolumn{1}{l|}{70} &
  \multicolumn{1}{l|}{EPS\_EarningsSurprise} &
  \multicolumn{1}{l|}{19} &
  \multicolumn{1}{l|}{EPS\_Earnings\_Surprise\_Backward\_Diff} &
  \multicolumn{1}{l|}{7} \\ \hline
\multicolumn{1}{|l|}{\textbf{F2}} &
  \multicolumn{1}{l|}{EPS\_EarningsSurprise} &
  \multicolumn{1}{l|}{36} &
  \multicolumn{1}{l|}{EPS\_Earnings\_Surprise\_Backward\_Ave\_Diff} &
  \multicolumn{1}{l|}{22} &
  \multicolumn{1}{l|}{EPS\_Earnings\_Surprise\_Backward\_Diff} &
  \multicolumn{1}{l|}{17} \\ \hline
\multicolumn{1}{|l|}{\textbf{F3}} &
  \multicolumn{1}{l|}{Return\_On\_Common\_Equity} &
  \multicolumn{1}{l|}{30} &
  \multicolumn{1}{l|}{EPS\_Earnings\_Surprise\_Backward\_Diff} &
  \multicolumn{1}{l|}{21} &
  \multicolumn{1}{l|}{EPS\_EarningsSurprise} &
  \multicolumn{1}{l|}{20} \\ \hline
\multicolumn{1}{|l|}{\textbf{F4}} &
  \multicolumn{1}{l|}{Return\_On\_Common\_Equity} &
  \multicolumn{1}{l|}{32} &
  \multicolumn{1}{l|}{EPS\_Earnings\_Surprise\_Backward\_Diff} &
  \multicolumn{1}{l|}{11} &
  \multicolumn{1}{l|}{Total\_Liabilities\_Q\_Change} &
  \multicolumn{1}{l|}{6} \\ \hline
\multicolumn{1}{|l|}{\textbf{F5}} &
  \multicolumn{1}{l|}{Total\_Liabilities\_Q\_Change} &
  \multicolumn{1}{l|}{11} &
  \multicolumn{1}{l|}{Net\_Income\_Y\_Change} &
  \multicolumn{1}{l|}{10} &
  \multicolumn{1}{l|}{PE\_Ratios} &
  \multicolumn{1}{l|}{9} \\ \hline
 &
   &
   &
   &
   &
   &
   \\ \hline
\multicolumn{1}{|l|}{\textbf{2016}} &
  \multicolumn{1}{l|}{\textbf{Highest Occurance}} &
  \multicolumn{1}{l|}{\textbf{Count}} &
  \multicolumn{1}{l|}{\textbf{Second Highest Occurance}} &
  \multicolumn{1}{l|}{\textbf{Count}} &
  \multicolumn{1}{l|}{\textbf{Third Highest Occurance}} &
  \multicolumn{1}{l|}{\textbf{Count}} \\ \hline
\multicolumn{1}{|l|}{\textbf{F1}} &
  \multicolumn{1}{l|}{EPS\_EarningsSurprise} &
  \multicolumn{1}{l|}{47} &
  \multicolumn{1}{l|}{EPS\_Earnings\_Surprise\_Backward\_Diff} &
  \multicolumn{1}{l|}{25} &
  \multicolumn{1}{l|}{EPS\_Earnings\_Surprise\_Backward\_Ave\_Diff} &
  \multicolumn{1}{l|}{21} \\ \hline
\multicolumn{1}{|l|}{\textbf{F2}} &
  \multicolumn{1}{l|}{EPS\_Earnings\_Surprise\_Backward\_Diff} &
  \multicolumn{1}{l|}{35} &
  \multicolumn{1}{l|}{EPS\_EarningsSurprise} &
  \multicolumn{1}{l|}{33} &
  \multicolumn{1}{l|}{EPS\_Earnings\_Surprise\_Backward\_Ave\_Diff} &
  \multicolumn{1}{l|}{11} \\ \hline
\multicolumn{1}{|l|}{\textbf{F3}} &
  \multicolumn{1}{l|}{EPS\_Earnings\_Surprise\_Backward\_Diff} &
  \multicolumn{1}{l|}{25} &
  \multicolumn{1}{l|}{EPS\_Earnings\_Surprise\_Backward\_Ave\_Diff} &
  \multicolumn{1}{l|}{21} &
  \multicolumn{1}{l|}{Return\_On\_Common\_Equity} &
  \multicolumn{1}{l|}{16} \\ \hline
\multicolumn{1}{|l|}{\textbf{F4}} &
  \multicolumn{1}{l|}{Return\_On\_Common\_Equity} &
  \multicolumn{1}{l|}{40} &
  \multicolumn{1}{l|}{Net\_Income\_Y\_Change} &
  \multicolumn{1}{l|}{11} &
  \multicolumn{1}{l|}{EPS\_Earnings\_Surprise\_Backward\_Ave\_Diff} &
  \multicolumn{1}{l|}{8} \\ \hline
\multicolumn{1}{|l|}{\textbf{F5}} &
  \multicolumn{1}{l|}{Net\_Income\_Y\_Change} &
  \multicolumn{1}{l|}{19} &
  \multicolumn{1}{l|}{Return\_On\_Common\_Equity} &
  \multicolumn{1}{l|}{12} &
  \multicolumn{1}{l|}{Total\_Liabilities\_Q\_Change} &
  \multicolumn{1}{l|}{11} \\ \hline
 &
   &
   &
   &
   &
   &
   \\ \hline
\multicolumn{1}{|l|}{\textbf{2015}} &
  \multicolumn{1}{l|}{\textbf{Highest Occurance}} &
  \multicolumn{1}{l|}{\textbf{Count}} &
  \multicolumn{1}{l|}{\textbf{Second Highest Occurance}} &
  \multicolumn{1}{l|}{\textbf{Count}} &
  \multicolumn{1}{l|}{\textbf{Third Highest Occurance}} &
  \multicolumn{1}{l|}{\textbf{Count}} \\ \hline
\multicolumn{1}{|l|}{\textbf{F1}} &
  \multicolumn{1}{l|}{EPS\_EarningsSurprise} &
  \multicolumn{1}{l|}{61} &
  \multicolumn{1}{l|}{EPS\_Earnings\_Surprise\_Backward\_Ave\_Diff} &
  \multicolumn{1}{l|}{16} &
  \multicolumn{1}{l|}{Return\_On\_Common\_Equity} &
  \multicolumn{1}{l|}{12} \\ \hline
\multicolumn{1}{|l|}{\textbf{F2}} &
  \multicolumn{1}{l|}{Return\_On\_Common\_Equity} &
  \multicolumn{1}{l|}{38} &
  \multicolumn{1}{l|}{EPS\_EarningsSurprise} &
  \multicolumn{1}{l|}{20} &
  \multicolumn{1}{l|}{EPS\_Earnings\_Surprise\_Backward\_Ave\_Diff} &
  \multicolumn{1}{l|}{17} \\ \hline
\multicolumn{1}{|l|}{\textbf{F3}} &
  \multicolumn{1}{l|}{EPS\_Earnings\_Surprise\_Backward\_Diff} &
  \multicolumn{1}{l|}{29} &
  \multicolumn{1}{l|}{Return\_On\_Common\_Equity} &
  \multicolumn{1}{l|}{21} &
  \multicolumn{1}{l|}{EPS\_Earnings\_Surprise\_Backward\_Ave\_Diff} &
  \multicolumn{1}{l|}{16} \\ \hline
\multicolumn{1}{|l|}{\textbf{F4}} &
  \multicolumn{1}{l|}{Net\_Income\_Y\_Change} &
  \multicolumn{1}{l|}{27} &
  \multicolumn{1}{l|}{Return\_On\_Common\_Equity} &
  \multicolumn{1}{l|}{19} &
  \multicolumn{1}{l|}{EPS\_Earnings\_Surprise\_Backward\_Ave\_Diff} &
  \multicolumn{1}{l|}{16} \\ \hline
\multicolumn{1}{|l|}{\textbf{F5}} &
  \multicolumn{1}{l|}{Net\_Income\_Y\_Change} &
  \multicolumn{1}{l|}{32} &
  \multicolumn{1}{l|}{PE\_Ratios} &
  \multicolumn{1}{l|}{12} &
  \multicolumn{1}{l|}{EPS\_Earnings\_Surprise\_Backward\_Ave\_Diff} &
  \multicolumn{1}{l|}{8} \\ \hline
 &
   &
   &
   &
   &
   &
   \\ \hline
\multicolumn{1}{|l|}{\textbf{2014}} &
  \multicolumn{1}{l|}{\textbf{Highest Occurance}} &
  \multicolumn{1}{l|}{\textbf{Count}} &
  \multicolumn{1}{l|}{\textbf{Second Highest Occurance}} &
  \multicolumn{1}{l|}{\textbf{Count}} &
  \multicolumn{1}{l|}{\textbf{Third Highest Occurance}} &
  \multicolumn{1}{l|}{\textbf{Count}} \\ \hline
\multicolumn{1}{|l|}{\textbf{F1}} &
  \multicolumn{1}{l|}{EPS\_Earnings\_Surprise\_Backward\_Diff} &
  \multicolumn{1}{l|}{42} &
  \multicolumn{1}{l|}{EPS\_EarningsSurprise} &
  \multicolumn{1}{l|}{28} &
  \multicolumn{1}{l|}{EPS\_EarningsSurprise} &
  \multicolumn{1}{l|}{28} \\ \hline
\multicolumn{1}{|l|}{\textbf{F2}} &
  \multicolumn{1}{l|}{EPS\_EarningsSurprise} &
  \multicolumn{1}{l|}{41} &
  \multicolumn{1}{l|}{EPS\_Earnings\_Surprise\_Backward\_Diff} &
  \multicolumn{1}{l|}{27} &
  \multicolumn{1}{l|}{EPS\_Earnings\_Surprise\_Backward\_Ave\_Diff} &
  \multicolumn{1}{l|}{22} \\ \hline
\multicolumn{1}{|l|}{\textbf{F3}} &
  \multicolumn{1}{l|}{Return\_On\_Common\_Equity} &
  \multicolumn{1}{l|}{22} &
  \multicolumn{1}{l|}{EPS\_Earnings\_Surprise\_Backward\_Diff} &
  \multicolumn{1}{l|}{21} &
  \multicolumn{1}{l|}{EPS\_EarningsSurprise} &
  \multicolumn{1}{l|}{18} \\ \hline
\multicolumn{1}{|l|}{\textbf{F4}} &
  \multicolumn{1}{l|}{Return\_On\_Common\_Equity} &
  \multicolumn{1}{l|}{26} &
  \multicolumn{1}{l|}{Net\_Income\_Y\_Change} &
  \multicolumn{1}{l|}{25} &
  \multicolumn{1}{l|}{EPS\_Earnings\_Surprise\_Backward\_Ave\_Diff} &
  \multicolumn{1}{l|}{8} \\ \hline
\multicolumn{1}{|l|}{\textbf{F5}} &
  \multicolumn{1}{l|}{Net\_Income\_Y\_Change} &
  \multicolumn{1}{l|}{29} &
  \multicolumn{1}{l|}{Return\_On\_Common\_Equity} &
  \multicolumn{1}{l|}{24} &
  \multicolumn{1}{l|}{EPS\_Earnings\_Surprise\_Backward\_Ave\_Diff} &
  \multicolumn{1}{l|}{10} \\ \hline
\end{tabular}%
}
\caption{Top five driving factors for the \textbf{Consumer Cyclical} stocks in each financial reporting year from 2014 to 2018. }
\label{tab:FactorOccuranceCount_ConsumerCyclical}
\end{table*}

% \usepackage{graphicx}
\begin{table*}[]
\centering
\resizebox{0.9\textwidth}{!}{%
\begin{tabular}{lllllll}
\hline
\multicolumn{1}{|l|}{\textbf{2018}} &
  \multicolumn{1}{l|}{\textbf{Highest Occurance}} &
  \multicolumn{1}{l|}{\textbf{Count}} &
  \multicolumn{1}{l|}{\textbf{\begin{tabular}[c]{@{}l@{}}Second Highest\\   Occurance\end{tabular}}} &
  \multicolumn{1}{l|}{\textbf{Count}} &
  \multicolumn{1}{l|}{\textbf{Third Highest Occurance}} &
  \multicolumn{1}{l|}{\textbf{Count}} \\ \hline
\multicolumn{1}{|l|}{\textbf{F1}} &
  \multicolumn{1}{l|}{EPS\_Earnings\_Surprise\_Backward\_Diff} &
  \multicolumn{1}{l|}{41} &
  \multicolumn{1}{l|}{EPS\_Earnings\_Surprise\_Backward\_Ave\_Diff} &
  \multicolumn{1}{l|}{28} &
  \multicolumn{1}{l|}{EPS\_EarningsSurprise} &
  \multicolumn{1}{l|}{20} \\ \hline
\multicolumn{1}{|l|}{\textbf{F2}} &
  \multicolumn{1}{l|}{EPS\_Earnings\_Surprise\_Backward\_Ave\_Diff} &
  \multicolumn{1}{l|}{27} &
  \multicolumn{1}{l|}{EPS\_EarningsSurprise} &
  \multicolumn{1}{l|}{26} &
  \multicolumn{1}{l|}{EPS\_Earnings\_Surprise\_Backward\_Diff} &
  \multicolumn{1}{l|}{23} \\ \hline
\multicolumn{1}{|l|}{\textbf{F3}} &
  \multicolumn{1}{l|}{EPS\_EarningsSurprise} &
  \multicolumn{1}{l|}{32} &
  \multicolumn{1}{l|}{EPS\_Earnings\_Surprise\_Backward\_Ave\_Diff} &
  \multicolumn{1}{l|}{15} &
  \multicolumn{1}{l|}{EPS\_Earnings\_Surprise\_Backward\_Diff} &
  \multicolumn{1}{l|}{11} \\ \hline
\multicolumn{1}{|l|}{\textbf{F4}} &
  \multicolumn{1}{l|}{Return\_On\_Common\_Equity} &
  \multicolumn{1}{l|}{19} &
  \multicolumn{1}{l|}{EPS\_EarningsSurprise} &
  \multicolumn{1}{l|}{10} &
  \multicolumn{1}{l|}{Operating\_Margin\_Y\_Change} &
  \multicolumn{1}{l|}{8} \\ \hline
\multicolumn{1}{|l|}{\textbf{F5}} &
  \multicolumn{1}{l|}{Return\_On\_Common\_Equity} &
  \multicolumn{1}{l|}{12} &
  \multicolumn{1}{l|}{Operating\_Margin\_Y\_Change} &
  \multicolumn{1}{l|}{7} &
  \multicolumn{1}{l|}{Operating\_Margin\_Y\_Change} &
  \multicolumn{1}{l|}{7} \\ \hline
 &
   &
   &
   &
   &
   &
   \\ \hline
\multicolumn{1}{|l|}{\textbf{2017}} &
  \multicolumn{1}{l|}{\textbf{Highest Occurance}} &
  \multicolumn{1}{l|}{\textbf{Count}} &
  \multicolumn{1}{l|}{\textbf{Second Highest Occurance}} &
  \multicolumn{1}{l|}{\textbf{Count}} &
  \multicolumn{1}{l|}{\textbf{Third Highest Occurance}} &
  \multicolumn{1}{l|}{\textbf{Count}} \\ \hline
\multicolumn{1}{|l|}{\textbf{F1}} &
  \multicolumn{1}{l|}{EPS\_Earnings\_Surprise\_Backward\_Ave\_Diff} &
  \multicolumn{1}{l|}{51} &
  \multicolumn{1}{l|}{EPS\_Earnings\_Surprise\_Backward\_Diff} &
  \multicolumn{1}{l|}{34} &
  \multicolumn{1}{l|}{EPS\_EarningsSurprise} &
  \multicolumn{1}{l|}{10} \\ \hline
\multicolumn{1}{|l|}{\textbf{F2}} &
  \multicolumn{1}{l|}{EPS\_Earnings\_Surprise\_Backward\_Diff} &
  \multicolumn{1}{l|}{35} &
  \multicolumn{1}{l|}{EPS\_Earnings\_Surprise\_Backward\_Ave\_Diff} &
  \multicolumn{1}{l|}{34} &
  \multicolumn{1}{l|}{EPS\_EarningsSurprise} &
  \multicolumn{1}{l|}{16} \\ \hline
\multicolumn{1}{|l|}{\textbf{F3}} &
  \multicolumn{1}{l|}{EPS\_EarningsSurprise} &
  \multicolumn{1}{l|}{53} &
  \multicolumn{1}{l|}{EPS\_Earnings\_Surprise\_Backward\_Diff} &
  \multicolumn{1}{l|}{17} &
  \multicolumn{1}{l|}{EPS\_Earnings\_Surprise\_Backward\_Ave\_Diff} &
  \multicolumn{1}{l|}{8} \\ \hline
\multicolumn{1}{|l|}{\textbf{F4}} &
  \multicolumn{1}{l|}{Operating\_Income\_Y\_Change} &
  \multicolumn{1}{l|}{22} &
  \multicolumn{1}{l|}{EPS\_EarningsSurprise} &
  \multicolumn{1}{l|}{9} &
  \multicolumn{1}{l|}{Return\_On\_Common\_Equity} &
  \multicolumn{1}{l|}{6} \\ \hline
\multicolumn{1}{|l|}{\textbf{F5}} &
  \multicolumn{1}{l|}{Operating\_Income\_Y\_Change} &
  \multicolumn{1}{l|}{14} &
  \multicolumn{1}{l|}{Return\_On\_Common\_Equity} &
  \multicolumn{1}{l|}{13} &
  \multicolumn{1}{l|}{Cash\_Q\_Change} &
  \multicolumn{1}{l|}{11} \\ \hline
 &
   &
   &
   &
   &
   &
   \\ \hline
\multicolumn{1}{|l|}{\textbf{2016}} &
  \multicolumn{1}{l|}{\textbf{Highest Occurance}} &
  \multicolumn{1}{l|}{\textbf{Count}} &
  \multicolumn{1}{l|}{\textbf{Second Highest Occurance}} &
  \multicolumn{1}{l|}{\textbf{Count}} &
  \multicolumn{1}{l|}{\textbf{Third Highest Occurance}} &
  \multicolumn{1}{l|}{\textbf{Count}} \\ \hline
\multicolumn{1}{|l|}{\textbf{F1}} &
  \multicolumn{1}{l|}{EPS\_Earnings\_Surprise\_Backward\_Ave\_Diff} &
  \multicolumn{1}{l|}{52} &
  \multicolumn{1}{l|}{EPS\_Earnings\_Surprise\_Backward\_Diff} &
  \multicolumn{1}{l|}{35} &
  \multicolumn{1}{l|}{EPS\_EarningsSurprise} &
  \multicolumn{1}{l|}{12} \\ \hline
\multicolumn{1}{|l|}{\textbf{F2}} &
  \multicolumn{1}{l|}{EPS\_Earnings\_Surprise\_Backward\_Diff} &
  \multicolumn{1}{l|}{45} &
  \multicolumn{1}{l|}{EPS\_Earnings\_Surprise\_Backward\_Ave\_Diff} &
  \multicolumn{1}{l|}{34} &
  \multicolumn{1}{l|}{EPS\_EarningsSurprise} &
  \multicolumn{1}{l|}{13} \\ \hline
\multicolumn{1}{|l|}{\textbf{F3}} &
  \multicolumn{1}{l|}{EPS\_EarningsSurprise} &
  \multicolumn{1}{l|}{46} &
  \multicolumn{1}{l|}{EPS\_Earnings\_Surprise\_Backward\_Ave\_Diff} &
  \multicolumn{1}{l|}{10} &
  \multicolumn{1}{l|}{EPS\_Earnings\_Surprise\_Backward\_Diff} &
  \multicolumn{1}{l|}{8} \\ \hline
\multicolumn{1}{|l|}{\textbf{F4}} &
  \multicolumn{1}{l|}{Return\_On\_Common\_Equity} &
  \multicolumn{1}{l|}{21} &
  \multicolumn{1}{l|}{EPS\_EarningsSurprise} &
  \multicolumn{1}{l|}{7} &
  \multicolumn{1}{l|}{EPS\_EarningsSurprise} &
  \multicolumn{1}{l|}{7} \\ \hline
\multicolumn{1}{|l|}{\textbf{F5}} &
  \multicolumn{1}{l|}{Return\_On\_Common\_Equity} &
  \multicolumn{1}{l|}{15} &
  \multicolumn{1}{l|}{Gross\_Profit\_Y\_Change} &
  \multicolumn{1}{l|}{13} &
  \multicolumn{1}{l|}{PS\_Ratios\_Y\_Change} &
  \multicolumn{1}{l|}{8} \\ \hline
 &
   &
   &
   &
   &
   &
   \\ \hline
\multicolumn{1}{|l|}{\textbf{2015}} &
  \multicolumn{1}{l|}{\textbf{Highest Occurance}} &
  \multicolumn{1}{l|}{\textbf{Count}} &
  \multicolumn{1}{l|}{\textbf{Second Highest Occurance}} &
  \multicolumn{1}{l|}{\textbf{Count}} &
  \multicolumn{1}{l|}{\textbf{Third Highest Occurance}} &
  \multicolumn{1}{l|}{\textbf{Count}} \\ \hline
\multicolumn{1}{|l|}{\textbf{F1}} &
  \multicolumn{1}{l|}{EPS\_EarningsSurprise} &
  \multicolumn{1}{l|}{67} &
  \multicolumn{1}{l|}{EPS\_Earnings\_Surprise\_Backward\_Ave\_Diff} &
  \multicolumn{1}{l|}{23} &
  \multicolumn{1}{l|}{Dividend\_Payout\_Ratio\_Q\_Change} &
  \multicolumn{1}{l|}{2} \\ \hline
\multicolumn{1}{|l|}{\textbf{F2}} &
  \multicolumn{1}{l|}{EPS\_Earnings\_Surprise\_Backward\_Ave\_Diff} &
  \multicolumn{1}{l|}{37} &
  \multicolumn{1}{l|}{EPS\_EarningsSurprise} &
  \multicolumn{1}{l|}{22} &
  \multicolumn{1}{l|}{EPS\_Earnings\_Surprise\_Backward\_Diff} &
  \multicolumn{1}{l|}{13} \\ \hline
\multicolumn{1}{|l|}{\textbf{F3}} &
  \multicolumn{1}{l|}{EPS\_Earnings\_Surprise\_Backward\_Diff} &
  \multicolumn{1}{l|}{47} &
  \multicolumn{1}{l|}{EPS\_Earnings\_Surprise\_Backward\_Ave\_Diff} &
  \multicolumn{1}{l|}{10} &
  \multicolumn{1}{l|}{EPS\_EarningsSurprise} &
  \multicolumn{1}{l|}{5} \\ \hline
\multicolumn{1}{|l|}{\textbf{F4}} &
  \multicolumn{1}{l|}{Return\_On\_Common\_Equity} &
  \multicolumn{1}{l|}{12} &
  \multicolumn{1}{l|}{Total\_Liabilities\_Q\_Change} &
  \multicolumn{1}{l|}{9} &
  \multicolumn{1}{l|}{Dividend\_Yield\_Q\_Change} &
  \multicolumn{1}{l|}{7} \\ \hline
\multicolumn{1}{|l|}{\textbf{F5}} &
  \multicolumn{1}{l|}{EPS\_Earnings\_Surprise\_Backward\_Ave\_Diff} &
  \multicolumn{1}{l|}{8} &
  \multicolumn{1}{l|}{EPS\_Earnings\_Surprise\_Backward\_Ave\_Diff} &
  \multicolumn{1}{l|}{8} &
  \multicolumn{1}{l|}{EPS\_Earnings\_Surprise\_Backward\_Ave\_Diff} &
  \multicolumn{1}{l|}{8} \\ \hline
 &
   &
   &
   &
   &
   &
   \\ \hline
\multicolumn{1}{|l|}{\textbf{2014}} &
  \multicolumn{1}{l|}{\textbf{Highest Occurance}} &
  \multicolumn{1}{l|}{\textbf{Count}} &
  \multicolumn{1}{l|}{\textbf{Second Highest Occurance}} &
  \multicolumn{1}{l|}{\textbf{Count}} &
  \multicolumn{1}{l|}{\textbf{Third Highest Occurance}} &
  \multicolumn{1}{l|}{\textbf{Count}} \\ \hline
\multicolumn{1}{|l|}{\textbf{F1}} &
  \multicolumn{1}{l|}{EPS\_Earnings\_Surprise\_Backward\_Ave\_Diff} &
  \multicolumn{1}{l|}{50} &
  \multicolumn{1}{l|}{EPS\_Earnings\_Surprise\_Backward\_Diff} &
  \multicolumn{1}{l|}{37} &
  \multicolumn{1}{l|}{EPS\_EarningsSurprise} &
  \multicolumn{1}{l|}{13} \\ \hline
\multicolumn{1}{|l|}{\textbf{F2}} &
  \multicolumn{1}{l|}{EPS\_Earnings\_Surprise\_Backward\_Ave\_Diff} &
  \multicolumn{1}{l|}{35} &
  \multicolumn{1}{l|}{EPS\_Earnings\_Surprise\_Backward\_Diff} &
  \multicolumn{1}{l|}{33} &
  \multicolumn{1}{l|}{EPS\_EarningsSurprise} &
  \multicolumn{1}{l|}{26} \\ \hline
\multicolumn{1}{|l|}{\textbf{F3}} &
  \multicolumn{1}{l|}{EPS\_EarningsSurprise} &
  \multicolumn{1}{l|}{49} &
  \multicolumn{1}{l|}{EPS\_Earnings\_Surprise\_Backward\_Diff} &
  \multicolumn{1}{l|}{23} &
  \multicolumn{1}{l|}{EPS\_Earnings\_Surprise\_Backward\_Ave\_Diff} &
  \multicolumn{1}{l|}{9} \\ \hline
\multicolumn{1}{|l|}{\textbf{F4}} &
  \multicolumn{1}{l|}{Return\_On\_Common\_Equity} &
  \multicolumn{1}{l|}{16} &
  \multicolumn{1}{l|}{Cash\_Q\_Change} &
  \multicolumn{1}{l|}{11} &
  \multicolumn{1}{l|}{EPS\_EarningsSurprise} &
  \multicolumn{1}{l|}{5} \\ \hline
\multicolumn{1}{|l|}{\textbf{F5}} &
  \multicolumn{1}{l|}{Return\_On\_Common\_Equity} &
  \multicolumn{1}{l|}{16} &
  \multicolumn{1}{l|}{Cash\_Q\_Change} &
  \multicolumn{1}{l|}{6} &
  \multicolumn{1}{l|}{EPS\_EarningsSurprise} &
  \multicolumn{1}{l|}{5} \\ \hline
\end{tabular}%
}
\caption{Top five driving factors for the \textbf{Consumer Non-Cyclical} stocks in each financial reporting year from 2014 to 2018.}
\label{tab:FactorOccuranceCount_ConsumerNonCyclical}
\end{table*}

% \usepackage{graphicx}
\begin{table*}[]
\centering
\resizebox{0.9\textwidth}{!}{%
\begin{tabular}{lllllll}
\hline
\multicolumn{1}{|l|}{\textbf{2018}} &
  \multicolumn{1}{l|}{\textbf{Highest Occurance}} &
  \multicolumn{1}{l|}{\textbf{Count}} &
  \multicolumn{1}{l|}{\textbf{\begin{tabular}[c]{@{}l@{}}Second Highest\\   Occurance\end{tabular}}} &
  \multicolumn{1}{l|}{\textbf{Count}} &
  \multicolumn{1}{l|}{\textbf{Third Highest Occurance}} &
  \multicolumn{1}{l|}{\textbf{Count}} \\ \hline
\multicolumn{1}{|l|}{\textbf{F1}} &
  \multicolumn{1}{l|}{EPS\_Earnings\_Surprise\_Backward\_Ave\_Diff} &
  \multicolumn{1}{l|}{95} &
  \multicolumn{1}{l|}{EPS\_Earnings\_Surprise\_Backward\_Diff} &
  \multicolumn{1}{l|}{5} &
  \multicolumn{1}{l|}{EPS\_EarningsSurprise} &
  \multicolumn{1}{l|}{0} \\ \hline
\multicolumn{1}{|l|}{\textbf{F2}} &
  \multicolumn{1}{l|}{EPS\_Earnings\_Surprise\_Backward\_Diff} &
  \multicolumn{1}{l|}{62} &
  \multicolumn{1}{l|}{EPS\_EarningsSurprise} &
  \multicolumn{1}{l|}{11} &
  \multicolumn{1}{l|}{EPS\_Earnings\_Surprise\_Backward\_Ave\_Diff} &
  \multicolumn{1}{l|}{5} \\ \hline
\multicolumn{1}{|l|}{\textbf{F3}} &
  \multicolumn{1}{l|}{EPS\_EarningsSurprise} &
  \multicolumn{1}{l|}{28} &
  \multicolumn{1}{l|}{EPS\_EarningsSurprise} &
  \multicolumn{1}{l|}{28} &
  \multicolumn{1}{l|}{EPS\_Earnings\_Surprise\_Backward\_Diff} &
  \multicolumn{1}{l|}{20} \\ \hline
\multicolumn{1}{|l|}{\textbf{F4}} &
  \multicolumn{1}{l|}{PS\_Ratios} &
  \multicolumn{1}{l|}{29} &
  \multicolumn{1}{l|}{EPS\_EarningsSurprise} &
  \multicolumn{1}{l|}{22} &
  \multicolumn{1}{l|}{Operating\_Margin\_Y\_Change} &
  \multicolumn{1}{l|}{10} \\ \hline
\multicolumn{1}{|l|}{\textbf{F5}} &
  \multicolumn{1}{l|}{PS\_Ratios} &
  \multicolumn{1}{l|}{19} &
  \multicolumn{1}{l|}{PS\_Ratios\_Y\_Change} &
  \multicolumn{1}{l|}{14} &
  \multicolumn{1}{l|}{EPS\_EarningsSurprise} &
  \multicolumn{1}{l|}{11} \\ \hline
 &
   &
   &
   &
   &
   &
   \\ \hline
\multicolumn{1}{|l|}{\textbf{2017}} &
  \multicolumn{1}{l|}{\textbf{Highest Occurance}} &
  \multicolumn{1}{l|}{\textbf{Count}} &
  \multicolumn{1}{l|}{\textbf{Second Highest Occurance}} &
  \multicolumn{1}{l|}{\textbf{Count}} &
  \multicolumn{1}{l|}{\textbf{Third Highest Occurance}} &
  \multicolumn{1}{l|}{\textbf{Count}} \\ \hline
\multicolumn{1}{|l|}{\textbf{F1}} &
  \multicolumn{1}{l|}{EPS\_Earnings\_Surprise\_Backward\_Diff} &
  \multicolumn{1}{l|}{52} &
  \multicolumn{1}{l|}{EPS\_Earnings\_Surprise\_Backward\_Ave\_Diff} &
  \multicolumn{1}{l|}{46} &
  \multicolumn{1}{l|}{EPS\_EarningsSurprise} &
  \multicolumn{1}{l|}{1} \\ \hline
\multicolumn{1}{|l|}{\textbf{F2}} &
  \multicolumn{1}{l|}{EPS\_Earnings\_Surprise\_Backward\_Ave\_Diff} &
  \multicolumn{1}{l|}{47} &
  \multicolumn{1}{l|}{EPS\_Earnings\_Surprise\_Backward\_Diff} &
  \multicolumn{1}{l|}{34} &
  \multicolumn{1}{l|}{EPS\_EarningsSurprise} &
  \multicolumn{1}{l|}{8} \\ \hline
\multicolumn{1}{|l|}{\textbf{F3}} &
  \multicolumn{1}{l|}{PS\_Ratios} &
  \multicolumn{1}{l|}{31} &
  \multicolumn{1}{l|}{EPS\_EarningsSurprise} &
  \multicolumn{1}{l|}{27} &
  \multicolumn{1}{l|}{Income\_from\_Continued\_Operations\_Q\_Change} &
  \multicolumn{1}{l|}{10} \\ \hline
\multicolumn{1}{|l|}{\textbf{F4}} &
  \multicolumn{1}{l|}{PS\_Ratios} &
  \multicolumn{1}{l|}{37} &
  \multicolumn{1}{l|}{PS\_Ratios\_Y\_Change} &
  \multicolumn{1}{l|}{13} &
  \multicolumn{1}{l|}{EPS\_EarningsSurprise} &
  \multicolumn{1}{l|}{9} \\ \hline
\multicolumn{1}{|l|}{\textbf{F5}} &
  \multicolumn{1}{l|}{EPS\_EarningsSurprise} &
  \multicolumn{1}{l|}{13} &
  \multicolumn{1}{l|}{PS\_Ratios} &
  \multicolumn{1}{l|}{11} &
  \multicolumn{1}{l|}{PS\_Ratios\_Y\_Change} &
  \multicolumn{1}{l|}{8} \\ \hline
 &
   &
   &
   &
   &
   &
   \\ \hline
\multicolumn{1}{|l|}{\textbf{2016}} &
  \multicolumn{1}{l|}{\textbf{Highest Occurance}} &
  \multicolumn{1}{l|}{\textbf{Count}} &
  \multicolumn{1}{l|}{\textbf{Second Highest Occurance}} &
  \multicolumn{1}{l|}{\textbf{Count}} &
  \multicolumn{1}{l|}{\textbf{Third Highest Occurance}} &
  \multicolumn{1}{l|}{\textbf{Count}} \\ \hline
\multicolumn{1}{|l|}{\textbf{F1}} &
  \multicolumn{1}{l|}{EPS\_Earnings\_Surprise\_Backward\_Diff} &
  \multicolumn{1}{l|}{50} &
  \multicolumn{1}{l|}{EPS\_Earnings\_Surprise\_Backward\_Ave\_Diff} &
  \multicolumn{1}{l|}{47} &
  \multicolumn{1}{l|}{EPS\_EarningsSurprise} &
  \multicolumn{1}{l|}{3} \\ \hline
\multicolumn{1}{|l|}{\textbf{F2}} &
  \multicolumn{1}{l|}{EPS\_Earnings\_Surprise\_Backward\_Diff} &
  \multicolumn{1}{l|}{43} &
  \multicolumn{1}{l|}{EPS\_Earnings\_Surprise\_Backward\_Ave\_Diff} &
  \multicolumn{1}{l|}{35} &
  \multicolumn{1}{l|}{EPS\_EarningsSurprise} &
  \multicolumn{1}{l|}{9} \\ \hline
\multicolumn{1}{|l|}{\textbf{F3}} &
  \multicolumn{1}{l|}{PS\_Ratios} &
  \multicolumn{1}{l|}{33} &
  \multicolumn{1}{l|}{EPS\_EarningsSurprise} &
  \multicolumn{1}{l|}{27} &
  \multicolumn{1}{l|}{Return\_On\_Common\_Equity} &
  \multicolumn{1}{l|}{4} \\ \hline
\multicolumn{1}{|l|}{\textbf{F4}} &
  \multicolumn{1}{l|}{PS\_Ratios} &
  \multicolumn{1}{l|}{38} &
  \multicolumn{1}{l|}{EPS\_EarningsSurprise} &
  \multicolumn{1}{l|}{18} &
  \multicolumn{1}{l|}{PS\_Ratios\_Y\_Change} &
  \multicolumn{1}{l|}{8} \\ \hline
\multicolumn{1}{|l|}{\textbf{F5}} &
  \multicolumn{1}{l|}{EPS\_EarningsSurprise} &
  \multicolumn{1}{l|}{15} &
  \multicolumn{1}{l|}{PS\_Ratios} &
  \multicolumn{1}{l|}{11} &
  \multicolumn{1}{l|}{Return\_On\_Common\_Equity} &
  \multicolumn{1}{l|}{10} \\ \hline
 &
   &
   &
   &
   &
   &
   \\ \hline
\multicolumn{1}{|l|}{\textbf{2015}} &
  \multicolumn{1}{l|}{\textbf{Highest Occurance}} &
  \multicolumn{1}{l|}{\textbf{Count}} &
  \multicolumn{1}{l|}{\textbf{Second Highest Occurance}} &
  \multicolumn{1}{l|}{\textbf{Count}} &
  \multicolumn{1}{l|}{\textbf{Third Highest Occurance}} &
  \multicolumn{1}{l|}{\textbf{Count}} \\ \hline
\multicolumn{1}{|l|}{\textbf{F1}} &
  \multicolumn{1}{l|}{EPS\_Earnings\_Surprise\_Backward\_Ave\_Diff} &
  \multicolumn{1}{l|}{92} &
  \multicolumn{1}{l|}{EPS\_Earnings\_Surprise\_Backward\_Diff} &
  \multicolumn{1}{l|}{6} &
  \multicolumn{1}{l|}{Inventory\_Turnover} &
  \multicolumn{1}{l|}{2} \\ \hline
\multicolumn{1}{|l|}{\textbf{F2}} &
  \multicolumn{1}{l|}{EPS\_Earnings\_Surprise\_Backward\_Diff} &
  \multicolumn{1}{l|}{78} &
  \multicolumn{1}{l|}{EPS\_Earnings\_Surprise\_Backward\_Ave\_Diff} &
  \multicolumn{1}{l|}{8} &
  \multicolumn{1}{l|}{EPS\_EarningsSurprise} &
  \multicolumn{1}{l|}{5} \\ \hline
\multicolumn{1}{|l|}{\textbf{F3}} &
  \multicolumn{1}{l|}{EPS\_EarningsSurprise} &
  \multicolumn{1}{l|}{49} &
  \multicolumn{1}{l|}{PS\_Ratios\_Y\_Change} &
  \multicolumn{1}{l|}{19} &
  \multicolumn{1}{l|}{PS\_Ratios} &
  \multicolumn{1}{l|}{13} \\ \hline
\multicolumn{1}{|l|}{\textbf{F4}} &
  \multicolumn{1}{l|}{PS\_Ratios} &
  \multicolumn{1}{l|}{31} &
  \multicolumn{1}{l|}{PS\_Ratios\_Y\_Change} &
  \multicolumn{1}{l|}{24} &
  \multicolumn{1}{l|}{EPS\_EarningsSurprise} &
  \multicolumn{1}{l|}{19} \\ \hline
\multicolumn{1}{|l|}{\textbf{F5}} &
  \multicolumn{1}{l|}{PS\_Ratios\_Y\_Change} &
  \multicolumn{1}{l|}{28} &
  \multicolumn{1}{l|}{PS\_Ratios} &
  \multicolumn{1}{l|}{21} &
  \multicolumn{1}{l|}{EPS\_EarningsSurprise} &
  \multicolumn{1}{l|}{12} \\ \hline
 &
   &
   &
   &
   &
   &
   \\ \hline
\multicolumn{1}{|l|}{\textbf{2014}} &
  \multicolumn{1}{l|}{\textbf{Highest Occurance}} &
  \multicolumn{1}{l|}{\textbf{Count}} &
  \multicolumn{1}{l|}{\textbf{Second Highest Occurance}} &
  \multicolumn{1}{l|}{\textbf{Count}} &
  \multicolumn{1}{l|}{\textbf{Third Highest Occurance}} &
  \multicolumn{1}{l|}{\textbf{Count}} \\ \hline
\multicolumn{1}{|l|}{\textbf{F1}} &
  \multicolumn{1}{l|}{EPS\_Earnings\_Surprise\_Backward\_Ave\_Diff} &
  \multicolumn{1}{l|}{86} &
  \multicolumn{1}{l|}{EPS\_Earnings\_Surprise\_Backward\_Diff} &
  \multicolumn{1}{l|}{13} &
  \multicolumn{1}{l|}{EPS\_EarningsSurprise} &
  \multicolumn{1}{l|}{1} \\ \hline
\multicolumn{1}{|l|}{\textbf{F2}} &
  \multicolumn{1}{l|}{EPS\_Earnings\_Surprise\_Backward\_Diff} &
  \multicolumn{1}{l|}{71} &
  \multicolumn{1}{l|}{EPS\_Earnings\_Surprise\_Backward\_Ave\_Diff} &
  \multicolumn{1}{l|}{12} &
  \multicolumn{1}{l|}{PS\_Ratios} &
  \multicolumn{1}{l|}{4} \\ \hline
\multicolumn{1}{|l|}{\textbf{F3}} &
  \multicolumn{1}{l|}{PS\_Ratios} &
  \multicolumn{1}{l|}{52} &
  \multicolumn{1}{l|}{EPS\_EarningsSurprise} &
  \multicolumn{1}{l|}{12} &
  \multicolumn{1}{l|}{EPS\_Earnings\_Surprise\_Backward\_Diff} &
  \multicolumn{1}{l|}{8} \\ \hline
\multicolumn{1}{|l|}{\textbf{F4}} &
  \multicolumn{1}{l|}{PS\_Ratios} &
  \multicolumn{1}{l|}{26} &
  \multicolumn{1}{l|}{PS\_Ratios\_Y\_Change} &
  \multicolumn{1}{l|}{18} &
  \multicolumn{1}{l|}{EPS\_EarningsSurprise} &
  \multicolumn{1}{l|}{15} \\ \hline
\multicolumn{1}{|l|}{\textbf{F5}} &
  \multicolumn{1}{l|}{PS\_Ratios\_Y\_Change} &
  \multicolumn{1}{l|}{12} &
  \multicolumn{1}{l|}{PS\_Ratios} &
  \multicolumn{1}{l|}{7} &
  \multicolumn{1}{l|}{Short\_Term\_Debt\_Y\_Change} &
  \multicolumn{1}{l|}{6} \\ \hline
\end{tabular}%
}
\caption{Top five driving factors for the \textbf{Financial} stocks in each financial reporting year from 2014 to 2018. }
\label{tab:FactorOccuranceCount_Financial}
\end{table*}

% \usepackage{graphicx}
\begin{table*}[]
\centering
\resizebox{0.9\textwidth}{!}{%
\begin{tabular}{lllllll}
\hline
\multicolumn{1}{|l|}{\textbf{2018}} &
  \multicolumn{1}{l|}{\textbf{Highest Occurance}} &
  \multicolumn{1}{l|}{\textbf{Count}} &
  \multicolumn{1}{l|}{\textbf{\begin{tabular}[c]{@{}l@{}}Second Highest\\   Occurance\end{tabular}}} &
  \multicolumn{1}{l|}{\textbf{Count}} &
  \multicolumn{1}{l|}{\textbf{Third Highest Occurance}} &
  \multicolumn{1}{l|}{\textbf{Count}} \\ \hline
\multicolumn{1}{|l|}{\textbf{F1}} &
  \multicolumn{1}{l|}{EPS\_Earnings\_Surprise\_Backward\_Ave\_Diff} &
  \multicolumn{1}{l|}{57} &
  \multicolumn{1}{l|}{EPS\_Earnings\_Surprise\_Backward\_Diff} &
  \multicolumn{1}{l|}{14} &
  \multicolumn{1}{l|}{EPS\_EarningsSurprise} &
  \multicolumn{1}{l|}{13} \\ \hline
\multicolumn{1}{|l|}{\textbf{F2}} &
  \multicolumn{1}{l|}{EPS\_EarningsSurprise} &
  \multicolumn{1}{l|}{26} &
  \multicolumn{1}{l|}{EPS\_Earnings\_Surprise\_Backward\_Ave\_Diff} &
  \multicolumn{1}{l|}{18} &
  \multicolumn{1}{l|}{Return\_On\_Assets\_Q\_Change} &
  \multicolumn{1}{l|}{11} \\ \hline
\multicolumn{1}{|l|}{\textbf{F3}} &
  \multicolumn{1}{l|}{Return\_On\_Assets\_Q\_Change} &
  \multicolumn{1}{l|}{24} &
  \multicolumn{1}{l|}{EPS\_Earnings\_Surprise\_Backward\_Diff} &
  \multicolumn{1}{l|}{16} &
  \multicolumn{1}{l|}{EPS\_EarningsSurprise} &
  \multicolumn{1}{l|}{15} \\ \hline
\multicolumn{1}{|l|}{\textbf{F4}} &
  \multicolumn{1}{l|}{Return\_On\_Assets\_Q\_Change} &
  \multicolumn{1}{l|}{18} &
  \multicolumn{1}{l|}{EPS\_Earnings\_Surprise\_Backward\_Diff} &
  \multicolumn{1}{l|}{11} &
  \multicolumn{1}{l|}{EPS\_EarningsSurprise} &
  \multicolumn{1}{l|}{10} \\ \hline
\multicolumn{1}{|l|}{\textbf{F5}} &
  \multicolumn{1}{l|}{Return\_On\_Assets\_Q\_Change} &
  \multicolumn{1}{l|}{12} &
  \multicolumn{1}{l|}{EPS\_EarningsSurprise} &
  \multicolumn{1}{l|}{11} &
  \multicolumn{1}{l|}{Short\_Term\_Debt\_Q\_Change} &
  \multicolumn{1}{l|}{8} \\ \hline
 &
   &
   &
   &
   &
   &
   \\ \hline
\multicolumn{1}{|l|}{\textbf{2017}} &
  \multicolumn{1}{l|}{\textbf{Highest Occurance}} &
  \multicolumn{1}{l|}{\textbf{Count}} &
  \multicolumn{1}{l|}{\textbf{Second Highest Occurance}} &
  \multicolumn{1}{l|}{\textbf{Count}} &
  \multicolumn{1}{l|}{\textbf{Third Highest Occurance}} &
  \multicolumn{1}{l|}{\textbf{Count}} \\ \hline
\multicolumn{1}{|l|}{\textbf{F1}} &
  \multicolumn{1}{l|}{EPS\_Earnings\_Surprise\_Backward\_Ave\_Diff} &
  \multicolumn{1}{l|}{53} &
  \multicolumn{1}{l|}{EPS\_Earnings\_Surprise\_Backward\_Diff} &
  \multicolumn{1}{l|}{15} &
  \multicolumn{1}{l|}{EPS\_EarningsSurprise} &
  \multicolumn{1}{l|}{11} \\ \hline
\multicolumn{1}{|l|}{\textbf{F2}} &
  \multicolumn{1}{l|}{EPS\_EarningsSurprise} &
  \multicolumn{1}{l|}{37} &
  \multicolumn{1}{l|}{EPS\_Earnings\_Surprise\_Backward\_Ave\_Diff} &
  \multicolumn{1}{l|}{17} &
  \multicolumn{1}{l|}{Dividend\_Payout\_Ratio\_Y\_Change} &
  \multicolumn{1}{l|}{7} \\ \hline
\multicolumn{1}{|l|}{\textbf{F3}} &
  \multicolumn{1}{l|}{Return\_On\_Assets\_Q\_Change} &
  \multicolumn{1}{l|}{13} &
  \multicolumn{1}{l|}{EPS\_EarningsSurprise} &
  \multicolumn{1}{l|}{11} &
  \multicolumn{1}{l|}{Dividend\_Payout\_Ratio\_Y\_Change} &
  \multicolumn{1}{l|}{10} \\ \hline
\multicolumn{1}{|l|}{\textbf{F4}} &
  \multicolumn{1}{l|}{Return\_On\_Assets\_Q\_Change} &
  \multicolumn{1}{l|}{18} &
  \multicolumn{1}{l|}{EPS\_EarningsSurprise} &
  \multicolumn{1}{l|}{12} &
  \multicolumn{1}{l|}{Operating\_Income\_Y\_Change} &
  \multicolumn{1}{l|}{11} \\ \hline
\multicolumn{1}{|l|}{\textbf{F5}} &
  \multicolumn{1}{l|}{Dividend\_Payout\_Ratio\_Y\_Change} &
  \multicolumn{1}{l|}{9} &
  \multicolumn{1}{l|}{Return\_On\_Assets\_Q\_Change} &
  \multicolumn{1}{l|}{7} &
  \multicolumn{1}{l|}{DMA\_5D/200D} &
  \multicolumn{1}{l|}{6} \\ \hline
 &
   &
   &
   &
   &
   &
   \\ \hline
\multicolumn{1}{|l|}{\textbf{2016}} &
  \multicolumn{1}{l|}{\textbf{Highest Occurance}} &
  \multicolumn{1}{l|}{\textbf{Count}} &
  \multicolumn{1}{l|}{\textbf{Second Highest Occurance}} &
  \multicolumn{1}{l|}{\textbf{Count}} &
  \multicolumn{1}{l|}{\textbf{Third Highest Occurance}} &
  \multicolumn{1}{l|}{\textbf{Count}} \\ \hline
\multicolumn{1}{|l|}{\textbf{F1}} &
  \multicolumn{1}{l|}{EPS\_EarningsSurprise} &
  \multicolumn{1}{l|}{31} &
  \multicolumn{1}{l|}{EPS\_Earnings\_Surprise\_Backward\_Ave\_Diff} &
  \multicolumn{1}{l|}{21} &
  \multicolumn{1}{l|}{Operating\_Income\_Y\_Change} &
  \multicolumn{1}{l|}{9} \\ \hline
\multicolumn{1}{|l|}{\textbf{F2}} &
  \multicolumn{1}{l|}{EPS\_EarningsSurprise} &
  \multicolumn{1}{l|}{22} &
  \multicolumn{1}{l|}{EPS\_Earnings\_Surprise\_Backward\_Ave\_Diff} &
  \multicolumn{1}{l|}{14} &
  \multicolumn{1}{l|}{EPS\_Earnings\_Surprise\_Backward\_Ave\_Diff} &
  \multicolumn{1}{l|}{14} \\ \hline
\multicolumn{1}{|l|}{\textbf{F3}} &
  \multicolumn{1}{l|}{EPS\_EarningsSurprise} &
  \multicolumn{1}{l|}{18} &
  \multicolumn{1}{l|}{EPS\_Earnings\_Surprise\_Backward\_Diff} &
  \multicolumn{1}{l|}{11} &
  \multicolumn{1}{l|}{EPS\_Earnings\_Surprise\_Backward\_Ave\_Diff} &
  \multicolumn{1}{l|}{9} \\ \hline
\multicolumn{1}{|l|}{\textbf{F4}} &
  \multicolumn{1}{l|}{Operating\_Income\_Y\_Change} &
  \multicolumn{1}{l|}{11} &
  \multicolumn{1}{l|}{Short\_Term\_Debt\_Q\_Change} &
  \multicolumn{1}{l|}{10} &
  \multicolumn{1}{l|}{Short\_Term\_Debt\_Q\_Change} &
  \multicolumn{1}{l|}{10} \\ \hline
\multicolumn{1}{|l|}{\textbf{F5}} &
  \multicolumn{1}{l|}{Operating\_Income\_Y\_Change} &
  \multicolumn{1}{l|}{15} &
  \multicolumn{1}{l|}{EPS\_Earnings\_Surprise\_Backward\_Diff} &
  \multicolumn{1}{l|}{8} &
  \multicolumn{1}{l|}{Return\_On\_Assets\_Q\_Change} &
  \multicolumn{1}{l|}{7} \\ \hline
 &
   &
   &
   &
   &
   &
   \\ \hline
\multicolumn{1}{|l|}{\textbf{2015}} &
  \multicolumn{1}{l|}{\textbf{Highest Occurance}} &
  \multicolumn{1}{l|}{\textbf{Count}} &
  \multicolumn{1}{l|}{\textbf{Second Highest Occurance}} &
  \multicolumn{1}{l|}{\textbf{Count}} &
  \multicolumn{1}{l|}{\textbf{Third Highest Occurance}} &
  \multicolumn{1}{l|}{\textbf{Count}} \\ \hline
\multicolumn{1}{|l|}{\textbf{F1}} &
  \multicolumn{1}{l|}{EPS\_Earnings\_Surprise\_Backward\_Ave\_Diff} &
  \multicolumn{1}{l|}{61} &
  \multicolumn{1}{l|}{EPS\_EarningsSurprise} &
  \multicolumn{1}{l|}{11} &
  \multicolumn{1}{l|}{EPS\_EarningsSurprise} &
  \multicolumn{1}{l|}{11} \\ \hline
\multicolumn{1}{|l|}{\textbf{F2}} &
  \multicolumn{1}{l|}{EPS\_EarningsSurprise} &
  \multicolumn{1}{l|}{18} &
  \multicolumn{1}{l|}{EPS\_Earnings\_Surprise\_Backward\_Diff} &
  \multicolumn{1}{l|}{16} &
  \multicolumn{1}{l|}{EPS\_Earnings\_Surprise\_Backward\_Ave\_Diff} &
  \multicolumn{1}{l|}{15} \\ \hline
\multicolumn{1}{|l|}{\textbf{F3}} &
  \multicolumn{1}{l|}{EPS\_EarningsSurprise} &
  \multicolumn{1}{l|}{19} &
  \multicolumn{1}{l|}{EPS\_Earnings\_Surprise\_Backward\_Diff} &
  \multicolumn{1}{l|}{11} &
  \multicolumn{1}{l|}{Return\_On\_Assets\_Q\_Change} &
  \multicolumn{1}{l|}{9} \\ \hline
\multicolumn{1}{|l|}{\textbf{F4}} &
  \multicolumn{1}{l|}{EPS\_EarningsSurprise} &
  \multicolumn{1}{l|}{12} &
  \multicolumn{1}{l|}{EPS\_EarningsSurprise} &
  \multicolumn{1}{l|}{12} &
  \multicolumn{1}{l|}{Short\_Term\_Debt\_Q\_Change} &
  \multicolumn{1}{l|}{11} \\ \hline
\multicolumn{1}{|l|}{\textbf{F5}} &
  \multicolumn{1}{l|}{EPS\_EarningsSurprise} &
  \multicolumn{1}{l|}{9} &
  \multicolumn{1}{l|}{EPS\_EarningsSurprise} &
  \multicolumn{1}{l|}{9} &
  \multicolumn{1}{l|}{EPS\_Earnings\_Surprise\_Backward\_Diff} &
  \multicolumn{1}{l|}{8} \\ \hline
 &
   &
   &
   &
   &
   &
   \\ \hline
\multicolumn{1}{|l|}{\textbf{2014}} &
  \multicolumn{1}{l|}{\textbf{Highest Occurance}} &
  \multicolumn{1}{l|}{\textbf{Count}} &
  \multicolumn{1}{l|}{\textbf{Second Highest Occurance}} &
  \multicolumn{1}{l|}{\textbf{Count}} &
  \multicolumn{1}{l|}{\textbf{Third Highest Occurance}} &
  \multicolumn{1}{l|}{\textbf{Count}} \\ \hline
\multicolumn{1}{|l|}{\textbf{F1}} &
  \multicolumn{1}{l|}{EPS\_Earnings\_Surprise\_Backward\_Ave\_Diff} &
  \multicolumn{1}{l|}{55} &
  \multicolumn{1}{l|}{EPS\_Earnings\_Surprise\_Backward\_Diff} &
  \multicolumn{1}{l|}{14} &
  \multicolumn{1}{l|}{EPS\_EarningsSurprise} &
  \multicolumn{1}{l|}{6} \\ \hline
\multicolumn{1}{|l|}{\textbf{F2}} &
  \multicolumn{1}{l|}{EPS\_Earnings\_Surprise\_Backward\_Diff} &
  \multicolumn{1}{l|}{18} &
  \multicolumn{1}{l|}{EPS\_EarningsSurprise} &
  \multicolumn{1}{l|}{17} &
  \multicolumn{1}{l|}{EPS\_Earnings\_Surprise\_Backward\_Ave\_Diff} &
  \multicolumn{1}{l|}{12} \\ \hline
\multicolumn{1}{|l|}{\textbf{F3}} &
  \multicolumn{1}{l|}{EPS\_EarningsSurprise} &
  \multicolumn{1}{l|}{20} &
  \multicolumn{1}{l|}{EPS\_Earnings\_Surprise\_Backward\_Diff} &
  \multicolumn{1}{l|}{10} &
  \multicolumn{1}{l|}{Short\_Term\_Debt\_Q\_Change} &
  \multicolumn{1}{l|}{9} \\ \hline
\multicolumn{1}{|l|}{\textbf{F4}} &
  \multicolumn{1}{l|}{EPS\_Earnings\_Surprise\_Backward\_Diff} &
  \multicolumn{1}{l|}{11} &
  \multicolumn{1}{l|}{EPS\_EarningsSurprise} &
  \multicolumn{1}{l|}{8} &
  \multicolumn{1}{l|}{Return\_On\_Assets\_Q\_Change} &
  \multicolumn{1}{l|}{7} \\ \hline
\multicolumn{1}{|l|}{\textbf{F5}} &
  \multicolumn{1}{l|}{Operating\_Income\_Y\_Change} &
  \multicolumn{1}{l|}{9} &
  \multicolumn{1}{l|}{EPS\_EarningsSurprise} &
  \multicolumn{1}{l|}{8} &
  \multicolumn{1}{l|}{Short\_Term\_Debt\_Q\_Change} &
  \multicolumn{1}{l|}{7} \\ \hline
\end{tabular}%
}
\caption{Top five driving factors for the \textbf{Technology} stocks in each financial reporting year from 2014 to 2018. }
\label{tab:FactorOccuranceCount_Technology}
\end{table*}

% \usepackage{graphicx}
\begin{table*}[]
\centering
\resizebox{0.9\textwidth}{!}{%
\begin{tabular}{lllllll}
\hline
\multicolumn{1}{|l|}{\textbf{2018}} &
  \multicolumn{1}{l|}{\textbf{Highest Occurance}} &
  \multicolumn{1}{l|}{\textbf{Count}} &
  \multicolumn{1}{l|}{\textbf{\begin{tabular}[c]{@{}l@{}}Second Highest\\   Occurance\end{tabular}}} &
  \multicolumn{1}{l|}{\textbf{Count}} &
  \multicolumn{1}{l|}{\textbf{Third Highest Occurance}} &
  \multicolumn{1}{l|}{\textbf{Count}} \\ \hline
\multicolumn{1}{|l|}{\textbf{F1}} &
  \multicolumn{1}{l|}{Total\_Liabilities\_Q\_Change} &
  \multicolumn{1}{l|}{27} &
  \multicolumn{1}{l|}{Net\_Income\_Y\_Change} &
  \multicolumn{1}{l|}{18} &
  \multicolumn{1}{l|}{PE\_Ratios} &
  \multicolumn{1}{l|}{13} \\ \hline
\multicolumn{1}{|l|}{\textbf{F2}} &
  \multicolumn{1}{l|}{Total\_Liabilities\_Q\_Change} &
  \multicolumn{1}{l|}{26} &
  \multicolumn{1}{l|}{Net\_Income\_Y\_Change} &
  \multicolumn{1}{l|}{18} &
  \multicolumn{1}{l|}{Operating\_Income\_Y\_Change} &
  \multicolumn{1}{l|}{8} \\ \hline
\multicolumn{1}{|l|}{\textbf{F3}} &
  \multicolumn{1}{l|}{Operating\_Income\_Y\_Change} &
  \multicolumn{1}{l|}{15} &
  \multicolumn{1}{l|}{Total\_Liabilities\_Q\_Change} &
  \multicolumn{1}{l|}{13} &
  \multicolumn{1}{l|}{PE\_Ratios} &
  \multicolumn{1}{l|}{11} \\ \hline
\multicolumn{1}{|l|}{\textbf{F4}} &
  \multicolumn{1}{l|}{EPS\_Earnings\_Surprise\_Backward\_Ave\_Diff} &
  \multicolumn{1}{l|}{11} &
  \multicolumn{1}{l|}{PC\_Ratios} &
  \multicolumn{1}{l|}{6} &
  \multicolumn{1}{l|}{PC\_Ratios} &
  \multicolumn{1}{l|}{6} \\ \hline
\multicolumn{1}{|l|}{\textbf{F5}} &
  \multicolumn{1}{l|}{Net\_Income\_Y\_Change} &
  \multicolumn{1}{l|}{13} &
  \multicolumn{1}{l|}{Operating\_Income\_Y\_Change} &
  \multicolumn{1}{l|}{9} &
  \multicolumn{1}{l|}{PE\_Ratios} &
  \multicolumn{1}{l|}{7} \\ \hline
 &
   &
   &
   &
   &
   &
   \\ \hline
\multicolumn{1}{|l|}{\textbf{2017}} &
  \multicolumn{1}{l|}{\textbf{Highest Occurance}} &
  \multicolumn{1}{l|}{\textbf{Count}} &
  \multicolumn{1}{l|}{\textbf{Second Highest Occurance}} &
  \multicolumn{1}{l|}{\textbf{Count}} &
  \multicolumn{1}{l|}{\textbf{Third Highest Occurance}} &
  \multicolumn{1}{l|}{\textbf{Count}} \\ \hline
\multicolumn{1}{|l|}{\textbf{F1}} &
  \multicolumn{1}{l|}{PE\_Ratios} &
  \multicolumn{1}{l|}{37} &
  \multicolumn{1}{l|}{PB\_Ratios\_Y\_Change} &
  \multicolumn{1}{l|}{8} &
  \multicolumn{1}{l|}{PB\_Ratios\_Y\_Change} &
  \multicolumn{1}{l|}{8} \\ \hline
\multicolumn{1}{|l|}{\textbf{F2}} &
  \multicolumn{1}{l|}{Net\_Income\_Y\_Change} &
  \multicolumn{1}{l|}{17} &
  \multicolumn{1}{l|}{PE\_Ratios} &
  \multicolumn{1}{l|}{14} &
  \multicolumn{1}{l|}{PB\_Ratios\_Y\_Change} &
  \multicolumn{1}{l|}{8} \\ \hline
\multicolumn{1}{|l|}{\textbf{F3}} &
  \multicolumn{1}{l|}{PB\_Ratios\_Y\_Change} &
  \multicolumn{1}{l|}{12} &
  \multicolumn{1}{l|}{PB\_Ratios\_Y\_Change} &
  \multicolumn{1}{l|}{12} &
  \multicolumn{1}{l|}{Net\_Income\_Y\_Change} &
  \multicolumn{1}{l|}{10} \\ \hline
\multicolumn{1}{|l|}{\textbf{F4}} &
  \multicolumn{1}{l|}{Net\_Income\_Y\_Change} &
  \multicolumn{1}{l|}{17} &
  \multicolumn{1}{l|}{Cost\_Of\_Revenue\_Y\_Change} &
  \multicolumn{1}{l|}{9} &
  \multicolumn{1}{l|}{PB\_Ratios\_Y\_Change} &
  \multicolumn{1}{l|}{8} \\ \hline
\multicolumn{1}{|l|}{\textbf{F5}} &
  \multicolumn{1}{l|}{Total\_Liabilities\_Q\_Change} &
  \multicolumn{1}{l|}{9} &
  \multicolumn{1}{l|}{EPS\_Earnings\_Surprise\_Backward\_Ave\_Diff} &
  \multicolumn{1}{l|}{8} &
  \multicolumn{1}{l|}{Cost\_Of\_Revenue\_Y\_Change} &
  \multicolumn{1}{l|}{7} \\ \hline
 &
   &
   &
   &
   &
   &
   \\ \hline
\multicolumn{1}{|l|}{\textbf{2016}} &
  \multicolumn{1}{l|}{\textbf{Highest Occurance}} &
  \multicolumn{1}{l|}{\textbf{Count}} &
  \multicolumn{1}{l|}{\textbf{Second Highest Occurance}} &
  \multicolumn{1}{l|}{\textbf{Count}} &
  \multicolumn{1}{l|}{\textbf{Third Highest Occurance}} &
  \multicolumn{1}{l|}{\textbf{Count}} \\ \hline
\multicolumn{1}{|l|}{\textbf{F1}} &
  \multicolumn{1}{l|}{Total\_Liabilities\_Q\_Change} &
  \multicolumn{1}{l|}{21} &
  \multicolumn{1}{l|}{PS\_Ratios\_Y\_Change} &
  \multicolumn{1}{l|}{12} &
  \multicolumn{1}{l|}{EPS\_Earnings\_Surprise\_Backward\_Ave\_Diff} &
  \multicolumn{1}{l|}{6} \\ \hline
\multicolumn{1}{|l|}{\textbf{F2}} &
  \multicolumn{1}{l|}{Total\_Liabilities\_Q\_Change} &
  \multicolumn{1}{l|}{11} &
  \multicolumn{1}{l|}{PS\_Ratios\_Y\_Change} &
  \multicolumn{1}{l|}{10} &
  \multicolumn{1}{l|}{PE\_Ratios} &
  \multicolumn{1}{l|}{7} \\ \hline
\multicolumn{1}{|l|}{\textbf{F3}} &
  \multicolumn{1}{l|}{Total\_Liabilities\_Q\_Change} &
  \multicolumn{1}{l|}{12} &
  \multicolumn{1}{l|}{EPS\_Earnings\_Surprise\_Backward\_Ave\_Diff} &
  \multicolumn{1}{l|}{10} &
  \multicolumn{1}{l|}{PS\_Ratios\_Y\_Change} &
  \multicolumn{1}{l|}{8} \\ \hline
\multicolumn{1}{|l|}{\textbf{F4}} &
  \multicolumn{1}{l|}{Total\_Liabilities\_Q\_Change} &
  \multicolumn{1}{l|}{8} &
  \multicolumn{1}{l|}{PS\_Ratios\_Y\_Change} &
  \multicolumn{1}{l|}{7} &
  \multicolumn{1}{l|}{PE\_Ratios} &
  \multicolumn{1}{l|}{6} \\ \hline
\multicolumn{1}{|l|}{\textbf{F5}} &
  \multicolumn{1}{l|}{Total\_Liabilities\_Q\_Change} &
  \multicolumn{1}{l|}{13} &
  \multicolumn{1}{l|}{EPS\_Earnings\_Surprise\_Backward\_Ave\_Diff} &
  \multicolumn{1}{l|}{8} &
  \multicolumn{1}{l|}{PE\_Ratios} &
  \multicolumn{1}{l|}{7} \\ \hline
 &
   &
   &
   &
   &
   &
   \\ \hline
\multicolumn{1}{|l|}{\textbf{2015}} &
  \multicolumn{1}{l|}{\textbf{Highest Occurance}} &
  \multicolumn{1}{l|}{\textbf{Count}} &
  \multicolumn{1}{l|}{\textbf{Second Highest Occurance}} &
  \multicolumn{1}{l|}{\textbf{Count}} &
  \multicolumn{1}{l|}{\textbf{Third Highest Occurance}} &
  \multicolumn{1}{l|}{\textbf{Count}} \\ \hline
\multicolumn{1}{|l|}{\textbf{F1}} &
  \multicolumn{1}{l|}{Total\_Liabilities\_Q\_Change} &
  \multicolumn{1}{l|}{18} &
  \multicolumn{1}{l|}{PE\_Ratios} &
  \multicolumn{1}{l|}{12} &
  \multicolumn{1}{l|}{Operating\_Income\_Y\_Change} &
  \multicolumn{1}{l|}{8} \\ \hline
\multicolumn{1}{|l|}{\textbf{F2}} &
  \multicolumn{1}{l|}{Total\_Liabilities\_Q\_Change} &
  \multicolumn{1}{l|}{14} &
  \multicolumn{1}{l|}{EPS\_Earnings\_Surprise\_Backward\_Ave\_Diff} &
  \multicolumn{1}{l|}{11} &
  \multicolumn{1}{l|}{PE\_Ratios} &
  \multicolumn{1}{l|}{10} \\ \hline
\multicolumn{1}{|l|}{\textbf{F3}} &
  \multicolumn{1}{l|}{Operating\_Income\_Y\_Change} &
  \multicolumn{1}{l|}{10} &
  \multicolumn{1}{l|}{PE\_Ratios} &
  \multicolumn{1}{l|}{8} &
  \multicolumn{1}{l|}{EPS\_Earnings\_Surprise\_Backward\_Ave\_Diff} &
  \multicolumn{1}{l|}{7} \\ \hline
\multicolumn{1}{|l|}{\textbf{F4}} &
  \multicolumn{1}{l|}{Total\_Liabilities\_Q\_Change} &
  \multicolumn{1}{l|}{11} &
  \multicolumn{1}{l|}{EPS\_Earnings\_Surprise\_Backward\_Ave\_Diff} &
  \multicolumn{1}{l|}{8} &
  \multicolumn{1}{l|}{PE\_Ratios} &
  \multicolumn{1}{l|}{7} \\ \hline
\multicolumn{1}{|l|}{\textbf{F5}} &
  \multicolumn{1}{l|}{Total\_Liabilities\_Q\_Change} &
  \multicolumn{1}{l|}{14} &
  \multicolumn{1}{l|}{PE\_Ratios} &
  \multicolumn{1}{l|}{5} &
  \multicolumn{1}{l|}{DMA\_50D/200D} &
  \multicolumn{1}{l|}{4} \\ \hline
 &
   &
   &
   &
   &
   &
   \\ \hline
\multicolumn{1}{|l|}{\textbf{2014}} &
  \multicolumn{1}{l|}{\textbf{Highest Occurance}} &
  \multicolumn{1}{l|}{\textbf{Count}} &
  \multicolumn{1}{l|}{\textbf{Second Highest Occurance}} &
  \multicolumn{1}{l|}{\textbf{Count}} &
  \multicolumn{1}{l|}{\textbf{Third Highest Occurance}} &
  \multicolumn{1}{l|}{\textbf{Count}} \\ \hline
\multicolumn{1}{|l|}{\textbf{F1}} &
  \multicolumn{1}{l|}{Total\_Liabilities\_Q\_Change} &
  \multicolumn{1}{l|}{41} &
  \multicolumn{1}{l|}{PE\_Ratios} &
  \multicolumn{1}{l|}{8} &
  \multicolumn{1}{l|}{PS\_Ratios\_Y\_Change} &
  \multicolumn{1}{l|}{7} \\ \hline
\multicolumn{1}{|l|}{\textbf{F2}} &
  \multicolumn{1}{l|}{Total\_Liabilities\_Q\_Change} &
  \multicolumn{1}{l|}{20} &
  \multicolumn{1}{l|}{PS\_Ratios\_Y\_Change} &
  \multicolumn{1}{l|}{16} &
  \multicolumn{1}{l|}{PE\_Ratios} &
  \multicolumn{1}{l|}{7} \\ \hline
\multicolumn{1}{|l|}{\textbf{F3}} &
  \multicolumn{1}{l|}{Total\_Liabilities\_Q\_Change} &
  \multicolumn{1}{l|}{12} &
  \multicolumn{1}{l|}{PS\_Ratios\_Y\_Change} &
  \multicolumn{1}{l|}{10} &
  \multicolumn{1}{l|}{PE\_Ratios} &
  \multicolumn{1}{l|}{8} \\ \hline
\multicolumn{1}{|l|}{\textbf{F4}} &
  \multicolumn{1}{l|}{Operating\_Income\_Y\_Change} &
  \multicolumn{1}{l|}{7} &
  \multicolumn{1}{l|}{Operating\_Income\_Y\_Change} &
  \multicolumn{1}{l|}{7} &
  \multicolumn{1}{l|}{PE\_Ratios} &
  \multicolumn{1}{l|}{6} \\ \hline
\multicolumn{1}{|l|}{\textbf{F5}} &
  \multicolumn{1}{l|}{PE\_Ratios} &
  \multicolumn{1}{l|}{12} &
  \multicolumn{1}{l|}{Cost\_Of\_Revenue\_Y\_Change} &
  \multicolumn{1}{l|}{6} &
  \multicolumn{1}{l|}{Cost\_Of\_Revenue\_Y\_Change} &
  \multicolumn{1}{l|}{6} \\ \hline
\end{tabular}%
}
\caption{Top five driving factors for the \textbf{Communications} stocks in each financial reporting year from 2014 to 2018. }
\label{tab:FactorOccuranceCount_Communications}
\end{table*}

% \usepackage{graphicx}

\begin{table*}[]
\centering
\resizebox{0.9\textwidth}{!}{%
\begin{tabular}{lllllll}
\hline
\multicolumn{1}{|l|}{\textbf{2018}} &
  \multicolumn{1}{l|}{\textbf{Highest Occurance}} &
  \multicolumn{1}{l|}{\textbf{Count}} &
  \multicolumn{1}{l|}{\textbf{\begin{tabular}[c]{@{}l@{}}Second Highest\\   Occurance\end{tabular}}} &
  \multicolumn{1}{l|}{\textbf{Count}} &
  \multicolumn{1}{l|}{\textbf{Third Highest Occurance}} &
  \multicolumn{1}{l|}{\textbf{Count}} \\ \hline
\multicolumn{1}{|l|}{\textbf{F1}} &
  \multicolumn{1}{l|}{EPS\_EarningsSurprise} &
  \multicolumn{1}{l|}{33} &
  \multicolumn{1}{l|}{DMA\_50D/200D} &
  \multicolumn{1}{l|}{15} &
  \multicolumn{1}{l|}{DMA\_5D/200D} &
  \multicolumn{1}{l|}{5} \\ \hline
\multicolumn{1}{|l|}{\textbf{F2}} &
  \multicolumn{1}{l|}{PS\_Ratios} &
  \multicolumn{1}{l|}{15} &
  \multicolumn{1}{l|}{EPS\_EarningsSurprise} &
  \multicolumn{1}{l|}{13} &
  \multicolumn{1}{l|}{DMA\_50D/200D} &
  \multicolumn{1}{l|}{8} \\ \hline
\multicolumn{1}{|l|}{\textbf{F3}} &
  \multicolumn{1}{l|}{EPS\_EarningsSurprise} &
  \multicolumn{1}{l|}{12} &
  \multicolumn{1}{l|}{DMA\_50D/200D} &
  \multicolumn{1}{l|}{8} &
  \multicolumn{1}{l|}{EPS\_Earnings\_Surprise\_Backward\_Ave\_Diff} &
  \multicolumn{1}{l|}{5} \\ \hline
\multicolumn{1}{|l|}{\textbf{F4}} &
  \multicolumn{1}{l|}{DMA\_50D/200D} &
  \multicolumn{1}{l|}{9} &
  \multicolumn{1}{l|}{PC\_Ratios\_Q\_Change} &
  \multicolumn{1}{l|}{8} &
  \multicolumn{1}{l|}{PC\_Ratios\_Q\_Change} &
  \multicolumn{1}{l|}{8} \\ \hline
\multicolumn{1}{|l|}{\textbf{F5}} &
  \multicolumn{1}{l|}{EPS\_EarningsSurprise} &
  \multicolumn{1}{l|}{6} &
  \multicolumn{1}{l|}{EPS\_EarningsSurprise} &
  \multicolumn{1}{l|}{6} &
  \multicolumn{1}{l|}{DMA\_5D/200D} &
  \multicolumn{1}{l|}{5} \\ \hline
 &
  &
  &
  &
  &
  &
  \\ \hline
\multicolumn{1}{|l|}{\textbf{2017}} &
  \multicolumn{1}{l|}{\textbf{Highest Occurance}} &
  \multicolumn{1}{l|}{\textbf{Count}} &
  \multicolumn{1}{l|}{\textbf{Second Highest Occurance}} &
  \multicolumn{1}{l|}{\textbf{Count}} &
  \multicolumn{1}{l|}{\textbf{Third Highest Occurance}} &
  \multicolumn{1}{l|}{\textbf{Count}} \\ \hline
\multicolumn{1}{|l|}{\textbf{F1}} &
  \multicolumn{1}{l|}{EPS\_EarningsSurprise} &
  \multicolumn{1}{l|}{34} &
  \multicolumn{1}{l|}{DMA\_5D/200D} &
  \multicolumn{1}{l|}{8} &
  \multicolumn{1}{l|}{DMA\_5D/200D} &
  \multicolumn{1}{l|}{8} \\ \hline
\multicolumn{1}{|l|}{\textbf{F2}} &
  \multicolumn{1}{l|}{EPS\_EarningsSurprise} &
  \multicolumn{1}{l|}{12} &
  \multicolumn{1}{l|}{EPS\_EarningsSurprise} &
  \multicolumn{1}{l|}{12} &
  \multicolumn{1}{l|}{DMA\_50D/200D} &
  \multicolumn{1}{l|}{9} \\ \hline
\multicolumn{1}{|l|}{\textbf{F3}} &
  \multicolumn{1}{l|}{Current\_Ratio} &
  \multicolumn{1}{l|}{12} &
  \multicolumn{1}{l|}{Return\_On\_Assets\_Y\_Change} &
  \multicolumn{1}{l|}{10} &
  \multicolumn{1}{l|}{EPS\_EarningsSurprise} &
  \multicolumn{1}{l|}{9} \\ \hline
\multicolumn{1}{|l|}{\textbf{F4}} &
  \multicolumn{1}{l|}{Return\_On\_Assets\_Y\_Change} &
  \multicolumn{1}{l|}{7} &
  \multicolumn{1}{l|}{EPS\_EarningsSurprise} &
  \multicolumn{1}{l|}{6} &
  \multicolumn{1}{l|}{EPS\_EarningsSurprise} &
  \multicolumn{1}{l|}{6} \\ \hline
\multicolumn{1}{|l|}{\textbf{F5}} &
  \multicolumn{1}{l|}{DMA\_50D/200D} &
  \multicolumn{1}{l|}{9} &
  \multicolumn{1}{l|}{Current\_Ratio} &
  \multicolumn{1}{l|}{8} &
  \multicolumn{1}{l|}{Return\_On\_Assets\_Y\_Change} &
  \multicolumn{1}{l|}{6} \\ \hline
 &
  &
  &
  &
  &
  &
  \\ \hline
\multicolumn{1}{|l|}{\textbf{2016}} &
  \multicolumn{1}{l|}{\textbf{Highest Occurance}} &
  \multicolumn{1}{l|}{\textbf{Count}} &
  \multicolumn{1}{l|}{\textbf{Second Highest Occurance}} &
  \multicolumn{1}{l|}{\textbf{Count}} &
  \multicolumn{1}{l|}{\textbf{Third Highest Occurance}} &
  \multicolumn{1}{l|}{\textbf{Count}} \\ \hline
\multicolumn{1}{|l|}{\textbf{F1}} &
  \multicolumn{1}{l|}{DMA\_50D/200D} &
  \multicolumn{1}{l|}{24} &
  \multicolumn{1}{l|}{DMA\_5D/200D} &
  \multicolumn{1}{l|}{13} &
  \multicolumn{1}{l|}{EPS\_Earnings\_Surprise\_Backward\_Ave\_Diff} &
  \multicolumn{1}{l|}{12} \\ \hline
\multicolumn{1}{|l|}{\textbf{F2}} &
  \multicolumn{1}{l|}{DMA\_50D/200D} &
  \multicolumn{1}{l|}{15} &
  \multicolumn{1}{l|}{PS\_Ratios} &
  \multicolumn{1}{l|}{14} &
  \multicolumn{1}{l|}{EPS\_Earnings\_Surprise\_Backward\_Ave\_Diff} &
  \multicolumn{1}{l|}{13} \\ \hline
\multicolumn{1}{|l|}{\textbf{F3}} &
  \multicolumn{1}{l|}{DMA\_50D/200D} &
  \multicolumn{1}{l|}{13} &
  \multicolumn{1}{l|}{DMA\_50D/200D} &
  \multicolumn{1}{l|}{13} &
  \multicolumn{1}{l|}{DMA\_5D/200D} &
  \multicolumn{1}{l|}{11} \\ \hline
\multicolumn{1}{|l|}{\textbf{F4}} &
  \multicolumn{1}{l|}{EPS\_Earnings\_Surprise\_Backward\_Ave\_Diff} &
  \multicolumn{1}{l|}{18} &
  \multicolumn{1}{l|}{DMA\_50D/200D} &
  \multicolumn{1}{l|}{12} &
  \multicolumn{1}{l|}{EPS\_EarningsSurprise} &
  \multicolumn{1}{l|}{8} \\ \hline
\multicolumn{1}{|l|}{\textbf{F5}} &
  \multicolumn{1}{l|}{DMA\_50D/200D} &
  \multicolumn{1}{l|}{11} &
  \multicolumn{1}{l|}{PS\_Ratios} &
  \multicolumn{1}{l|}{9} &
  \multicolumn{1}{l|}{EPS\_Earnings\_Surprise\_Backward\_Ave\_Diff} &
  \multicolumn{1}{l|}{8} \\ \hline
 &
  &
  &
  &
  &
  &
  \\ \hline
\multicolumn{1}{|l|}{\textbf{2015}} &
  \multicolumn{1}{l|}{\textbf{Highest Occurance}} &
  \multicolumn{1}{l|}{\textbf{Count}} &
  \multicolumn{1}{l|}{\textbf{Second Highest Occurance}} &
  \multicolumn{1}{l|}{\textbf{Count}} &
  \multicolumn{1}{l|}{\textbf{Third Highest Occurance}} &
  \multicolumn{1}{l|}{\textbf{Count}} \\ \hline
\multicolumn{1}{|l|}{\textbf{F1}} &
  \multicolumn{1}{l|}{EPS\_Earnings\_Surprise\_Backward\_Ave\_Diff} &
  \multicolumn{1}{l|}{69} &
  \multicolumn{1}{l|}{Gross\_Profit\_Q\_Change} &
  \multicolumn{1}{l|}{5} &
  \multicolumn{1}{l|}{DMA\_50D/200D} &
  \multicolumn{1}{l|}{4} \\ \hline
\multicolumn{1}{|l|}{\textbf{F2}} &
  \multicolumn{1}{l|}{DMA\_5D/200D} &
  \multicolumn{1}{l|}{16} &
  \multicolumn{1}{l|}{EPS\_Earnings\_Surprise\_Backward\_Ave\_Diff} &
  \multicolumn{1}{l|}{11} &
  \multicolumn{1}{l|}{EPS\_EarningsSurprise} &
  \multicolumn{1}{l|}{9} \\ \hline
\multicolumn{1}{|l|}{\textbf{F3}} &
  \multicolumn{1}{l|}{PS\_Ratios} &
  \multicolumn{1}{l|}{14} &
  \multicolumn{1}{l|}{EPS\_Earnings\_Surprise\_Backward\_Ave\_Diff} &
  \multicolumn{1}{l|}{11} &
  \multicolumn{1}{l|}{EPS\_EarningsSurprise} &
  \multicolumn{1}{l|}{9} \\ \hline
\multicolumn{1}{|l|}{\textbf{F4}} &
  \multicolumn{1}{l|}{DMA\_5D/200D} &
  \multicolumn{1}{l|}{9} &
  \multicolumn{1}{l|}{PS\_Ratios} &
  \multicolumn{1}{l|}{7} &
  \multicolumn{1}{l|}{EPS\_EarningsSurprise} &
  \multicolumn{1}{l|}{6} \\ \hline
\multicolumn{1}{|l|}{\textbf{F5}} &
  \multicolumn{1}{l|}{DMA\_5D/200D} &
  \multicolumn{1}{l|}{12} &
  \multicolumn{1}{l|}{PS\_Ratios} &
  \multicolumn{1}{l|}{11} &
  \multicolumn{1}{l|}{Gross\_Profit\_Q\_Change} &
  \multicolumn{1}{l|}{6} \\ \hline
 &
  &
  &
  &
  &
  &
  \\ \hline
\multicolumn{1}{|l|}{\textbf{2014}} &
  \multicolumn{1}{l|}{\textbf{Highest Occurance}} &
  \multicolumn{1}{l|}{\textbf{Count}} &
  \multicolumn{1}{l|}{\textbf{Second Highest Occurance}} &
  \multicolumn{1}{l|}{\textbf{Count}} &
  \multicolumn{1}{l|}{\textbf{Third Highest Occurance}} &
  \multicolumn{1}{l|}{\textbf{Count}} \\ \hline
\multicolumn{1}{|l|}{\textbf{F1}} &
  \multicolumn{1}{l|}{PS\_Ratios} &
  \multicolumn{1}{l|}{20} &
  \multicolumn{1}{l|}{DMA\_5D/200D} &
  \multicolumn{1}{l|}{15} &
  \multicolumn{1}{l|}{EPS\_EarningsSurprise} &
  \multicolumn{1}{l|}{12} \\ \hline
\multicolumn{1}{|l|}{\textbf{F2}} &
  \multicolumn{1}{l|}{PS\_Ratios} &
  \multicolumn{1}{l|}{17} &
  \multicolumn{1}{l|}{EPS\_EarningsSurprise} &
  \multicolumn{1}{l|}{14} &
  \multicolumn{1}{l|}{DMA\_50D/200D} &
  \multicolumn{1}{l|}{11} \\ \hline
\multicolumn{1}{|l|}{\textbf{F3}} &
  \multicolumn{1}{l|}{PS\_Ratios} &
  \multicolumn{1}{l|}{18} &
  \multicolumn{1}{l|}{EPS\_Earnings\_Surprise\_Backward\_Ave\_Diff} &
  \multicolumn{1}{l|}{15} &
  \multicolumn{1}{l|}{EPS\_EarningsSurprise} &
  \multicolumn{1}{l|}{9} \\ \hline
\multicolumn{1}{|l|}{\textbf{F4}} &
  \multicolumn{1}{l|}{DMA\_50D/200D} &
  \multicolumn{1}{l|}{11} &
  \multicolumn{1}{l|}{Return\_On\_Assets\_Y\_Change} &
  \multicolumn{1}{l|}{10} &
  \multicolumn{1}{l|}{EPS\_EarningsSurprise} &
  \multicolumn{1}{l|}{9} \\ \hline
\multicolumn{1}{|l|}{\textbf{F5}} &
  \multicolumn{1}{l|}{DMA\_50D/200D} &
  \multicolumn{1}{l|}{11} &
  \multicolumn{1}{l|}{Cash\_Q\_Change} &
  \multicolumn{1}{l|}{10} &
  \multicolumn{1}{l|}{PS\_Ratios} &
  \multicolumn{1}{l|}{8} \\ \hline
\end{tabular}%
}
\caption{Top five driving factors for the \textbf{Energy} stocks in each financial reporting year from 2014 to 2018. }
\label{tab:FactorOccuranceCount_Energy}
\end{table*}

\begin{table*}[ht]
\centering
\resizebox{0.9\textwidth}{!}{%
\begin{tabular}{lllllll}
\hline
\multicolumn{1}{|l|}{\textbf{2018}} &
  \multicolumn{1}{l|}{\textbf{Highest Occurance}} &
  \multicolumn{1}{l|}{\textbf{Count}} &
  \multicolumn{1}{l|}{\textbf{\begin{tabular}[c]{@{}l@{}}Second Highest\\   Occurance\end{tabular}}} &
  \multicolumn{1}{l|}{\textbf{Count}} &
  \multicolumn{1}{l|}{\textbf{Third Highest Occurance}} &
  \multicolumn{1}{l|}{\textbf{Count}} \\ \hline
\multicolumn{1}{|l|}{\textbf{F1}} &
  \multicolumn{1}{l|}{Short\_Term\_Debt\_Q\_Change} &
  \multicolumn{1}{l|}{38} &
  \multicolumn{1}{l|}{Return\_On\_Assets\_Y\_Change} &
  \multicolumn{1}{l|}{12} &
  \multicolumn{1}{l|}{Return\_On\_Assets} &
  \multicolumn{1}{l|}{3} \\ \hline
\multicolumn{1}{|l|}{\textbf{F2}} &
  \multicolumn{1}{l|}{Return\_On\_Assets\_Y\_Change} &
  \multicolumn{1}{l|}{19} &
  \multicolumn{1}{l|}{Short\_Term\_Debt\_Q\_Change} &
  \multicolumn{1}{l|}{18} &
  \multicolumn{1}{l|}{DMA\_50D/200D} &
  \multicolumn{1}{l|}{6} \\ \hline
\multicolumn{1}{|l|}{\textbf{F3}} &
  \multicolumn{1}{l|}{Return\_On\_Assets\_Y\_Change} &
  \multicolumn{1}{l|}{10} &
  \multicolumn{1}{l|}{Net\_Debt\_to\_EBIT\_Y\_Change} &
  \multicolumn{1}{l|}{6} &
  \multicolumn{1}{l|}{Net\_Debt\_to\_EBIT\_Y\_Change} &
  \multicolumn{1}{l|}{6} \\ \hline
\multicolumn{1}{|l|}{\textbf{F4}} &
  \multicolumn{1}{l|}{Short\_Term\_Debt\_Y\_Change} &
  \multicolumn{1}{l|}{8} &
  \multicolumn{1}{l|}{Return\_On\_Assets\_Y\_Change} &
  \multicolumn{1}{l|}{7} &
  \multicolumn{1}{l|}{Return\_On\_Assets\_Y\_Change} &
  \multicolumn{1}{l|}{7} \\ \hline
\multicolumn{1}{|l|}{\textbf{F5}} &
  \multicolumn{1}{l|}{Short\_Term\_Debt\_Y\_Change} &
  \multicolumn{1}{l|}{7} &
  \multicolumn{1}{l|}{Return\_On\_Assets\_Y\_Change} &
  \multicolumn{1}{l|}{6} &
  \multicolumn{1}{l|}{Short\_Term\_Debt\_Q\_Change} &
  \multicolumn{1}{l|}{5} \\ \hline
 &
  &
  &
  &
  &
  &
  \\ \hline
\multicolumn{1}{|l|}{\textbf{2017}} &
  \multicolumn{1}{l|}{\textbf{Highest Occurance}} &
  \multicolumn{1}{l|}{\textbf{Count}} &
  \multicolumn{1}{l|}{\textbf{Second Highest Occurance}} &
  \multicolumn{1}{l|}{\textbf{Count}} &
  \multicolumn{1}{l|}{\textbf{Third Highest Occurance}} &
  \multicolumn{1}{l|}{\textbf{Count}} \\ \hline
\multicolumn{1}{|l|}{\textbf{F1}} &
  \multicolumn{1}{l|}{Return\_On\_Assets\_Y\_Change} &
  \multicolumn{1}{l|}{36} &
  \multicolumn{1}{l|}{Dividend\_Yield\_Y\_Change} &
  \multicolumn{1}{l|}{28} &
  \multicolumn{1}{l|}{Short\_Term\_Debt\_Q\_Change} &
  \multicolumn{1}{l|}{8} \\ \hline
\multicolumn{1}{|l|}{\textbf{F2}} &
  \multicolumn{1}{l|}{Dividend\_Yield\_Y\_Change} &
  \multicolumn{1}{l|}{24} &
  \multicolumn{1}{l|}{Return\_On\_Assets\_Y\_Change} &
  \multicolumn{1}{l|}{18} &
  \multicolumn{1}{l|}{Short\_Term\_Debt\_Q\_Change} &
  \multicolumn{1}{l|}{12} \\ \hline
\multicolumn{1}{|l|}{\textbf{F3}} &
  \multicolumn{1}{l|}{Short\_Term\_Debt\_Q\_Change} &
  \multicolumn{1}{l|}{20} &
  \multicolumn{1}{l|}{Return\_On\_Assets\_Y\_Change} &
  \multicolumn{1}{l|}{11} &
  \multicolumn{1}{l|}{Dividend\_Yield\_Y\_Change} &
  \multicolumn{1}{l|}{8} \\ \hline
\multicolumn{1}{|l|}{\textbf{F4}} &
  \multicolumn{1}{l|}{Short\_Term\_Debt\_Q\_Change} &
  \multicolumn{1}{l|}{11} &
  \multicolumn{1}{l|}{Dividend\_Yield\_Y\_Change} &
  \multicolumn{1}{l|}{7} &
  \multicolumn{1}{l|}{Free\_Cash\_Flow\_Q\_Change} &
  \multicolumn{1}{l|}{6} \\ \hline
\multicolumn{1}{|l|}{\textbf{F5}} &
  \multicolumn{1}{l|}{Free\_Cash\_Flow\_Q\_Change} &
  \multicolumn{1}{l|}{11} &
  \multicolumn{1}{l|}{Short\_Term\_Debt\_Q\_Change} &
  \multicolumn{1}{l|}{9} &
  \multicolumn{1}{l|}{PC\_Ratios\_Y\_Change} &
  \multicolumn{1}{l|}{6} \\ \hline
 &
  &
  &
  &
  &
  &
  \\ \hline
\multicolumn{1}{|l|}{\textbf{2016}} &
  \multicolumn{1}{l|}{\textbf{Highest Occurance}} &
  \multicolumn{1}{l|}{\textbf{Count}} &
  \multicolumn{1}{l|}{\textbf{Second Highest Occurance}} &
  \multicolumn{1}{l|}{\textbf{Count}} &
  \multicolumn{1}{l|}{\textbf{Third Highest Occurance}} &
  \multicolumn{1}{l|}{\textbf{Count}} \\ \hline
\multicolumn{1}{|l|}{\textbf{F1}} &
  \multicolumn{1}{l|}{Short\_Term\_Debt\_Q\_Change} &
  \multicolumn{1}{l|}{20} &
  \multicolumn{1}{l|}{Operating\_Margin\_Y\_Change} &
  \multicolumn{1}{l|}{18} &
  \multicolumn{1}{l|}{Return\_On\_Assets\_Y\_Change} &
  \multicolumn{1}{l|}{14} \\ \hline
\multicolumn{1}{|l|}{\textbf{F2}} &
  \multicolumn{1}{l|}{Return\_On\_Assets\_Y\_Change} &
  \multicolumn{1}{l|}{18} &
  \multicolumn{1}{l|}{Operating\_Margin\_Y\_Change} &
  \multicolumn{1}{l|}{10} &
  \multicolumn{1}{l|}{Short\_Term\_Debt\_Q\_Change} &
  \multicolumn{1}{l|}{7} \\ \hline
\multicolumn{1}{|l|}{\textbf{F3}} &
  \multicolumn{1}{l|}{Return\_On\_Assets\_Y\_Change} &
  \multicolumn{1}{l|}{13} &
  \multicolumn{1}{l|}{Operating\_Margin\_Y\_Change} &
  \multicolumn{1}{l|}{8} &
  \multicolumn{1}{l|}{DMA\_5D/50D} &
  \multicolumn{1}{l|}{7} \\ \hline
\multicolumn{1}{|l|}{\textbf{F4}} &
  \multicolumn{1}{l|}{Operating\_Margin\_Y\_Change} &
  \multicolumn{1}{l|}{10} &
  \multicolumn{1}{l|}{Operating\_Margin\_Y\_Change} &
  \multicolumn{1}{l|}{10} &
  \multicolumn{1}{l|}{PC\_Ratios\_Y\_Change} &
  \multicolumn{1}{l|}{6} \\ \hline
\multicolumn{1}{|l|}{\textbf{F5}} &
  \multicolumn{1}{l|}{Short\_Term\_Debt\_Q\_Change} &
  \multicolumn{1}{l|}{6} &
  \multicolumn{1}{l|}{Short\_Term\_Debt\_Q\_Change} &
  \multicolumn{1}{l|}{6} &
  \multicolumn{1}{l|}{PC\_Ratios} &
  \multicolumn{1}{l|}{5} \\ \hline
 &
  &
  &
  &
  &
  &
  \\ \hline
\multicolumn{1}{|l|}{\textbf{2015}} &
  \multicolumn{1}{l|}{\textbf{Highest Occurance}} &
  \multicolumn{1}{l|}{\textbf{Count}} &
  \multicolumn{1}{l|}{\textbf{Second Highest Occurance}} &
  \multicolumn{1}{l|}{\textbf{Count}} &
  \multicolumn{1}{l|}{\textbf{Third Highest Occurance}} &
  \multicolumn{1}{l|}{\textbf{Count}} \\ \hline
\multicolumn{1}{|l|}{\textbf{F1}} &
  \multicolumn{1}{l|}{Short\_Term\_Debt\_Q\_Change} &
  \multicolumn{1}{l|}{21} &
  \multicolumn{1}{l|}{Return\_On\_Assets\_Y\_Change} &
  \multicolumn{1}{l|}{13} &
  \multicolumn{1}{l|}{DMA\_5D/50D} &
  \multicolumn{1}{l|}{12} \\ \hline
\multicolumn{1}{|l|}{\textbf{F2}} &
  \multicolumn{1}{l|}{Short\_Term\_Debt\_Q\_Change} &
  \multicolumn{1}{l|}{19} &
  \multicolumn{1}{l|}{Return\_On\_Assets\_Y\_Change} &
  \multicolumn{1}{l|}{12} &
  \multicolumn{1}{l|}{DMA\_5D/50D} &
  \multicolumn{1}{l|}{5} \\ \hline
\multicolumn{1}{|l|}{\textbf{F3}} &
  \multicolumn{1}{l|}{Return\_On\_Assets\_Y\_Change} &
  \multicolumn{1}{l|}{13} &
  \multicolumn{1}{l|}{DMA\_5D/50D} &
  \multicolumn{1}{l|}{9} &
  \multicolumn{1}{l|}{Short\_Term\_Debt\_Q\_Change} &
  \multicolumn{1}{l|}{8} \\ \hline
\multicolumn{1}{|l|}{\textbf{F4}} &
  \multicolumn{1}{l|}{DMA\_5D/50D} &
  \multicolumn{1}{l|}{12} &
  \multicolumn{1}{l|}{Short\_Term\_Debt\_Q\_Change} &
  \multicolumn{1}{l|}{7} &
  \multicolumn{1}{l|}{Short\_Term\_Debt\_Q\_Change} &
  \multicolumn{1}{l|}{7} \\ \hline
\multicolumn{1}{|l|}{\textbf{F5}} &
  \multicolumn{1}{l|}{PC\_Ratios} &
  \multicolumn{1}{l|}{6} &
  \multicolumn{1}{l|}{PC\_Ratios} &
  \multicolumn{1}{l|}{6} &
  \multicolumn{1}{l|}{PC\_Ratios} &
  \multicolumn{1}{l|}{6} \\ \hline
 &
  &
  &
  &
  &
  &
  \\ \hline
\multicolumn{1}{|l|}{\textbf{2014}} &
  \multicolumn{1}{l|}{\textbf{Highest Occurance}} &
  \multicolumn{1}{l|}{\textbf{Count}} &
  \multicolumn{1}{l|}{\textbf{Second Highest Occurance}} &
  \multicolumn{1}{l|}{\textbf{Count}} &
  \multicolumn{1}{l|}{\textbf{Third Highest Occurance}} &
  \multicolumn{1}{l|}{\textbf{Count}} \\ \hline
\multicolumn{1}{|l|}{\textbf{F1}} &
  \multicolumn{1}{l|}{PC\_Ratios} &
  \multicolumn{1}{l|}{12} &
  \multicolumn{1}{l|}{Short\_Term\_Debt\_Q\_Change} &
  \multicolumn{1}{l|}{10} &
  \multicolumn{1}{l|}{Free\_Cash\_Flow\_Q\_Change} &
  \multicolumn{1}{l|}{8} \\ \hline
\multicolumn{1}{|l|}{\textbf{F2}} &
  \multicolumn{1}{l|}{Short\_Term\_Debt\_Q\_Change} &
  \multicolumn{1}{l|}{7} &
  \multicolumn{1}{l|}{Free\_Cash\_Flow\_Q\_Change} &
  \multicolumn{1}{l|}{5} &
  \multicolumn{1}{l|}{Free\_Cash\_Flow\_Q\_Change} &
  \multicolumn{1}{l|}{5} \\ \hline
\multicolumn{1}{|l|}{\textbf{F3}} &
  \multicolumn{1}{l|}{Free\_Cash\_Flow\_Q\_Change} &
  \multicolumn{1}{l|}{5} &
  \multicolumn{1}{l|}{Free\_Cash\_Flow\_Q\_Change} &
  \multicolumn{1}{l|}{5} &
  \multicolumn{1}{l|}{Net\_Debt\_to\_EBIT\_Y\_Change} &
  \multicolumn{1}{l|}{4} \\ \hline
\multicolumn{1}{|l|}{\textbf{F4}} &
  \multicolumn{1}{l|}{Short\_Term\_Debt\_Q\_Change} &
  \multicolumn{1}{l|}{6} &
  \multicolumn{1}{l|}{Short\_Term\_Debt\_Q\_Change} &
  \multicolumn{1}{l|}{6} &
  \multicolumn{1}{l|}{Cost\_Of\_Revenue\_Q\_Change} &
  \multicolumn{1}{l|}{5} \\ \hline
\multicolumn{1}{|l|}{\textbf{F5}} &
  \multicolumn{1}{l|}{Short\_Term\_Debt\_Q\_Change} &
  \multicolumn{1}{l|}{5} &
  \multicolumn{1}{l|}{PC\_Ratios} &
  \multicolumn{1}{l|}{4} &
  \multicolumn{1}{l|}{PC\_Ratios} &
  \multicolumn{1}{l|}{4} \\ \hline
\end{tabular}%
}
\caption{Top five driving factors for the \textbf{Utilities} stocks in each financial reporting year from 2014 to 2018. }
\label{tab:FactorOccuranceCount_Utilities}
\end{table*}

%Filters bibliography
%\printbibliography[heading=subbibintoc,type=article,title={Articles only}]
%\printbibliography[type=book,title={Books only}]

%\printbibliography[keyword={physics},title={Physics-related only}]
%\printbibliography[keyword={latex},title={\LaTeX-related only}]

\end{document}